\documentclass[
    reprint,twocolumn,
    aps,prx,10pt,amsmath,amssymb,
    longbibliography,superscriptaddress,
    showpacs,preprintnumbers,
    nofootinbib,
]{revtex4-2}

\usepackage{jmlstyle}
\usepackage{jmltikz}

\myexternaldocument{supp}{}

\allowdisplaybreaks

\makeatletter
\def\maketitle{
\@author@finish
\title@column\titleblock@produce
\suppressfloats[t]}
\makeatother

\newcommand{\RA}{{\mathrm{R\!/\! A}}}
\newcommand{\nsp}[0]{\mspace{-2mu}}  
\newcommand{\Cpr}[0]{C'\nsp}  
\newcommand{\cpr}[0]{c'\nsp}

\newcommand{\eph}[0]{e-ph\xspace}
\newcommand{\EPjl}[0]{\textsc{ElectronPhonon.jl}\xspace}
\newcommand{\scGD}[0]{sc$GD_0$\xspace}
\newcommand{\GD}[0]{$G_0 D_0$\xspace}
\newcommand{\opmbJ}[0]{\hat{\mathbf{J}}}

\begin{document}

\title{Beyond-quasiparticle transport with vertex correction:\\ self-consistent ladder formalism for electron-phonon interactions}

\author{Jae-Mo Lihm}
\email{jaemo.lihm@gmail.com}
\affiliation{%
European Theoretical Spectroscopy Facility and Institute of Condensed Matter and Nanosciences, Université catholique de Louvain, Chemin des Étoiles 8, B-1348 Louvain-la-Neuve, Belgium
}%
\author{Samuel Ponc\'e}
\email{samuel.ponce@uclouvain.be}
\affiliation{%
European Theoretical Spectroscopy Facility and Institute of Condensed Matter and Nanosciences, Université catholique de Louvain, Chemin des Étoiles 8, B-1348 Louvain-la-Neuve, Belgium
}%
\affiliation{%
WEL Research Institute, avenue Pasteur 6, 1300 Wavre, Belgium
}%
\date{March 9, 2026}

\begin{abstract}
We present a self-consistent many-body framework for computing phonon-limited electronic transport from first principles, incorporating both beyond-quasiparticle effects and vertex corrections.
Using the recently developed first-principles \scGD method, we calculate spectral functions with nonperturbative effects such as broadening, satellites, and energy-dependent renormalization.
We show that the \scGD spectral functions outperform one-shot \GD and cumulant approximations in model Hamiltonians and real materials, eliminating unphysical spectral kinks and correctly predicting the phonon emission continuum.
Building on this, we introduce the self-consistent ladder formalism for transport, which captures vertex corrections due to electron-phonon interactions.
This approach unifies and improves upon the two state-of-the-art approaches for first-principles phonon-limited transport: the bubble approximation and the Boltzmann transport equation.
Moreover, as a charge-conserving approximation, it enables consistent calculations of the optical conductivity and dielectric function.
We validate the developed method against numerically exact results for model Hamiltonians in the dilute polaronic limit and apply it to real materials.
Our results show quantitative agreement with the experimental dc conductivities in intrinsic semiconductors Si and ZnO and the SrVO$_3$ metal, as well as excellent agreement with the experimental THz optical and dielectric properties of Si and ZnO.
This work unifies first-principles and many-body approaches for studying transport, opening new directions for applying many-body theory to materials with strong electron-phonon interactions.
\end{abstract}

\maketitle

\section{Introduction}

Electrical conductivity is a fundamental material property with direct implications for electronic, optical, thermoelectric, and photovoltaic applications.
At room temperature, electron-phonon (\eph) coupling often limits charge transport, placing an upper bound on a material's intrinsic conductivity~\cite{Giustino2017, Ponce2020Review}.
Predictive calculations of phonon-limited transport are essential for understanding the fundamental physics of materials and for designing compounds with tailored properties.

First-principles calculations with the \textit{ab initio} Boltzmann transport equation (BTE)~\cite{Ponce2020Review} have become a powerful tool for computing phonon-limited conductivities in semiconductors~\cite{Restrepo2009, Zhou2016, Ponce2018, Macheda2018, Zhou2018, Schlip2018MAPI, Brunin2020PRB, Macheda2020, Ponce2020Review, Park2020, Ponce2021, Desai2021, Claes2022, Ponce2023PRL, Ponce2023PRB, Lihm2024Piezo, Roisin2024, Zhou2024} and metals~\cite{Mustafa2016, Desai2023, Goudreault2024, Ponce2016EPW}.
This formalism has been implemented in several open-source software packages~\cite{Ponce2016EPW, Lee2023EPW, Zhou2021Perturbo, Gonze2020Abinit, Cepellotti2022Phoebe, Protik2022elphbolt} and is widely used in the computational materials community~\cite{Claes2025, Park2025Review}.
The BTE describes transport through occupation functions determined by a balance between electric-field-driving, scattering-out, and scattering-in processes.
Crucially, the scattering-in term generates vertex corrections, which significantly influence carrier mobility in real materials~\cite{Ponce2020Review, Ponce2021, Desai2021, Claes2022}.

Despite these successes, the assumption of a well-defined quasiparticle band structure underlying the BTE breaks down in many materials.
Spectral functions measured by angle-resolved photoemission spectroscopy (ARPES) reveal photoemission kinks in metals~\cite{Valla1999, Eiguren2003} and satellite peaks in polar semiconductors~\cite{Moser2013, Chen2015, Cancellieri2016, Wang2016, Riley2018}, often interpreted as signatures of large polarons~\cite{Franchini2021}.
Even in the absence of satellites, quasiparticle peaks may be substantially broadened, necessitating a treatment that goes beyond the semiclassical picture~\cite{Zhou2019STO, Pickem2022}.
First-principles calculations based on Green's functions have successfully captured these \eph-induced spectral renormalizations~\cite{Verdi2017, Nery2018, Antonius2020, Caruso2020Review, DeAbreu2022, Lihm2025, Nery2022, Houtput2025, Zappacosta2025}.

However, computing transport properties in the regime beyond long-lived quasiparticles, while incorporating vertex corrections, remains a major challenge in \textit{ab initio} \eph physics.
Recent efforts using \textit{ab initio} spectral functions to compute conductivity~\cite{Zhou2019STO, Chang2022, Abramovitch2023, Abramovitch2024} rely on the bubble approximation, which neglects vertex corrections.
Although nonperturbative~\cite{Troisi2006, Fratini2016, Frost2017, Fetherolf2020} or numerically exact~\cite{Marsiglio1993, Fehske2000, Hohenadler2004, Schubert2005, Fehske2007, Mishchenko2015DMC, Mischenko2019DMC, Wang2022, Jankovic2023, Miladic2023, Jankovic2024Holstein, Rammal2024, Jankovic2025a, Jankovic2025b} methods can address this problem, they are mostly limited to simplified models, with a very recent exception of Ref.~\cite{Luo2025DiagMC}.
Finding a balance between computational feasibility and physical accuracy is key to predictive transport calculations in real materials.

\begin{figure*}[tb]
\centering
\includegraphics[width=0.99\linewidth]{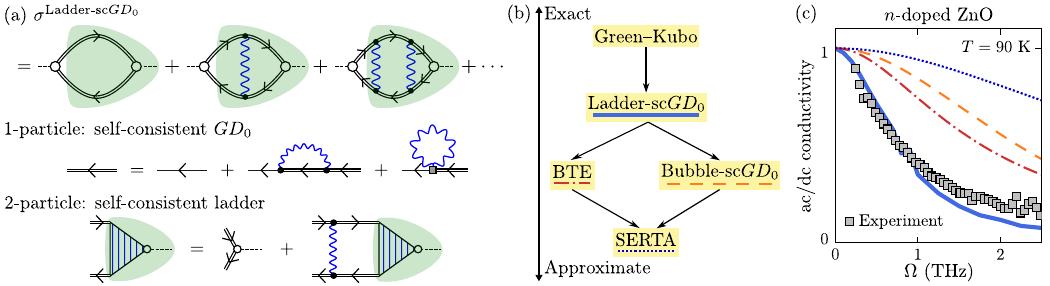}
\caption{
    Overview of the ladder-\scGD transport formalism.
    (a) Feynman diagrams for the self-consistent $GD_0$ (\scGD) self-energy and the self-consistent ladder equation.
    Single solid lines show bare electron Green's functions; double solid lines show dressed ones.
    Blue wavy lines indicate the bare phonon propagator, black dots represent the \eph coupling, gray squares indicate Debye--Waller coupling, and white circles represent coupling to external electric fields.
    Green shaded regions indicate the renormalized current vertex function.
    (b) Hierarchy of linear-response \eph transport methods.
    The Boltzmann transport equation (BTE)~\cite{Ponce2020Review}, self-energy relaxation time approximation (SERTA)~\cite{Zhou2016, Ponce2018}, and the bubble approximation~\cite{Basov2011RMP, Zhou2019STO} can be derived as approximations to ladder-\scGD.
    (c) Real part of the normalized ac conductivity of $n$-doped ZnO at $T=90~\mathrm{K}$ with calculations compared to experimental data~\cite{Baxter2009} (see also \Fig{fig:ZnO_ac}).
    Ladder-\scGD accurately captures the frequency dependence, whereas other methods yield a decay that is too slow with frequency.
}
\label{fig:overview}
\end{figure*}

Bridging this gap requires a framework that unifies the Green's functions formalism and BTE while remaining computationally tractable.
Formally, vertex corrections to conductivity can be obtained from the summation of ladder-type Feynman diagrams~\cite{Edwards1958, Engelsberg1963, Holstein1964, Prange1964, MahanBook}.
This ladder approximation formally reconciles the BTE with many-body theory and yields the BTE in the limit of weak \eph renormalization~\cite{Holstein1964, Prange1964, Hansch1983, MahanBook, Kim2019Vertex}.
Despite its conceptual appeal, the ladder approximation has not yet been applied in first-principles transport due to several hurdles.
The vertex function, which is the fundamental quantity in the ladder approximation, is a two-particle response function that goes beyond single-particle descriptions of the spectral functions.
This formal complexity translates into significant numerical challenges in computing the vertex functions.
Furthermore, realistic \textit{ab initio} calculations must account for multiband electronic structures with nonlocal and long-ranged \eph interactions, which are features not present in model Hamiltonians.

In this work, we address these longstanding challenges by developing an \textit{ab initio} many-body framework for calculating phonon-limited transport.
Building on our previous work employing self-consistent $GD_0$ (\scGD) Green's functions~\cite{Lihm2025}, we introduce the ladder approximation within the equilibrium Keldysh formalism, leading to what we term the ladder-\scGD method illustrated in \Fig{fig:overview}(a).
The ladder-\scGD method simultaneously captures vertex corrections and beyond-quasiparticle effects, thereby advancing beyond the state-of-the-art methods, namely the BTE and the bubble approximation.
This hierarchy of linear-response \eph transport methods is shown in \Fig{fig:overview}(b).
We also derive a relation between vertex-corrected conductivity and dielectric function, revealing an often overlooked phonon-assisted current arising from nonlocal \eph interactions.
The ladder-\scGD method produces accurate conductivities, as illustrated for the ac conductivity of ZnO in \Fig{fig:overview}(c); further applications to model Hamiltonians, intrinsic Si and ZnO, and metallic SrVO$_3$ are also presented in this work.
In addition, our method accurately captures the THz optical absorption and refractive index in Si and ZnO, even when previous state-of-the-art methods fall short.
As a charge-conserving approximation, ladder-\scGD constitutes the first \textit{ab initio} approach to fulfill the Ward identity.

Although the ladder vertex correction from \eph coupling is a well-established concept~\cite{Edwards1958, Engelsberg1963, Holstein1964, Prange1964, MahanBook, MahanBook}, its application to real-frequency calculations has been limited by the complexity of its analytic structure and has generally required additional weak-coupling and isotropic approximations~\cite{Scher1970, Allen1971, Kupcic2014, Kupcic2015}.
The explicit Keldysh formulation presented in this work considerably simplifies the equations and makes their implementation more straightforward.
Within this framework, the phonon-assisted current emerges naturally; contrary to studies based on single-band models~\cite{Gosar1966, Gosar1970, Hannewald2004}, our derivation is fully general and clarifies the role of covariant derivatives and their connection to the velocity matrix.
Based on our theoretical and numerical developments, we evaluate the real-frequency ladder equation without additional approximations, providing a complete calculation of the ladder conductivity for both model Hamiltonians and real materials.
Our ladder-\scGD method and its open-source implementation~\cite{EPjl, MaterialsCloudArchive} offer an accurate and practical tool for both model studies and material applications.

This paper is structured as follows.
In \Sec{sec:selfen}, we introduce the \eph Hamiltonian, the Keldysh Green's functions, and the \scGD self-energy.
We compute the \scGD spectral functions for several model Hamiltonians and real materials, which will serve as the starting point for the transport calculations.
We derive the ladder-\scGD formalism as a linear response to the \scGD self-energy in \Sec{sec:lr} and apply it to the transport problem in \Sec{sec:transport}.
We validate our method against numerically exact benchmarks for model systems and experimental data for real materials.
In \Sec{sec:dielectric}, we study dielectric functions using ladder-\scGD and demonstrate their consistency with optical conductivities.
In \Sec{sec:conclusion}, we summarize our findings and discuss future directions.

\vspace{-0.1em}

\section{Self-consistent electron self-energy} \label{sec:selfen}
\vspace{-0.1em}

In this section, we introduce the \eph Hamiltonian and Keldysh Green's functions for electrons and phonons.
Then, we review the \scGD method for the self-energy~\cite{Capone2003, Bauer2011, Sakkinen2015A, Sakkinen2015B, Rademaker2016SCBA, Chen2016SCBA, Esterlis2018SCBA, Dee2019SCBA, Dee2020SCBA, Mitric2022, Mitric2023, Stefanucci2023} that we introduced to \textit{ab initio} \eph calculations in a previous work~\cite{Lihm2025}.
We apply this method to model Hamiltonians and real materials.
These results serve as the starting point for the self-consistent theory of transport developed in the following sections.

\vspace{-0.3em}

\subsection{The electron-phonon Hamiltonian}
\vspace{-0.3em}

We consider the standard \eph Hamiltonian~\cite{Giustino2017}
\begin{subequations} \label{eq:H_def}
\begin{align}
    \hat{H}   &= \hat{H}_0 + \hat{H}^{\rm ep}\,,\\
    \hat{H}_0 &= \sum_\nk (\veps_\nk - \mu) \hat{c}_\nk^\dagger \hat{c}_\nk
    + \sum_\nuq \omega_\nuq (\hat{a}^\dagger_\nuq \hat{a}_\nuq \!+\! \tfrac{1}{2}) \,, \\
    \hat{H}^{\rm ep}
    &= \frac{1}{\sqrt{N_\bk}} \sum_{\substack{\bk\bq \\ mn\nu}} g_{mn\nu}(\bk, \bq) \,
    \hat{c}^\dagger_\mkq \, \hat{c}_\nk \, \hat{x}_\nuq \,,
\end{align}
\end{subequations}
where $\veps_\nk$ is the electron energy at wavevector $\bk$ and band $n$,
$\omega_\nuq$ the phonon energy at wavevector $\bq$ and mode $\nu$,
$g_{mn\nu}(\bk, \bq)$ the \eph coupling,
$\mu$ the chemical potential,
$N_\bk$ the number of $\bk$ points,
$\hat{c}_\nk$ ($\hat{a}_\nuq$) the electron (phonon) annihilation operator,
and
\begin{equation} \label{eq:u_def}
    \hat{x}_\nuq = \hat{a}_\nuq + \hat{a}^\dagger_{\nu -\bq}
\end{equation}
the phonon position operator in the eigenmode basis.
We set $k_{\rm B} = \hbar = 1$ throughout.

We write the Bose--Einstein occupation function as
\begin{equation}
    n(\omega) = \frac{1}{e^{\beta\omega} - 1} \,,
\end{equation}
where $\beta = 1 / T$ is the inverse temperature,
and the Fermi--Dirac electron and hole occupation functions as
\begin{equation}
    f^+(\veps) = \frac{1}{e^{\beta(\veps - \mu)} + 1},\ f^-(\veps) = 1 - f^+(\veps)\,.
\end{equation}
We define the bare phonon and electron occupation as
\begin{equation}
    n_\nuq \equiv n(\omega_\nuq)\,, \ %
    f^\pm_\nk \equiv f^\pm(\veps_\nk) \,.
\end{equation}

Throughout this work, we frequently treat energy and momentum as composite variables for notational simplicity.
Specifically, we define $k \equiv (\veps, \bk)$ for electrons, $q \equiv (\omega, \bq)$ for phonons, and $Q \equiv (\Omega, \bQ)$ for external perturbations, and use them exclusively for these respective contexts.
For convenience, we also define the Brillouin-zone integral $\intbk$ (and analogously $\intbq$) as
\begin{equation}
    \intbk f(\bk) \equiv \nint \frac{\dd \bk}{V^{\rm BZ}} f(\bk) = \frac{1}{N_\bk} \sum_\bk f(\bk) \,,
\end{equation}
where $V^{\rm BZ}$ is the volume of the Brillouin zone.
When integrating both energy and momentum, we use the shorthand $\intkk$ (and similarly $\intqq$), defined as
\begin{equation} \label{eq:intkk_def}
    \intkk \equiv \nint \frac{\dd\veps}{2\pi i} \, \nint \frac{\dd\bk}{V^{\rm BZ}} \,.
\end{equation}

\subsection{Keldysh Green's functions and self-energies} \label{sec:selfen_keldysh}

\begin{figure}[tb]
    \centering
    \includegraphics[width=0.99\linewidth]{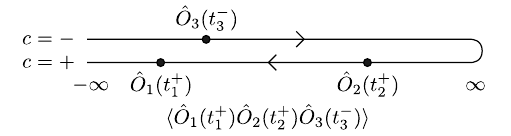}
    \caption{
    Time ordering on the Schwinger--Keldysh contour.
    }
    \label{fig:contour}
\end{figure}

The spectral properties of an \eph system are captured by its single-particle Green's functions and self-energies.
Although these functions are often calculated in the imaginary-time domain via the Matsubara formalism, accessing real-frequency observables such as spectral functions and conductivities requires a numerically ill-posed analytic continuation~\cite{Gunnarsson2010,Goulko2017,Miladic2023}.
To avoid this difficulty, we instead adopt the equilibrium Keldysh formalism~\cite{Schwinger1960, Keldysh1965}, which enables direct calculations at real frequencies and finite temperatures.
In this framework, operators are ordered along the Schwinger--Keldysh contour, a time path consisting of two branches: a forward branch ($c = -$) and a backward branch ($c = +$).
Operators on the forward branch are time-ordered, while those on the backward branch are anti-time-ordered, as illustrated in \Fig{fig:contour}.

In the Keldysh formalism, the electron Green's function is defined as a contour-ordered two-point correlation function of annihilation and creation operators with contour-time arguments $t_1^{c_1}$ and $t_2^{c_2}$~\cite{Kugler2021}:
\begin{equation} \label{eq:G_k_kp}
    G_{n_1\bk_1, n_2\bk_2}^{c_1c_2}(t_1, t_2)
    \equiv -i \expval{ \mcT \, \opc_{n_1\bk_1}(t_1^{c_1}) \, \opcd_{n_2\bk_2}(t_2^{c_2})} \,.
\end{equation}
Here, $\mcT$ denotes the time ordering along the Schwinger--Keldysh contour.
The operators evolve in the Heisenberg picture, $\hat{c}_\nk(t^c) = e^{i\hat{H} t} \, \hat{c}_\nk \, e^{-i\hat{H} t}$,
and are identical on the two branches: $\hat{c}_\nk(t^+) = \hat{c}_\nk(t^-)$.
The expectation value in Eq.~\eqref{eq:G_k_kp} is taken with respect to the equilibrium density matrix at temperature $T = 1/\beta$:
\begin{equation}
    \expval{\hat{A}} = \frac{\mathrm{Tr} \, e^{-\beta \hat{H}} \hat{A}}{\mathrm{Tr} \, e^{-\beta \hat{H}}}\,.
\end{equation}
Assuming temporal and discrete spatial translational symmetries, the Green's function becomes
\begin{equation} \label{eq:G_t}
    G^{c_1 c_2}_{n_1 n_2 \bk}(t_1 - t_2) = -i \expval{\mcT \, \opc_{n_1\bk}(t_1^{c_1}) \, \opcd_{n_2\bk}(t_2^{c_2})} \,,
\end{equation}
with the Fourier transform
\begin{equation} \label{eq:G_w}
    G^{c_1 c_2}_{n_1 n_2 \bk}(\veps) = \nint \dd t \, G^{c_1 c_2}_{n_1 n_2 \bk}(t) \, e^{i (\veps - \mu) t} \,.
\end{equation}
We include the chemical potential $\mu$ in the Fourier factor to make the bare Green's functions peaked at the corresponding bare band energy [see \Eq{eq:Gbare}].
For brevity, we use numeric subscripts to denote the band index and the contour index together:
\begin{equation}
    G_{12}(k) \equiv G^{c_1 c_2}_{n_1 n_2 \bk}(\veps) \,.
\end{equation}

\begin{figure}[tb]
\centering
\includegraphics[width=0.99\linewidth]{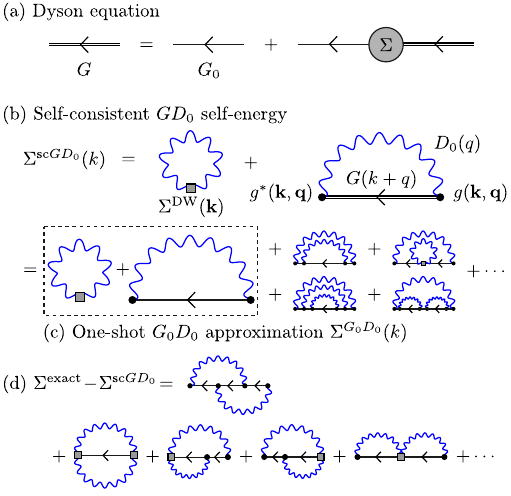}
\caption{
    Feynman diagrams of the (a) Dyson equation for the Green's function, (b) self-consistent $GD_0$ (\scGD) self-energy including the Debye--Waller (DW) term, (c) one-shot \GD self-energy, and (d) low-order diagrams not included in the \scGD approximation.
}
\label{fig:feynman_selfen}
\end{figure}

The four components of the contour-ordered Green's function are the time-ordered, lesser, greater, and anti-time-ordered Green's functions:
\begin{equation} \label{eq:G_matrix}
    \begin{pmatrix}
        G^{\rm T} & G^{<} \\ G^{>} & G^{\rm \bar{T}}
    \end{pmatrix}
    =
    \begin{pmatrix}
        G^{--} & G^{-+} \\ G^{+-} & G^{++}
    \end{pmatrix} \,,
\end{equation}
where we have omitted the subscripts and arguments.
Their linear combinations yield the retarded (R) and advanced (A) Green's functions
\begin{equation} \label{eq:G_RA}
    G^{\RA} = \tfrac{1}{2} \bigl( G^{--} \pm G^{+-} \mp G^{-+} - G^{++} \bigr) \, .
\end{equation}
The electron spectral function is defined as
\begin{equation} \label{eq:A}
    A_{n_1 n_2 \bk}(\veps) \equiv \frac{i}{2\pi} \Bigl[ G^{\rm R}_{n_1 n_2 \bk}(\veps) - G^{\rm A}_{n_1 n_2 \bk}(\veps) \Bigr]
\end{equation}
and satisfies
\begin{equation} \label{eq:sc_G_KK}
    G^{\rm R}_{n_1 n_2 \bk}(\veps)
    = \nint \dd\veps' \, \frac{
        A_{n_1 n_2 \bk}(\veps')
    }{\veps - \veps' + i0^+} \,,
\end{equation}
where $0^+$ denotes a positive infinitesimal.
All components of the Green's function can be written in terms of the spectral function using the fermionic fluctuation-dissipation relations~\cite{BruusBook, StefanucciBook}: see \Eq{eq:sc_fdt_G} for details.
The renormalized single-particle density matrix is defined as
\begin{equation} \label{eq:sc_rho_def}
    \rho_{n_1 n_2 \bk} \equiv \nint \dd\veps \, A_{n_1 n_2 \bk}(\veps) f^+(\veps) \,.
\end{equation}

When phonon-induced interband hybridization is negligible, the Green's function and self-energy become diagonal in the band index.
This approximation is commonly applied for \textit{ab initio} calculations~\cite{Giustino2017, Ponce2020Review}.
The diagonal self-energy directly renormalizes the quasiparticle energy as $\veps_n \to \veps_n + \Sigma_{nn}$.
In contrast, the off-diagonal self-energy modifies the quasiparticle energy only at second order, $\veps_n \to \veps_n + \abs{\Sigma_{mn}}^2 / (\veps_m - \veps_n)$.
Because the interband energy spacing is typically much larger than the phonon-induced self-energy, the diagonal approximation is valid in most systems.
The off-diagonal hybridization becomes important when the band gap is comparable to the self-energy, as in phonon-induced transitions between topological and normal insulators~\cite{Lihm2020}, or when multiple bands cross the Fermi level at the same $\bk$ point, as in Weyl semimetals~\cite{BrousseauCouture2020}.
Within the diagonal approximation, the spectral function and retarded Green's function simplify to
\begin{subequations} \label{eq:sc_G_diag_approx}
\begin{align}
    A_{nn\bk}(\veps)
    &= -\frac{1}{\pi} \Im G^{\rm R}_{nn\bk}(\veps) \,,
    \\
    G^{\rm R}_{nn\bk}(\veps)
    &= \nint \dd\veps' \, \frac{A_{nn\bk}(\veps')}{\veps - \veps' + i0^+} \,.
\end{align}
\end{subequations}

For the bare Green's function $G_0$, we only write a single band index as it is always diagonal in the band basis:
\begin{equation}
    G^{c_1 c_2}_{0, n_1 \bk}(\veps) = G^{c_1 c_2}_{0, n_1 n_2 \bk}(\veps) \, \delta_{n_1 n_2} \,.
\end{equation}
The bare spectral function and retarded Green's function read
\begin{subequations} \label{eq:Gbare}
\begin{align}
    A_{0,\nk}(\veps) &= \delta(\veps - \veps_\nk) \,,
    \\
    \label{eq:Gbare_GR}
    G^{\rm R}_{0,\nk}(\veps) &= \frac{1}{\veps - \veps_\nk + i0^+} \,.
\end{align}
\end{subequations}
Note that we included $\mu$ in the Fourier factor for Green's functions in \Eq{eq:G_w}, so that $\mu$ does not explicitly appear in \Eq{eq:Gbare_GR}.
The inverse bare Green's function reads
\begin{equation} \label{eq:Gbare_inv}
    G^{-1}_{0,12}(k)
    = (\veps - \veps_{n_1 \bk} \delta_{n_1 n_2}) Z^{c_1 c_2} \,,
\end{equation}
where $Z^{c_1 c_2} = \delta_{c_1 c_2} (-1)^{\delta_{c_1 +}}$ is the sign factor from the contour ordering.

The electron self-energy $\Sigma_{12}(k)$ relates the bare and renormalized Green's functions through the Dyson equation
\begin{equation} \label{eq:sc_Dyson}
    G_{12}(k)
    = G_{0, 12}(k)
    + \sum_{1' 2'} G_{0, 11'}(k) \, \Sigma_{1'2'}(k) \, G_{2'2}(k) \,,
\end{equation}
which is illustrated in \Fig{fig:feynman_selfen}(a).
This equation has the structure of a matrix-matrix product that involves summation over both the band and the contour indices.
Explicitly, the Dyson equation reads
\begin{multline} \label{eq:sc_Dyson_verbose}
    G^{c_1 c_2}_{n_1 n_2}(k)
    = G^{c_1 c_2}_{0, n_1}(k) \, \delta_{n_1 n_2}
    \\
    + \sum_{c_1' c_2' n_2'} G^{c_1 c_1'}_{0, n_1}(k) \, \Sigma^{c_1' c_2'}_{n_1 n_2'}(k) \, G^{c_2' c_2}_{n_2' n_2}(k) \,.
\end{multline}
The Dyson equation can also be written in the inverse form:
\begin{equation} \label{eq:sc_Dyson_inv}
    G^{-1}_{12}(k) = G^{-1}_{0, 12}(k) - \Sigma_{12}(k) \,.
\end{equation}

The greater, lesser, retarded, and advanced components of the electron self-energy are defined as
\begin{subequations} \label{eq:Sigma_RA}
\begin{align}
    \Sigma^{\gtrless} &= -\Sigma^{\pm \mp} \,, \\
    \Sigma^\RA &= \tfrac{1}{2} \bigl( \Sigma^{--} \mp \Sigma^{+-} \pm \Sigma^{-+} - \Sigma^{++} \bigr) \,,
\end{align}
\end{subequations}
where the subscripts and argument are omitted.
Note that the signs of $\Sigma^{+-}$ and $\Sigma^{-+}$ are opposite to those in the corresponding Green's function definitions in \Eqs{eq:G_matrix} and \eqref{eq:G_RA}.
The electron self-energy obeys the same fermionic fluctuation-dissipation relation [\Eq{eq:sc_fdt_G}] as the Green's function.
The Dyson equation for the retarded or advanced Green's functions are decoupled in the contour indices:
\begin{align} \label{eq:Dyson_RA}
    G^{\RA}_{n_1 n_2 \bk}(\veps)
    &= G^\RA_{0, n_1 \bk}(\veps) \, \delta_{n_1 n_2}
    \nnnl
    &+ \sum_{n_2'} G^\RA_{0, n_1 \bk}(\veps) \, \Sigma^{\RA}_{n_1 n_2' \bk}(\veps) \, G^{\RA}_{n_2' n_2 \bk}(\veps) \,.
\end{align}
For the other components, the Dyson equation retains its full $2 \times 2$ matrix structure in contour indices.

We similarly define the bare phonon Green's functions:
\begin{equation} \label{eq:D_t}
    D^{c_1 c_2}_{0,\nuq}(t_1 - t_2) = -i \expval{\mcT \, \hat{x}_\nuq(t_1^{c_1}) \hat{x}_\nuq^\dagger(t_2^{c_2})} \,.
\end{equation}
The Fourier transform and the definitions of the various Green's function components follow analogously to the electronic case, and the explicit expressions are given in \Eq{eq:D0}.
Throughout this work, we neglect the renormalization of the phonon Green's function and use only its bare form, computed from density functional perturbation theory (DFPT).
For lightly doped semiconductors, the number of free carriers is insufficient to renormalize the phonons, and the $D_0$ approximation is therefore valid.
This approximation may break down in heavily doped semiconductors, where free carriers form plasmons that can resonantly couple to phonons, leading to strong renormalization.
An \textit{ab initio} description of such plasmon-phonon hybridization remains an active area of research~\cite{Macheda2024, Lihm2024PlPh, Krsnik2024}.
For metals, the plasmon energy lies well above the phonon energy scale, so the adiabatic DFPT treatment of phonons is again justified.
It has been numerically shown that the nonadiabatic renormalization occurs only within small regions of the Brillouin zone~\cite{Berges2023}.

\subsection{Self-consistent \texorpdfstring{$GD_0$}{GD0} self-energy} \label{sec:selfen_theory}

With the Green's functions now defined, we can introduce approximations for the self-energy.
In this work, we adopt the self-consistent $GD_0$ (\scGD) approximation~\cite{Capone2003, Bauer2011, Sakkinen2015A, Sakkinen2015B, Rademaker2016SCBA, Chen2016SCBA, Esterlis2018SCBA, Dee2019SCBA, Dee2020SCBA, Mitric2022, Mitric2023, Stefanucci2023, Lihm2025}, where the self-energy is computed from a one-loop diagram involving a dressed electron Green's function and a bare phonon propagator [\Fig{fig:feynman_selfen}(b)].
This approximation neglects the crossing diagrams, which are relevant for electron lifetimes due to two-phonon scattering~\cite{Lee2020TwoPh, Hatanpaa2023, Esho2023, Sun2023}, as well as nonperturbative effects from nonlinear \eph coupling~\cite{Nery2022, Houtput2025, Zappacosta2025}.
For example, the one-electron-two-phonon matrix elements (gray squares in Fig.~3) are treated at the mean-field level [first term in Fig.~3(b)], while higher-order diagrams are neglected [second line of Fig.~3(d)], since the \scGD method treats only the linear \eph coupling $g$ nonperturbatively.

By evaluating the Feynman diagram in \Fig{fig:feynman_selfen}(b), one obtains the \scGD self-energy:
\begin{align} \label{eq:sc_Sigma_c}
    \Sigma^{{\rm sc}GD_0}_{12}(k)
    &= \Sigma^{\rm DW}_{12}(\bk) - Z^{c_1} Z^{c_2} \!\! \sum_{\nu n_3 n_4} \intqq G^{c_1 c_2}_{n_3 n_4}(k+q)
    \nnnl
    &\ \times g^*_{n_3 n_1 \nu}(\bk, \bq)  g_{n_4 n_2 \nu}(\bk, \bq) D^{c_2 c_1}_{0, \nu}(q) \,.
\end{align}
The first term
\begin{equation}
    \Sigma^{\rm DW}_{12}(\bk) = Z^{c_1 c_2} \, \Sigma^{\rm DW}_{n_1 n_2 \bk}
\end{equation}
is the Debye--Waller (DW) self-energy~\cite{Allen1976, Allen1981, Allen1983}, see Eqs.~(3)--(5) of Ref.~\cite{Lihm2020} for the full expression.
The second term is the self-consistent Fan--Migdal self-energy.
The $Z^{c} = (-1)^{\delta_{c +}}$ factors arise because the temporal integral on the backward branch ($c=+$) is taken from $+\infty$ to $-\infty$, rather than from $-\infty$ to $\infty$~\cite{StefanucciBook}.
Combining this with the Dyson equation [\Eq{eq:sc_Dyson}], we obtain the self-consistent equations for the Green's function and the self-energy.
Henceforth, we omit the superscript \scGD when referring to the \scGD self-energy.

Within the diagonal approximation [\Eq{eq:sc_G_diag_approx}], the \scGD self-energy expression simplifies to
\begin{multline} \label{eq:sc_Sigma_c_diag}
    \Sigma^{c_1 c_2}_{nn\bk}(\veps)
    = Z^{c_1 c_2} \, \Sigma^{{\rm DW}}_{nn\bk} - Z^{c_1} Z^{c_2} \sum_{\nu m} \intqq G^{c_1 c_2}_{mm}(k+q)
    \\
    \times \abs{g_{mn\nu}(\bk, \bq)}^2 D^{c_2 c_1}_{0, \nu}(q) \,.
\end{multline}
Since the four components of the Green's function and self-energy are related by the fluctuation-dissipation relations [\Eq{eq:sc_fdt_G}], it suffices to compute only one of the components.
It is most convenient to compute the retarded self-energy, first the imaginary part as~\cite{Abramovitch2023, Abramovitch2024, Lihm2025}
\begin{multline} \label{eq:sc_Sigma_R_Im}
    \Im \Sigma^{\rm R}_{nn\bk}(\veps)
    = -\pi \sum_{\nu m} \intbq \abs{g_{mn\nu}(\bk, \bq)}^2
    \\
    \times \sum_{\pm} \bigl[f^\pm(\veps \pm \omega_\nuq) + n_\nuq \bigr] A_\mkq(\veps \pm \omega_\nuq) \,,
\end{multline}
and the real part using the Kramers--Kronig relation
\begin{equation} \label{eq:sc_Sigma_R_Re}
    \Re \Sigma^{\rm R}_{nn\bk}(\veps)
    = \frac{1}{\pi} \,\mcP\! \nbint{-\infty}{\infty} \dd\veps' \,
    \frac{\Im \Sigma^{\rm R}_{nn\bk}(\veps')}{\veps' - \veps}
    + \Sigma^{\rm DW}_{nn\bk} \,.
\end{equation}
The complete $2\times 2$ self-energy matrix can be computed using the fluctuation-dissipation relations in \Eq{eq:sc_fdt_G}.
We define the quasiparticle energy as the peak position of the self-consistent spectral function:
\begin{equation} \label{eq:E_nk_def}
    E_\nk = \argmax_{\veps} A_{nn\bk}(\veps) \,.
\end{equation}

In \textit{ab initio} calculations, we truncate the infinite sum over frequency and bands using the static one-shot approximation for high-energy off-shell contributions, as detailed in the Supplemental Material (SM) of Ref.~\cite{Lihm2025}.
In short, we only self-consistently update the Green's functions and the self-energy for states inside the active space $\mc{A} = \{\nk : \veps^{\rm min} \leq \veps_\nk \leq \veps^{\rm max} \}$.
The contribution of the states outside the active space to the self-energy is treated within the static one-shot approximation.
It is computed using the linear Sternheimer equation on a coarse $\bk$ grid~\cite{Gonze2011, Ponce2014} and interpolated to a dense $\bk$ grid using Wannier function perturbation theory~\cite{Lihm2021WFPT}.
The full expression for the \textit{ab initio} \scGD self-energy reads~\cite{Lihm2025}
\begin{align}
    \label{eq:scGD0_Im_window}
    \Im \Sigma^{{\rm R},\, \mc{A}}_{nn\bk}(\veps)
    &= -\pi \sum_{\nu} \intbq \sum_{m}^{\mc{A}} \, \abs{g_{mn\nu}(\bk, \bq)}^2
    \nnnl
    &\hspace{-3em} \times \sum_{\pm} \bigl[f^\pm(\veps \pm \omega_\nuq) + n_\nuq \bigr] A_\mkq(\veps \pm \omega_\nuq) \,,
    \\
    \label{eq:scGD0_Re_window}
    \Re \Sigma^{{\rm R},\, \mc{A}}_{nn\bk}(\veps)
    &= \frac{1}{\pi} \,\mcP\! \nbint{-\infty}{\infty} \dd\veps'
    \frac{\Im \Sigma^{{\rm R},\, \mc{A}}_{nn\bk}(\veps')}{\veps' - \veps}
    + \Sigma^{\rm static}_{nn\bk} \,,
\end{align}
with the static self-energy given by
\begin{equation} \label{eq:scGD0_static}
    \Sigma^{\rm static}_{nn\bk}
    = \Sigma^{\rm DW}_{nn\bk}
    + \sum_{\nu} \intbq \sum_{m}^{\mc{A}^\mathsf{c}} \frac{\abs{g_{mn\nu}(\bk, \bq)}^2}{\veps_\nk \!-\! \veps_{\mkq}}
    (2 n_\nuq + 1),
\end{equation}
where $\mc{A}^\mathsf{c}$ is the complement of the active space $\mc{A}$.

\begin{figure}[t]
    \centering
    \begin{tikzpicture}[node distance = 1.2cm and 2cm]
    \node (start0) [startstop, align=center,]
        {Input: $\veps_\nk$, $\omega_\nuq$, $g_{mn\nu}(\bk, \bq)$};
    \node (start) [process, below of=start0]
        {Initial guess: $\Sigma^{\rm R}_\nk(\veps) = -i \eta$};
    \node (pro1) [process, below of=start]
        {$G^{\rm R}_\nk(\veps)$~[\Eq{eq:Dyson_RA}]};
    \node (pro2) [process, below of=pro1]
        {$A_\nk(\veps)$~[\Eq{eq:A}]};
    \node (pro3) [process, below of=pro2]
        {$\Im \Sigma^{\rm R}_\nk(\veps)$~[\Eq{eq:scGD0_Im_window}]};
    \node (pro4) [process, below of=pro3]
        {$\Re \Sigma^{\rm R}_\nk(\veps)$~[\Eq{eq:scGD0_Re_window}]};
    \node (dec) [decision, below of=pro4, aspect=3, align=center, inner sep=-0.1ex, yshift=-0.1cm]
        {$\textrm{max} \abs{\Sigma^{\rm in} - \Sigma^{\rm out}} < s$?};
    \node (mix) [process, right of=pro2, xshift=2.5cm, yshift=-0.75cm]
        {Mix $\Sigma^{\rm R}_\nk(\veps)$ [\Eq{eq:sc_mixing}]};
    \node (out) [io, below of=dec, yshift=-0.1cm]
        {Output: $\Sigma^{\rm R}_\nk(\veps)$, $G^{\rm R}_\nk(\veps)$};
    \draw [arrow] (start0) -- (start);
    \draw [arrow] (start) -- (pro1);
    \draw [arrow] (pro1) -- (pro2);
    \draw [arrow] (pro2) -- (pro3);
    \draw [arrow] (pro3) -- (pro4);
    \draw [arrow] (pro4) -- (dec);
    \draw [arrow] (dec) -| node[anchor=west] {no} (mix);
    \draw [arrow] (mix) |- (pro1);
    \draw [arrow] (dec) -- node[anchor=west] {yes} (out);
    \end{tikzpicture}
    \caption{Flowchart for the self-consistent calculation of the self-energy.}
    \label{fig:selfen_flowchart}
\end{figure}
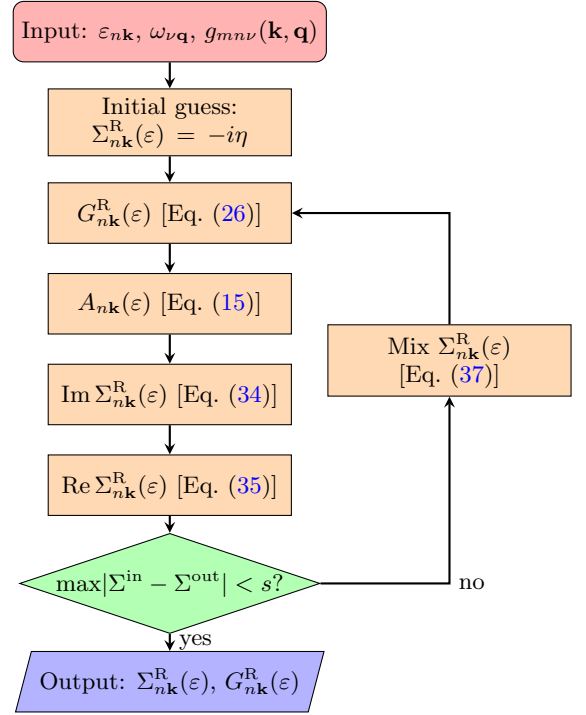

Figure~\ref{fig:selfen_flowchart} illustrates the workflow for the calculation of the \scGD self-energy and Green's function.
We initialize retarded self-energies with a constant imaginary part of $-i\eta$, with $\eta = 5$~meV.
We have verified that the converged results are insensitive to the choice of $\eta$.
After each iteration, we update the self-energy using a linear mixing scheme:
\begin{equation} \label{eq:sc_mixing}
    \Sigma^{\rm R,\,next}_{nn\bk}(\veps)
    = \lambda^{\rm mix} \, \Sigma^{\rm R,\,out}_{nn\bk}(\veps)
    + (1 - \lambda^{\rm mix}) \, \Sigma^{\rm R,\,in}_{nn\bk}(\veps) \,.
\end{equation}
The self-consistent iteration is stable for most cases; hence, we set $\lambda^{\rm mix} = 1$, and reduce it to $\lambda^{\rm mix} = 0.5$ only when the iteration does not converge.
The iteration continues until the maximum change in the self-energy for all states $\nk$ and energies $\veps$ is below a threshold of $s = 0.1$~meV.
We have implemented this workflow as an extension to the in-house developed \EPjl package~\cite{Lihm2024NLHE, EPjl}, which uses real-space electron-phonon matrix elements computed with the \textsc{EPW} code~\cite{Ponce2016EPW,Lee2023EPW}.
We also plan to implement the \scGD method in the \textsc{EPW} software~\cite{Ponce2016EPW, Lee2023EPW} in the future.

\begin{figure*}[tb]
\centering
\includegraphics[width=0.99\linewidth]{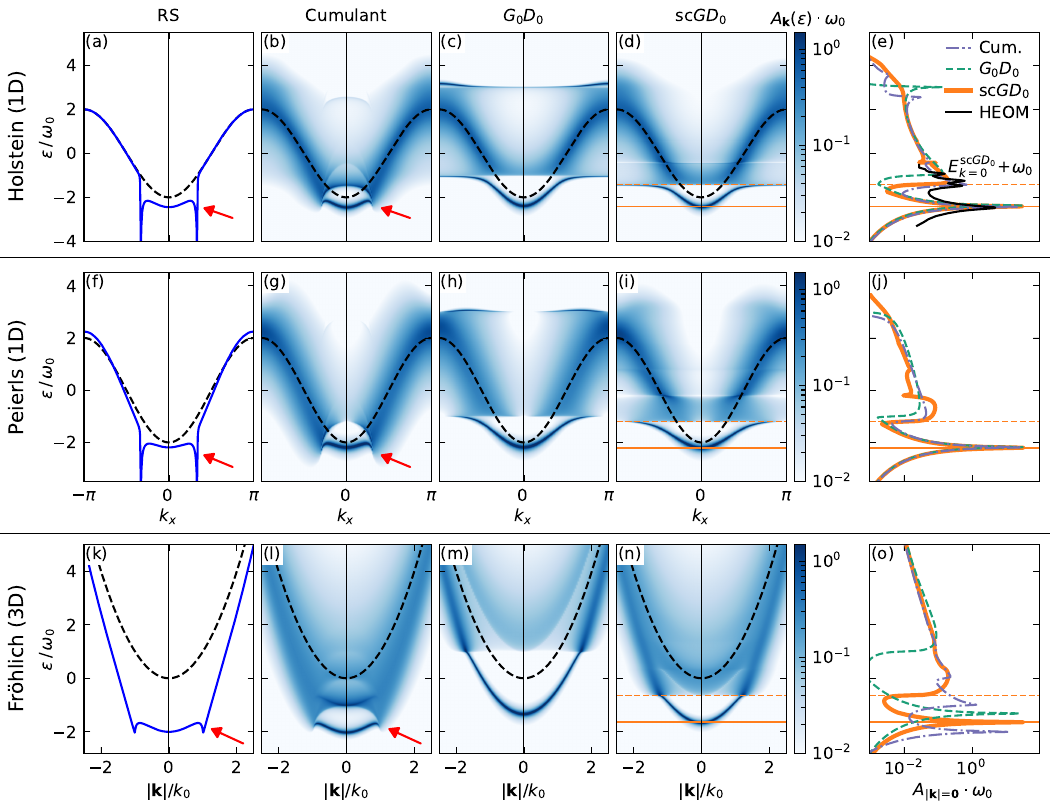}
\caption{
    (a)--(d) Spectral functions of the undoped 1D Holstein model with parameters $t = \omega_0 = 1$, $\lambda = 0.5$, and $T=0$ computed using the Rayleigh--Schr\"odinger perturbation theory (RS), cumulant approximation, one-shot \GD, and self-consistent $GD_0$ (\scGD) methods.
    Dashed black lines show the bare electron band.
    Red arrows mark nonphysical features in the RS dispersion and the cumulant spectral function.
    Solid horizontal line in (d) indicates the quasiparticle energy at $k_x=0$, and the dashed line indicates the energy above it by the phonon frequency $\omega_0$.
    (e) One-dimensional cuts of the spectral functions at $k_x=0$, shown on a logarithmic $x$ scale.
    We also show the exact result from the hierarchical equations of motion (HEOM) method~\cite{Mitric2022} as a thin black line.
    Note that the splitting of the satellite peak in the HEOM result around $\veps = -1.5 \omega_0$ is a finite-size artifact~\cite{Mitric2022}.
    (f)--(j) The same for the 1D Peierls model with $t = \omega_0 = 1$, $\lambda = 0.5$, and $T=0$.
    (k)--(o) The same for the 3D Fr\"ohlich model with $m_0 = \omega_0 = 1$, $\alpha = 2$, and $T=0$, where the wavevector $\abs{\bk}$ is given in units of $k_0 = \sqrt{2m_0 \omega_0}$.
    We used an artificial broadening of $\eta = 0.01$ for all calculations, which determines the width of the low-energy quasiparticle peaks.
}
\label{fig:spectral_models}
\end{figure*}

\subsection{The one-shot \texorpdfstring{\GD}{G0D0} and cumulant approximations}

In most first-principles studies, the self-consistency of the Green's function is neglected and the self-energy is computed using the bare Green's function.
This is the \GD approximation, where the self-energy reads~\cite{Giustino2017}
\begin{multline} \label{eq:sigma_G0D0}
    \Sigma_{nn\bk}^{{\rm R}, G_0 D_0}(\veps)
    = \Sigma^{\rm static}_{nn\bk}
    + \sum_{\nu } \intbq \sum_{m}^{\mc{A}}  \abs{g_{mn\nu}(\bk, \bq)}^2
    \\
    \times \sum_{\pm} \frac{f^\pm_\mkq + n_\nuq}{\veps - \veps_{\mkq} \pm \omega_\nuq + i\eta}
\end{multline}
with $\eta$ being a positive infinitesimal.

Starting from the \GD self-energy, one may approximately include the effect of higher-order diagrams via the cumulant approximation.
In the retarded cumulant approximation, one expands the retarded Green's function in terms of the cumulant function $C_\nk(t)$~\cite{Kas2014, Kas2017, Zhou2019STO}
\begin{equation} \label{eq:cum_G}
    G^{\rm R,\,Cum.}_\nk(\veps) =-i \nint \dd t \, \Theta(t) e^{i (\veps - \veps_\nk - \Sigma^{\rm static}_{nn\bk}) t} e^{C_\nk(t)} \,,
\end{equation}
where $\Theta(t)$ is the Heaviside function.
The cumulant function is then approximated using second-order perturbation theory~\cite{Story2014Cumulant}:
\begin{equation} \label{eq:cum_C}
    C_\nk(t) \approx
    \frac{1}{\pi} \nint \dd\veps \, \abs{\Im \Sigma_\nk^{{\rm R}, G_0 D_0}(\veps + \veps_\nk)} \frac{e^{-i\veps t} + i \veps t - 1}{\veps^2} \,.
\end{equation}
Details for the numerical implementation of this equation are presented in \App{app:cumulant}.

Finally, the Rayleigh--Schr\"odinger (RS) perturbation theory is obtained by evaluating the \GD self-energy at the bare eigenvalue:
\begin{equation}
    E^{\rm RS}_\nk = \veps_\nk + \Re \Sigma^{{\rm R}, G_0 D_0}_{nn\bk}(\veps_\nk) \,.
\end{equation}

\begin{figure*}[tb]
\centering
\includegraphics[width=0.99\linewidth]{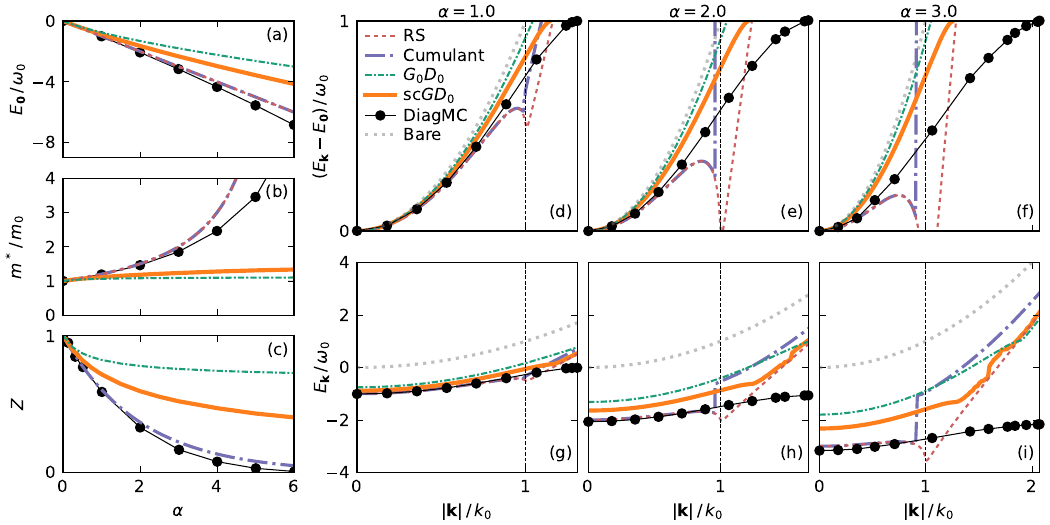}
\caption{
    (a) Quasiparticle energy, (b) mass renormalization, (c) quasiparticle weight, and (d)--(i) quasiparticle dispersion of the Fr\"ohlich model with $m_0 = \omega_0 = 1$ and $T=0$, computed using various methods.
    The quasiparticle energy is defined as the peak position of the spectral function [\Eq{eq:E_nk_def}].
    The wavevector $\abs{\bk}$ is given in units of $k_0 = \sqrt{2m_0 \omega_0}$.
    In (d)--(f), the quasiparticle dispersions are vertically shifted to align their minima.
    The DiagMC results are from Ref.~\cite{Hahn2018DMC}, except for (c), which uses data from Ref.~\cite{Mishchenko2000DMC}.
}
\label{fig:frohlich_polaron}
\end{figure*}

\subsection{Application to model Hamiltonians} \label{sec:spectral_models}

We first apply the \scGD method to three model Hamiltonians, the Holstein model~\cite{Holstein1959a}, the Peierls or Su--Schrieffer--Heeger model~\cite{Barisic1970, Su1979, Capone1997, Marchand2010}, and the Fr\"ohlich model~\cite{Frohlich1950}.
We compare the RS, cumulant, \GD, and \scGD approximations, using the numerically exact calculations with the hierarchical equations of motion (HEOM)~\cite{Mitric2022} and diagrammatic Monte Carlo (DiagMC)~\cite{Mishchenko2000DMC, Hahn2018DMC} methods as benchmarks.

The Holstein model~\cite{Holstein1959a} is one of the simplest models of \eph coupling that has been extensively studied in the literature~\cite{Fehske2007}.
In this model, electrons with nearest-neighbor hopping are coupled to a dispersionless phonon through a local \eph interaction.
We consider the 1D case, for which the \eph Hamiltonian parameters are given by~\cite{Jankovic2023}
\begin{equation} \label{eq:holstein_def}
\begin{gathered}
    \veps_\bk = -2t \cos k_x \,, \ %
    \omega_\bq = \omega_0 \,, \ %
    g(\bk, \bq) = \sqrt{2 \lambda \omega_0 t} \,,
\end{gathered}
\end{equation}
where $\lambda$ is the dimensionless parameter that characterizes the \eph interaction strength.

Figures~\ref{fig:spectral_models}(a)--(e) show the spectral functions of the Holstein model at $T=0$ with the interaction $\lambda = 0.5$ and $t = \omega_0 = 1$.
We find that RS perturbation theory yields a quasiparticle band with a large kink with an inverted curvature near $k_x = \pi/3$, where $\veps_\bk = \veps_{\mb{0}} + \omega_0$.
This nonphysical kink indicates the breakdown of the RS approximation arising from a resonant coupling with electrons at the bare band extrema.
The cumulant spectral function at the band edge, $k_x \approx 0$, shows a sharp quasiparticle peak, clearly separated from the continuum by a distance of $\omega_0$.
Still, a similar kink at $k_x \approx \pi/3$ is observed, consistent with the findings of Ref.~\cite{Mitric2023}.
The one-shot \GD method is free from this artifact but overestimates the separation between the quasiparticle peak and the satellite, as found in Ref.~\cite{Verdi2017} for the Fr\"ohlich model.
This issue stems from the fact that the onset of the phonon emission continuum in the \GD method is determined relative to the bare band minimum, not the renormalized one.
In contrast, the \scGD spectral function correctly places the onset of the continuum at the renormalized band minimum plus $\omega_0$ and does not exhibit any nonphysical kinks.
Below the onset, the spectral function displays a sharp quasiparticle peak, as phonon emission is forbidden.

Compared to the numerically exact HEOM result~\cite{Mitric2022}, \scGD quantitatively reproduces the position of the quasiparticle and satellite peaks.
We note that the splitting of the satellite peak in the HEOM result around $\veps = -1.5 \omega_0$ is due to finite-size effects~\cite{Mitric2022}, which our calculations do not have.
The \scGD spectral function is consistent with the absence of any peaks near the upper band edge ($\veps \sim 2t$) in exact diagonalization results~\cite{Fehske2007}, whereas the \GD and cumulant approximations generate a small peak there.

The Peierls model~\cite{Barisic1970, Su1979, Capone1997, Marchand2010} features the same dispersion as the Holstein model but incorporates a nonlocal \eph coupling that modifies the hopping energy rather than the onsite energy:
\begin{equation} \label{eq:peierls_def}
\begin{aligned}
    \veps_\bk &= -2t \cos k_x \,, \ %
    \omega_\bq = \omega_0 \,,
    \\
    g(\bk, \bq) &= -i \sqrt{2 \lambda \omega_0 t} [\sin(k_x+q_x) - \sin k_x].
\end{aligned}
\end{equation}
Figures~\ref{fig:spectral_models}(f)--(j) show the spectral functions of the Peierls model with $t = \omega_0 = 1$, $\lambda = 0.5$ and $T=0$.
The results are qualitatively similar to those of the Holstein model, with the \scGD method being the only one to consistently show a clear separation between well-defined quasiparticles and the phonon emission continuum by $\omega_0$.
One difference from the Holstein model is a small upward renormalization of states at $k_x \approx \pi$, which can be attributed to the $\bq$-dependent \eph coupling.

The Fr\"ohlich model~\cite{Frohlich1950} describes the coupling between electrons with a parabolic dispersion and phonons with a flat dispersion mediated by a dipolar \eph coupling~\cite{Verdi2015, Sjakste2015}:
\begin{equation} \label{eq:frohlich_def}
\begin{aligned}
    \veps_\bk = \frac{\abs{\bk}^2}{2m_0} \,, \ %
    \omega_\bq = \omega_0 \,, \ %
    g(\bk, \bq) = \frac{i}{\abs{\bq}} \sqrt{ \frac{4\pi \alpha}{\Omega} \sqrt{\tfrac{\omega_0^3}{2m_0}} } \,.
\end{aligned}
\end{equation}
Here, $\alpha$ is a dimensionless constant for the interaction strength and $\Omega$ is the volume of the system.
We consider a 3D continuum Fr\"ohlich model.
The spectral functions for the Fr\"ohlich model, shown in \Figs{fig:spectral_models}(k)--(o), exhibit qualitative similarities to those of the other models, except for the infinite bandwidth because of considering a continuum model.
We observe the same artifacts from the RS perturbation theory~\cite{Larsen1966} and the cumulant approximation~\cite{Kandolf2022}, including kinks and inverted curvatures away from the band edge, as highlighted by the red arrows in \Figs{fig:spectral_models}(k) and (l).

The \scGD spectral functions in Fig.~5(d) and (i) show additional features at $E_{k=0}^{\mathrm{sc}GD_0} + 2 \omega_0$, where the spectral function is strongly suppressed in a narrow energy region.
It remains to be understood whether this is an artifact of the noncrossing approximation or a generic property of these models.

To quantitatively assess the performance of the different approximations, we analyze the Fr\"ohlich polaron properties extracted from the quasiparticle peak of the spectral functions.
Figure~\ref{fig:frohlich_polaron} presents the Fr\"ohlich polaron properties computed using the different approximations and compares them with numerically exact DiagMC results~\cite{Mishchenko2000DMC, Hahn2018DMC}.
The \scGD method improves the agreement with the DiagMC results over the \GD method, but underestimates the renormalization compared to the RS and cumulant approximations.
In fact, the RS and cumulant approximations are known to yield very accurate results for the polaron properties at the band edge for $\alpha \leq 3$~\cite{Nery2018}.
A similar observation has been made for the Holstein model: the cumulant approximation overestimates the mass renormalization and the \scGD method underestimates it~\cite{Mitric2023}.
Moving away from the band edge, the RS and cumulant polaron dispersions show a strong kink at $k \approx k_0 = \sqrt{2m_0 \omega_0}$~\cite{Larsen1966}, while the \scGD method provides a polaron dispersion free from such an artifact.

Our benchmark calculations further show that the \scGD approach fails to capture the binding energy and mass renormalization of the Fr\"ohlich model in the strong-coupling regime.
In addition, in the strong-coupling limit $t \to 0$ of the Holstein model, the Hamiltonian is exactly solvable as a collection of isolated harmonic oscillators coupled to two-level systems.
In this regime, it can be shown that the crossing diagrams give the dominant contribution~\cite{Berciu2006}.
These considerations suggest that crossing diagrams (\eph vertex corrections), which are neglected in the \scGD approach but included in the DiagMC calculation, play a crucial role beyond weak coupling.

\begin{figure*}[tb]
\centering
\includegraphics[width=0.99\linewidth]{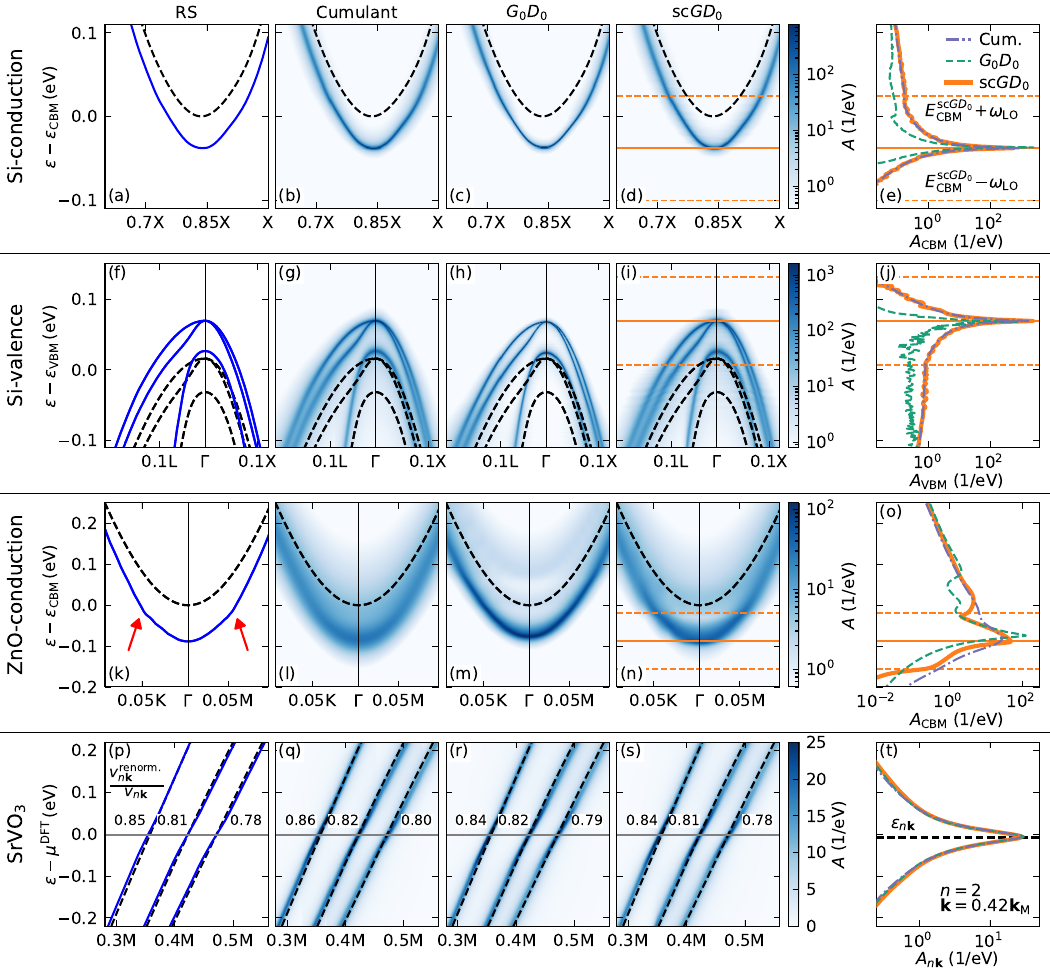}
\caption{
    Spectral functions, as in \Fig{fig:spectral_models}, for
    (a)--(e) the conduction band of Si at $300~\mathrm{K}$,
    (f)--(j) the valence bands of Si at $300~\mathrm{K}$,
    (k)--(o) the conduction band of ZnO at $150~\mathrm{K}$, and
    (p)--(t) the region near the Fermi level of SrVO$_3$ at $300~\mathrm{K}$.
    Red arrows in (k) mark a small kink in the RS dispersion.
    This kink is not visible in the cumulant spectral function in (l) due to broadening.
    In (t), the spectral function corresponds to the second ($n=2$) of the three $t_{2g}$ bands.
}
\label{fig:spectral_all}
\end{figure*}

\subsection{Application to real materials}

Having confirmed that the \scGD method accurately captures the expected behavior for the model Hamiltonians, we now apply it to real materials using \textit{ab initio} parameters.
In this work, we consider three materials: Si, ZnO, and SrVO$_3$ as representative classes of materials.
Si is a nonpolar semiconductor with weak \eph coupling, and ZnO is a polar semiconductor with stronger \eph coupling.
For both materials, experimental data on the dc mobility~\cite{Morin1954Si, Ludwig1956Si, Logan1960Si, Norton1973Si, Jacoboni1977Si, Hutson1959, Hutson1959, Wagner1974ZnO, Look1998ZnO, Makino2005} and the terahertz ac conductivity~\cite{vanExter1990Si, Jeon1997Si, Baxter2009} in the intrinsic (nondegenerate) regime are available.
Spectral functions of doped ZnO have been studied using the \GD and cumulant approximations~\cite{Antonius2020}.
SrVO$_3$ is a correlated metal in which a recent work has shown that \eph interactions play a dominant role in determining resistivity~\cite{Abramovitch2024}.
Transport in SrVO$_3$ has also been widely explored both experimentally~\cite{Nozaki1991SVO, Lan2003SVO, Itoh1995SVO, Inoue1998SVO, Reyes2000SVO, Brahlek2015SVO, Fouchet2016SVO, Zhang2016SVO, Xu2019SVO, Shoham2020SVO, Mirjolet2021SVO, Roth2021SVO, Ahn2022SVO, Brahlek2024SVO} and theoretically~\cite{Abramovitch2024, LeeHand2024SVO, LaBollita2025, Coulter2025}.
Therefore, these materials serve as useful test cases for the spectral and transport formalisms developed in this work.
In this section, we focus on the spectral functions of these materials.
These spectral functions are then used to compute the transport properties in the following sections.

For the valence band of Si and the conduction band of ZnO, we include spin-orbit coupling (SOC) in the electronic structure and phonon calculations.
For the conduction band of Si and SrVO$_3$, we neglect SOC~\cite{Ponce2021}.
For metals or doped semiconductors, we iteratively update the chemical potential to ensure charge neutrality.
The \GD and cumulant calculations require an artificial broadening.
Although the zero-broadening limit is commonly taken, it is not always well defined and can lead to divergences~\cite{Lihm2024Piezo}.
To ensure stability, we use a self-consistent broadening derived from the \scGD self-energy~\cite{Lihm2024Piezo}:
\begin{equation} \label{eq:gamma_sc}
    \gamma^{\rm sc} = -\Im \Sigma^{{\rm sc}GD_0,\,{\rm R}}_{n^* \bk^*}(E^{{\rm sc}GD_0}_{n^* \bk^*}) \,,
\end{equation}
where $E^{{\rm sc}GD_0}_{n^* \bk^*}$ is the \scGD quasiparticle energy [\Eq{eq:E_nk_def}].
For semiconductors, we use the self-energy from the conduction band minimum (CBM) (valence band maximum (VBM)) state when calculating the conduction (valence) band spectral functions, as they are the states with the largest occupation when $p$- ($n$-) doped.
For metals, we take an average over the states $n^* \bk^*$ within 5~meV of the Fermi level.
The \scGD calculations do not require the artificial broadening $\gamma^{\rm sc}$, except at $T=0$~K where the low-energy states have infinite lifetimes.

\Figus{fig:spectral_all}(a)--(j) show the spectral functions of the valence and conduction bands of Si at $T=300~\mathrm{K}$.
All methods yield similar dispersion and spectral functions.
In particular, we find almost exact agreement between the cumulant and \scGD spectral functions.
This result confirms that in the weak \eph coupling regime, the impact of self-consistency on the spectral functions is small.
The renormalizations of the CBM and VBM energies of Si at 300~K are $-38$~meV and $+68$~meV, respectively, across all methods.
The resulting band-gap renormalization of $-106$~meV is consistent with previous calculations ($-91.4$~meV~\cite{Ponce2015}) and experiment ($-102.3$~meV~\cite{Bludau1974}).

The results for polar materials, which exhibit a strong Fr\"ohlich \eph interaction~\cite{Verdi2015, Sjakste2015}, show significant differences.
\Figus{fig:spectral_all}(k)--(o) show the spectral functions of ZnO at $150~\mathrm{K}$.
The effective Fr\"ohlich parameter for ZnO is calculated to be $\alpha = 1.14$~\cite{deMelo2023}, which is within the regime where the \scGD method yields accurate results for the Fr\"ohlich model.
Similarly to the Fr\"ohlich model, the RS dispersion shows kinks at $\bk$ points where $\veps_\nk \approx \veps_{\rm CBM} + \omega_{\rm LO}$.
In the cumulant spectral function, this kink is not visible because of the finite broadening of the quasiparticle band.
The \scGD spectral function does not exhibit this artifact as well and produces the onset of the phonon emission continuum at $E_{\rm CBM} + \omega_{\rm LO}$.
Compared to the cumulant case, the quasiparticle peak and the satellite are better separated in the \scGD spectral function.
From \Fig{fig:spectral_all}(o), we find that the satellite peak of the \scGD spectral function is slightly above $E_{\rm CBM} + \omega_{\rm LO}$ (horizontal dashed line).
This shift occurs because $E_{\rm CBM} + \omega_{\rm LO}$ marks the onset of phonon emission, where the lower end of the satellite is positioned, and the peak lies at a higher energy.
To determine whether this shift is physically meaningful, it would need to be compared with experimental data or numerically exact calculations.

The broadening of the spectral function below the LO phonon emission threshold ($E_{\rm CBM} + \omega_{\rm LO}$, orange dashed line at $-0.02$~eV) arises from the interaction with low-energy acoustic phonons, combined with the finite temperature that allows both phonon absorption and emission.
This behavior is absent in the model calculations shown in Fig.~5, which include only optical phonons and are performed at $T = 0$.
We note that the CBM energy of ZnO at 150~K is renormalized by $-81$, $-74$, and $-88$~meV for the cumulant, \GD, and \scGD methods, respectively.

In addition, we find a shoulder of the spectral function due to phonon absorption at $E_{\rm CBM} - \omega_{\rm LO}$ (orange dashed line at $-0.15$~eV).
In the bandgap region below the phonon absorption threshold $E_{\rm CBM} - \omega_{\rm LO}$, the \scGD spectral function decays exponentially, whereas the \GD and cumulant spectral functions exhibit only a polynomial decay.
The slow decay arises from the self-consistent broadening parameter $\eta = 3.46~\mathrm{meV}$ required in one-shot calculations.
In Si, however, the effect is negligible due to the smaller broadening values of $0.332~\mathrm{meV}$ and $0.515~\mathrm{meV}$ for the conduction and valence bands, respectively.
The exponential decay is crucial for studying transport in the nondegenerate regime~\cite{Lihm2025}.

For definitiveness, we also study the spectral functions of NaCl and AlN, which are polar materials with a stronger \eph coupling (see \App{app:additional_spectral}).
As expected, we confirm that the difference between the \scGD and the cumulant spectral functions is more pronounced.

Finally, we apply the \scGD method to the metallic perovskite SrVO$_3$.
\Figus{fig:spectral_all}(p)--(t) show the spectral function of metallic SrVO$_3$ at 300~K along the $\Gamma$M path.
As a metal, the \eph interaction in SrVO$_3$ is screened and weak.
Thus, self-consistency has little effect on spectral functions, with all methods producing a velocity renormalization of $\sim$20\%, consistent with a previous study~\cite{Abramovitch2024}.
Electronic correlation, which is not considered in this work, has a stronger effect that reduces the bandwidth by $\sim$50\%~\cite{Nekrasov2006SVO}.

Having established the formalism for equilibrium Green's functions, we now shift our focus to the response properties required for studying transport.

\section{Linear response theory} \label{sec:lr}

In this section, we develop the linear response theory for the \scGD Green's functions and self-energies.
We begin by reviewing linear response theory in the Keldysh formalism in \Sec{sec:lr_keldysh}, with additional details provided in \App{sec:lr_detail}.
Then, we derive the self-consistent ladder equations for the linear response function from the \scGD self-energy formula in \Sec{sec:ladder}.
This result will be applied to electronic transport in \Sec{sec:transport} and to the calculation of the dielectric function in \Sec{sec:dielectric}.

\subsection{Linear response in the Keldysh formalism} \label{sec:lr_keldysh}

Consider a general linear response problem where the system is perturbed by a monochromatic single-particle operator $\hat{Y}_\bQ$, and we seek the expectation value of another single-particle operator $\hat{X}^\dagger_\bQ$.
In \Sec{sec:transport}, we consider the case where both $\hat{X}$ and $\hat{Y}$ correspond to the current operator, while in \Sec{sec:dielectric}, they represent a finite-$\bQ$ scalar-potential perturbation.
We write the operators as
\begin{equation} \label{eq:lr_X_def}
    \hat{X}^\dagger_{\bQ}
    = \hat{X}_{-\bQ}
    = \sum_\mnk X^*_{\mnk}(\bQ) \opcd_\nk \opc_\mkQ \,,
\end{equation}
and
\begin{equation} \label{eq:lr_Y_def}
    \hat{Y}_{\bQ} = \sum_\mnk Y_{\mnk}(\bQ) \opcd_{m\bkQ} \opc_\nk \,.
\end{equation}
To ensure Hermiticity, we have imposed the condition $\hat{X}_{-\bQ} = \hat{X}^\dagger_{\bQ}$ in Eq.~\eqref{eq:lr_X_def}, which implies
\begin{equation} \label{eq:lr_hermiticity}
    X_{mn\bkQ}(-\bQ) = X_{nm\bk}^*(\bQ) \,.
\end{equation}

We consider a monochromatic time dependence with frequency $\Omega$ and denote the strength of the external perturbation by $\delta Y^c(Q) = \delta Y^c(\bQ, \Omega)$.
The Hamiltonian, including the perturbation, is given by
\begin{equation} \label{eq:lr_H_pert}
    \hat{H}'(t^c) = \hat{H} + e^{-i\Omega t} \, Z^{c} \, \delta Y^c(Q) \, \hat{Y}_{\bQ}  + {\rm H.c.} \,,
\end{equation}
where H.c.\ denotes the Hermitian conjugate.
We allow the perturbation to differ on the forward and backward branches of the Schwinger--Keldysh contour by letting $\delta Y^c(Q)$ depend on the contour variable $c$.
The factor $Z^c$ will later cancel with the same factor during the integration over the contour [see \Eq{eq:lr_chi_response}].

We denote the response of the operator $\hat{X}^\dagger_{\bQ}$ induced by the $Y$ perturbation as $\delta_Y X_{\bQ}(t^c) \equiv \delta \expval{\hat{X}^\dagger_{\bQ}(t^c)}$.
Standard linear response theory yields~\cite{StefanucciBook}
\begin{equation} \label{eq:lr_chi_Q}
    \delta_Y X^{c}(Q) = \sum_{\cpr} \chi_{XY}^{c\cpr}(Q) \, \delta Y^{\cpr}(Q) \,,
\end{equation}
where $\delta_Y X^{c}(Q)$ is the Fourier transform of $\delta_Y X_{\bQ}(t^c)$, and $\chi_{XY}^{c\cpr}(Q)$ is the $XY$ susceptibility defined as the Fourier transform of the contour-ordered correlator between $\hat{X}^\dagger_\bQ$ and $\hat{Y}_\bQ$ [\Eq{eq:lr_chi_def}].

The experimentally relevant response function is the retarded susceptibility $\chi^{\rm R}_{XY}(Q)$.
The retarded, advanced, lesser, and greater components of the susceptibility are defined as in the electronic case [\Eq{eq:G_RA}].
The susceptibility satisfies the bosonic fluctuation-dissipation relations \Eq{eq:fdr_boson}~\cite{Kubo1957, Martin1959, StefanucciBook}.
The $XY$ and $YX$ susceptibilities are related by the symmetry
\begin{equation} \label{eq:chi_symmetry}
    \chi_{X Y}^{c\cpr}(Q)
    = \chi_{Y X}^{\cpr c}(-Q) \,.
\end{equation}

\begin{figure}[tb]
\centering
\includegraphics[width=0.99\linewidth]{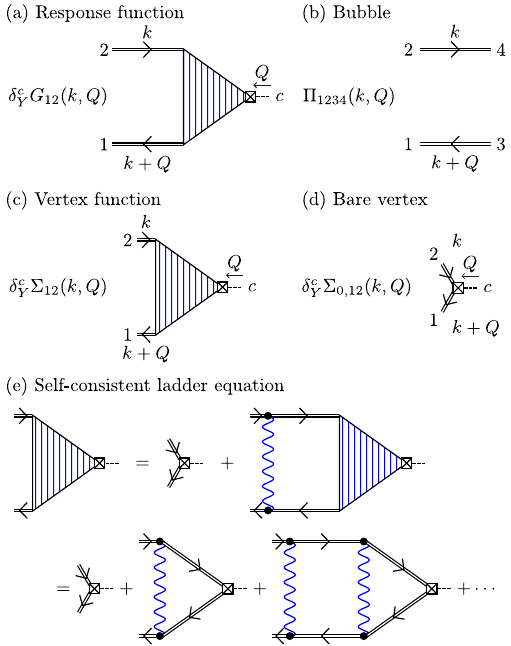}
\caption{
    Feynman diagrams of the linear response functions and the self-consistent ladder equation.
    The crossed square represents the perturbation $\delta Y$ and short double lines represent amputated propagators.
}
\label{fig:feynman_ladder}
\end{figure}

To evaluate the susceptibility, we compute the linear response of the electron Green's function to the perturbation $\hat{Y}_{\bQ}$.
This quantity, referred to as the response function and denoted as $\delta_Y^c G_{12}(k, Q)$, encapsulates all the information about the single-particle response of the system.
This allows for the calculation of the susceptibility of any single-particle operator with $\hat{Y}_{\bQ}$.
Since the perturbation breaks both spatial and temporal translational symmetries, we must consider the momentum- and time-off-diagonal Green's function [\Eq{eq:G_k_kp}].
For monochromatic perturbations, we can focus on the Fourier component
\begin{align} \label{eq:lr_R_def}
    &\delta_Y^c G_{12}(k, Q)
    \equiv \frac{\delta G_{12}(k, Q)}{\delta Y^c(Q)}
    \\
    &= \nint \dd t_1 \, \frac{\delta G_{n_1 \bkQ, n_2 \bk}(t_1^{c_1}, t_2^{c_2})}{\delta Y^c(Q)}
    \, e^{i (\veps + \Omega - \mu) t_1}
    \, e^{- i (\veps - \mu) t_2} \,.
    \nonumber
\end{align}
After the $t_1$ integration, the $t_2$ dependence vanishes due to time translational invariance.
This expression describes the emission of an electron-hole pair with energies $\veps+\Omega$ and $\veps$ at states $n_1\bkQ$ and $n_2\bk$, respectively.
The total energy and momentum of the pair are $\Omega$ and $\bQ$, originating from the external perturbation.
We use $c$ and $c'$ for external perturbations and $c_1$ and $c_2$ for electronic operators.

Given the response function, one can compute the $XY$ susceptibility as
\begin{equation} \label{eq:sigma_XY_from_R}
    \chi^{c\cpr}_{XY}(Q)
    = \sum_{mn} \intkk X_{\mnk}^*(\bQ) \, \delta_Y^{\cpr} G^{cc}_{mn}(k, Q) \,.
\end{equation}
This is the central equation of linear response theory in the Keldysh formalism.
Next, we discuss how to compute the response function.

\subsection{Self-consistent ladder equation} \label{sec:ladder}

To calculate the response function, we consider the linear response of the \scGD Green's function and self-energy~\cite{Edwards1958, Engelsberg1963, Holstein1964, Prange1964, MahanBook}.
From the functional derivative of the Dyson equation \eqref{eq:sc_Dyson_inv} and using $\delta G^{-1} = -G^{-1} (\delta G) G^{-1}$, we find the following expression for the response function:
\begin{multline} \label{eq:lr_delta_G_Dyson_long}
    \delta_Y^c G_{12}(k, Q)
    = - \sum_{34} G_{13}(k+Q)
    \\
    \times \biggl(
        \frac{\delta G^{-1}_{0,34}(k, Q)}{\delta Y^c(Q)}
        - \frac{\delta \Sigma_{34}(k, Q)}{\delta Y^c(Q)}
    \biggr)
    G_{42}(k) \,.
\end{multline}
Note that the left and right indices of $\delta_Y^c G_{12}(k, Q)$ correspond to states with energy-momentum $k+Q$ and $k$, respectively [see \Eq{eq:lr_R_def}].
For conciseness, we define the bubble function
\begin{equation} \label{eq:lr_bubble_def}
    \Pi_{1234}(k, Q)
    = G_{13}(k+Q) G_{42}(k) \,,
\end{equation}
and the vertex function
\begin{equation} \label{eq:lr_vertex_def}
    \delta_Y^c \Sigma_{34}(k, Q)
    \equiv -
    \biggl(
        \frac{\delta G^{-1}_{0,34}(k, Q)}{\delta Y^c(Q)}
        - \frac{\delta \Sigma_{34}(k, Q)}{\delta Y^c(Q)}
    \biggr) \,.
\end{equation}
We denote the multiplication between the bubble and vertex functions by $\circ$, defined as
\begin{equation} \label{eq:lr_Pi_X_def}
    (\Pi \circ \delta_Y^c \Sigma)_{12}(k, Q)
    \equiv \sum_{34} \Pi_{12 34}(k, Q) \, \delta_Y^c \Sigma_{34}(k, Q) \,.
\end{equation}
Thus, \Eq{eq:lr_delta_G_Dyson_long} can be compactly written as
\begin{equation} \label{eq:lr_delta_G_Dyson}
    \delta_Y^c G = \Pi \circ \delta_Y^c \Sigma \,.
\end{equation}
This equation is illustrated in \Fig{fig:feynman_ladder}(a), with the bubble function in \Fig{fig:feynman_ladder}(b) and the vertex function in \Fig{fig:feynman_ladder}(c).

Now, we evaluate the two terms in \Eq{eq:lr_vertex_def}.
For the first term, the derivative of the inverse bare Green's function is proportional to the change in the noninteracting Hamiltonian [see \Eq{eq:Gbare_inv}].
We refer to this quantity as the bare vertex $\delta_Y^c \Sigma_0$, which is given by
\begin{equation} \label{eq:lr_V0_def}
    \delta_Y^c \Sigma_{0,12}(k, Q)
    \equiv - \frac{\delta G^{-1}_{0,12}(k, Q)}{\delta Y^c(Q)}
    = Y_{n_1 n_2 \bk}(\bQ) \delta_{c_1 c} \delta_{c_2 c} \,,
\end{equation}
illustrated in \Fig{fig:feynman_ladder}(d).
Using $\delta_Y^c \Sigma_0$ instead of $\delta_Y^c \Sigma$ in \Eq{eq:lr_delta_G_Dyson} yields the bare response function
\begin{equation} \label{eq:lr_R0_def}
    \delta_Y^c G_0 = \Pi \circ \delta_Y^c \Sigma_0 \,.
\end{equation}

For the second term of \Eq{eq:lr_vertex_def}, by taking the functional derivative of the \scGD self-energy [\Eq{eq:sc_Sigma_c}], we find
\begin{multline} \label{eq:lr_dSigma_long}
    \frac{\delta \Sigma_{12}(k, Q)}{\delta Y^c(Q)}
    = - Z^{c_1} Z^{c_2} \!\! \sum_{\nu n_3 n_4} \intqq \, \frac{\delta G^{c_1 c_2}_{n_3 n_4}(k+q, Q)}{\delta Y^c(Q)}
    \\
    \times g^*_{n_3 n_1 \nu}(\bkQ, \bq)  g_{n_4 n_2 \nu}(\bk, \bq)
     D^{c_2 c_1}_{0, \nu}(q) \,.
\end{multline}
Note that the first \eph vertex is for the electron momentum $\bkQ$ and the second is for $\bk$.
Defining the phonon-mediated electron-electron interaction
\begin{multline} \label{eq:lr_W_def}
    W_{1234}(\bkQ, \bk, q)
    \equiv -Z^{c_1 c_3} Z^{c_2 c_4} \sum_{\nu} D^{c_2 c_1}_{0, \nu}(q) \\
    \times g^*_{n_3 n_1 \nu}(\bkQ, \bq) g_{n_4 n_2 \nu}(\bk, \bq) \,,
\end{multline}
we can compactly write \Eq{eq:lr_dSigma_long} as
\begin{equation} \label{eq:lr_dSigma}
    \frac{\delta \Sigma_{12}(k, Q)}{\delta Y^c(Q)}
    = \bigl( W \circ \delta_Y^c G \bigr)_{12}(k, Q) \,.
\end{equation}

By substituting \Eqs{eq:lr_delta_G_Dyson}, \eqref{eq:lr_V0_def}, and \eqref{eq:lr_dSigma} into \Eq{eq:lr_vertex_def}, we obtain a self-consistent equation for the vertex function:
\begin{equation} \label{eq:lr_ladder}
    \delta_Y^c \Sigma
    = \delta_Y^c \Sigma_0
    + W \circ \Pi \circ \delta_Y^c \Sigma \,.
\end{equation}
\Equ{eq:lr_ladder} is the self-consistent ladder equation that evaluates the infinite sum of ladder diagrams where the phonon lines connect the electron and hole propagators without crossing each other, as shown in \Fig{fig:feynman_ladder}(e).
Since the bubble function $\Pi$ involves the \scGD Green's functions, we call this approach the ladder-\scGD method.
With the response function, we can now compute the $XY$ susceptibility as
\begin{align} \label{eq:sigma_XY_R_Sigma}
    \chi^{c\cpr}_{XY}(Q)
    = \sum_{12} \! \intkk \! \delta_X^{c} \Sigma_{0, 21}(k+Q, -Q) \, \delta_Y^{\cpr} G_{12}(k, Q) \,.
\end{align}

\begin{figure}[tb]
\centering
\includegraphics[width=0.99\linewidth]{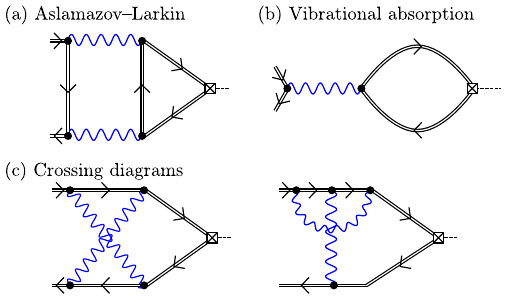}
\caption{
    Low-order Feynman diagrams that are excluded from the ladder-\scGD approach.
}
\label{fig:feynman_not_included}
\end{figure}

\Figu{fig:feynman_not_included} shows low-order diagrams that are not included in the ladder-\scGD approximation.
The Aslamazov--Larkin vertex correction~\cite{Aslamazov1968} with horizontal phonon lines and vertical electron lines [\Fig{fig:feynman_not_included}(a)] and vibrational absorption~\cite{Cappelluti2012, Bistoni2019, Binci2021, Marchese2024} [\Fig{fig:feynman_not_included}(b)] are not included.
These contributions could be incorporated by calculating the electron and phonon self-energies in the fully self-consistent $GD$ approximation.
Another class of neglected diagrams is those in which the phonon lines cross each other, as shown in \Fig{fig:feynman_not_included}(c).

The self-consistent ladder equation within the quasiparticle approximation has been recently used to study phonon-induced spin relaxation from first principles~\cite{Park2022PRB, Park2022PRL}.
Our formulation goes beyond the quasiparticle approximation by incorporating the \scGD Green's functions as the starting point, and calculating the vertex and response functions with their full frequency dependence.
This approach allows us to study the interplay between the beyond-quasiparticles spectral functions and vertex corrections, as demonstrated in the next section.

\section{Self-consistent spectral transport with vertex correction} \label{sec:transport}

\begin{figure}[b]
\centering
\begin{tikzpicture}[node distance=3.2cm]
    \node (st2) [smallitem] {Green--Kubo formula\\\Eq{eq:sigma_chiR}};
    \node (a)   [smallitem, yshift = 1.6cm, below of=st2]
        {Ladder-\scGD\\Eqs.~(\ref{eq:sigma_Lambda_diag}, \ref{eq:sigma_chiR_diag})};
    \node (b)   [smallitem, xshift =  0.6cm, yshift =  0.0cm, below left of=a]
        {(Linearized) BTE\\Eqs.~(\ref{eq:sigma_bte}, \ref{eq:sigma_bte_ac})};
    \node (c)   [smallitem, xshift = -0.6cm, yshift =  0.8cm, below right of=a]
        {Bubble-\scGD\\\Eq{eq:sigma_bubble_scGD_diag}};
    \node (e)   [smallitem, yshift = 1.6cm, below of=c]
        {Bubble-\GD\\\Eq{eq:sigma_bubble_G0D0_diag}};
    \node (d)   [smallitem, xshift =  0.6cm, yshift = 0.4cm, below left of=e]
        {SERTA\\Eqs.~(\ref{eq:sigma_serta}, \ref{eq:sigma_serta_ac})};
    \draw [arrow] (st2) -- node[anchor=east] {
        Neglect crossing diagrams
    } (a);
    \draw [arrow, color=orange!90!red!80!gray!90!black] (a) -- node[anchor=east, yshift=1ex, xshift=1ex, align=left] {
        Quasiparticle approx.\\
        No ph-assisted current\\
        No self-consistency
    } (b);
    \draw [arrow, color=blue!60!gray, align=left] (a) -- node[anchor=west, yshift=1ex, xshift=2ex] {
        Neglect vertex corr.\\
        No ph-assisted current
    } (c);
    \draw [arrow, color=blue!60!gray] (b) -- node[anchor=east, yshift=-1ex] {
        Neglect vertex corr.
    } (d);
    \begin{scope}[transform canvas={xshift=-2.5ex}]
    \draw [arrow, color=orange!90!red!80!gray!90!black] (c) -- node[anchor=west] {
        No self-consistency
    } (e);
    \end{scope}
    \draw [arrow, color=orange!90!red!80!gray!90!black] (e) -- node[anchor=west, yshift=-1ex, align=left] {
        Quasiparticle approx.
    } (d);
\end{tikzpicture}
\caption{
    The relations between the different linear response transport formalisms discussed in this paper.
}
\label{fig:sigma_methods}
\end{figure}
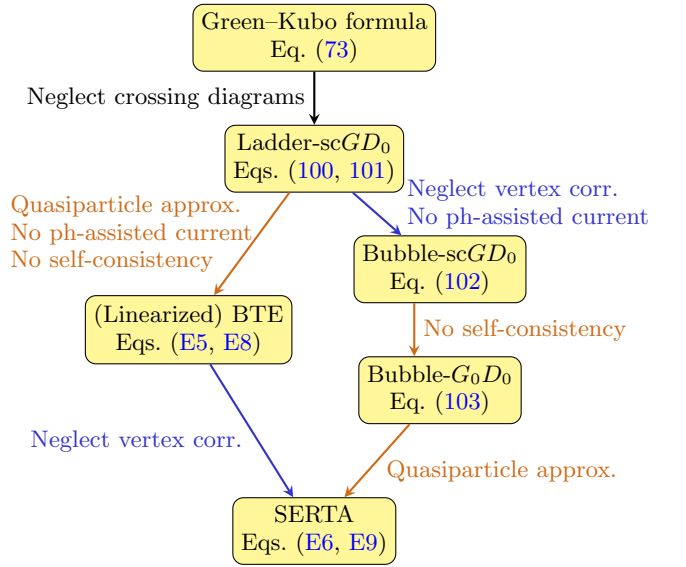

Building on the linear response theory from the previous section, we now address electronic transport under an external electric field and present our main theoretical result: the \textit{ab initio} ladder-\scGD method for phonon-limited transport.
\Figu{fig:sigma_methods} presents the relations between the different linear response transport methods considered in this paper, showing how the state-of-the-art BTE and bubble methods can be derived from the ladder-\scGD method.
In \Sec{sec:efield_gauge}, we introduce the current operator that describes the coupling between electrons and the external electric field.
Importantly, nonlocal \eph coupling generates an additional phonon-assisted contribution to the current~\cite{Gosar1966, Gosar1970, Hannewald2004, Hu2024, Jankovic2025a, Jankovic2025b}, which has not been considered in \textit{ab initio} calculations.
\Sect{sec:conductivity} outlines how dc and ac conductivities follow from the current-current susceptibility, and \Sec{sec:transport_el} and \Sec{sec:transport_ph} cover the specific case of electronic and phonon-assisted currents.
We summarize all equations for the ladder-\scGD transport calculation in \Sec{sec:transport_diagonal}.
Finally, we present applications to model Hamiltonians in \Sec{sec:transport_model}, and to Si, ZnO, and SrVO$_3$ in \Sec{sec:transport_materials}.

\subsection{The current operators} \label{sec:efield_gauge}

To study the response of the system to external electric fields, we use the velocity-gauge Hamiltonian
\begin{equation} \label{eq:gauge_H_A}
    \hat{H}^{\rm A}(t)
    = \hat{H}
    + e \mb{A}(t) \cdot \opmbJ
    + \frac{e^2}{2} \mb{A}(t) \cdot \opmbJ^{\rm D} \cdot \mb{A}(t)
    + O(\mathbf{A}^3) \,.
\end{equation}
Here, $-e$ is the electron charge, $\opmbJ$ is the (paramagnetic) current operator, and $\opmbJ^{\rm D}$ is the diamagnetic current operator.
The detailed derivation of the Hamiltonian and current operators is given in \App{app:velocity_gauge}.

\begin{figure}[tb]
\centering
\includegraphics[width=0.99\linewidth]{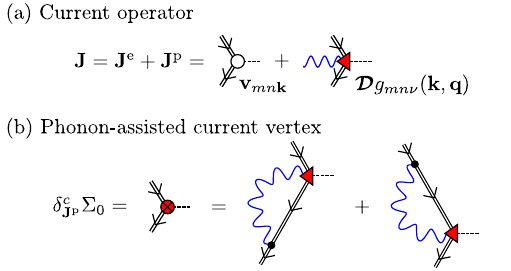}
\caption{
    (a) Feynman diagram of the current operator [\Eq{eq:current_J_def}].
    The white circle denotes the velocity matrix [\Eq{eq:current_v}].
    The red triangle denotes the covariant derivative of the \eph coupling [\Eq{eq:current_Dg}], which mediates the phonon-assisted current.
    (b) Feynman diagram of the phonon-assisted current vertex [\Eq{eq:lr_Vbare_ep}].
    The phonon line generated by the external photon directly connects to one of the two electronic lines.
}
\label{fig:feynman_current}
\end{figure}

Interestingly, the paramagnetic current operator has two contributions, electronic and phonon-assisted:
\begin{equation} \label{eq:current_J_def}
    \opmbJ
    = \opmbJ^{\rm e}
    + \opmbJ^{\rm p} \,.
\end{equation}
The electronic current has the form
\begin{equation} \label{eq:current_Je}
    \opmbJ^{\rm e}
    = \sum_{mn\bk} \mb{v}_\mnk \, \opcd_\mk \, \opc_\nk \,,
\end{equation}
where the velocity matrix is given by~\cite{Blount1962}
\begin{align} \label{eq:current_v}
    \mb{v}_\mnk
    &= \mel{u_\mk}{(\nabla_\bk H_\bk)}{u_\nk} \nnnl
    &= \delta_{mn} (\nabla_\bk \veps_\nk)
    -i \mb{\xi}_\mnk (\veps_\nk - \veps_\mk) \,.
\end{align}
Here, $u_\nk$ is the periodic part of the Bloch wavefunction, and $\mb{\xi}_\bk$ is the electron Berry connection defined as
\begin{equation} \label{eq:current_xi_def}
    \mb{\xi}_\mnk = i\braket{u_\mk | \mb{\nabla}_\bk u_\nk } \,.
\end{equation}
The phonon-assisted current operator is given by
\begin{equation} \label{eq:current_Jp}
    \opmbJ^{\rm p}
    = \frac{1}{\sqrt{N_\bk}} \sum_{\substack{\bk\bq \\ mn\nu}} (\mb{\mc{D}} g)_{mn\nu}(\bk, \bq) \, \opcd_\mkq \, \opc_\nk \, \hat{x}_\nuq \,,
\end{equation}
where $(\mb{\mc{D}} g)_{mn\nu}(\bk, \bq)$ is the covariant derivative of the \eph coupling defined as
\begin{align} \label{eq:current_Dg}
    &(\mb{\mc{D}} g)_{mn\nu}(\bk, \bq)\equiv \mel{u_\mkq}{\bigl( \nabla_\bk \frac{\partial H_\bk}{\partial x_\nuq} \bigr)}{u_\nk}\\
    &= \nabla_\bk g_{mn\nu}(\bk, \bq)
    -i [\mb{\xi}_\bkq g_{\nu}(\bk, \bq)]_{mn}
    +i [g_{\nu}(\bk, \bq) \mb{\xi}_\bk]_{mn} \,.
    \nonumber
\end{align}
This term can be understood as the change in the electron velocity matrix due to phonon displacement $\hat{x}_\nuq$, and is nonzero only when the \eph coupling is nonlocal.
In the single-band case, nonlocality refers to the $\bk$-dependence of $g(\bk, \bq)$, whereas in multiband systems it can also arise from the Berry connection.
In practice, we evaluate $\mb{\mc{D}} g$ using the tight-binding approximation, where the Berry connection is assumed to be diagonal in the Wannier function basis~\cite{Graf1995}: see \Eq{eq:current_Dg_tight-binding}.
The effects of phonon-assisted current on electronic conductivity have been studied for the Peierls model~\cite{Gosar1966, Gosar1970, Hannewald2004, Jankovic2025a, Jankovic2025b} and for a two-band model Hamiltonian~\cite{Hu2024}, but not for real materials.
The electronic and phonon-assisted current operators are diagrammatically illustrated in \Fig{fig:feynman_current}(a).

Finally, we note that the diamagnetic current~[\Eq{eq:current_JD}] can be indirectly computed from the current-current correlator: see \Sec{sec:supp_diamagnetic} of the SM~\cite{supplemental}.
Yet, since the diamagnetic current contributes only to the imaginary part of the conductivity, which can be obtained from the real part by the Kramers--Kronig relation, we do not explicitly evaluate the diamagnetic current.

\subsection{Current-current susceptibility and conductivity} \label{sec:conductivity}

\begin{figure}[tb]
\centering
\includegraphics[width=0.99\linewidth]{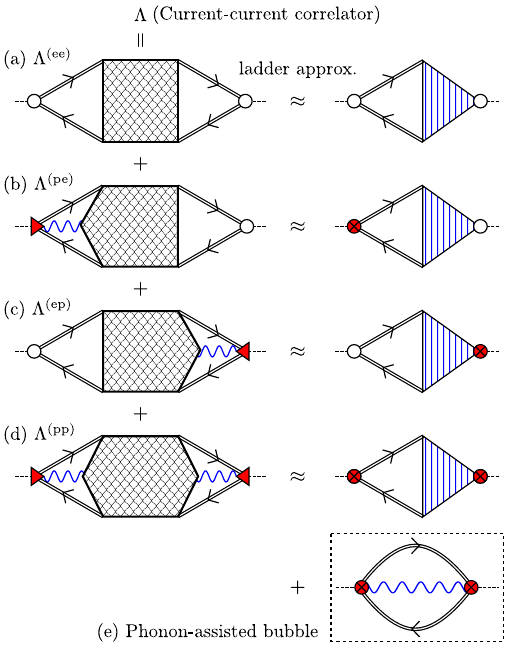}
\caption{
    Feynman diagrams for the current-current susceptibility: (a) electron-electron, (b) phonon-electron, (c) electron-phonon, and (d) phonon-phonon contributions [\Eq{eq:Lambda_def}].
    The cross-hatched polygons on the left denote the sum of all Feynman diagrams connecting the electron and phonon lines.
    The diagrams on the right correspond to the solution of the self-consistent ladder equation [\Fig{fig:feynman_ladder}(e)], with an additional contribution to the $\Lambda^{\rm (pp)}$ susceptibility from the phonon-assisted bubble diagram [panel (e), \Eq{eq:sigma_pp_bubble}].
}
\label{fig:feynman_conductivity}
\end{figure}

The electronic conductivity for electric fields of frequency $\Omega$ can be written as~\cite{MahanBook, Basov2011RMP}
\begin{equation} \label{eq:sigma_chiR}
    \sigma_{\alpha\beta}(\Omega)
    = \frac{i e^2}{\Omega V^{\rm uc}} \bigl[
        \Lambda^{\rm R}_{\alpha\beta}(\Omega) - \expval{\hat{J}^{\rm D}_{\alpha\beta}}
    \bigr] \,,
\end{equation}
where $V^{\rm uc}$ is the unit cell volume and $\Lambda^{\rm R}_{\alpha\beta}(\Omega)$ the retarded component of the current-current susceptibility
\begin{equation} \label{eq:chi_def}
    \Lambda^{c\cpr}_{\alpha\beta}(\Omega)
    \equiv \chi^{c\cpr}_{J_{\alpha} J_{\beta}}(\Omega, \bQ = \mb{0}) \,.
\end{equation}
For $n$-doped semiconductors, mobility is computed by dividing the conductivity by the carrier density
\begin{equation} \label{eq:sigma_density}
    n_{\rm c}^{n\tbar{\rm doped}} = \frac{1}{V^{\rm uc}} \sum_{n \in {\rm CB}} \intbk \nint \dd \veps \, A_{nn\bk}(\veps) f^+(\veps) \,,
\end{equation}
where one only counts the free carriers in the conduction bands (CB).
In $p$-doped systems, carriers are instead counted in the valence bands (VB):
\begin{equation} \label{eq:sigma_density_hole}
    n_{\rm c}^{p\tbar{\rm doped}} = \frac{1}{V^{\rm uc}} \sum_{n \in {\rm VB}} \intbk \nint \dd \veps \, A_{nn\bk}(\veps) f^-(\veps) \,.
\end{equation}

In this work, we focus on time-reversal invariant systems, where the conductivity is purely longitudinal.
The transverse linear conductivity corresponds to the anomalous Hall effect~\cite{Nagaosa2010}, which can be computed from the Berry curvature~\cite{Wang2006} without considering the \eph scattering.
The longitudinal conductivity is the symmetric part of $\sigma$:
\begin{equation}
    \sigma^{\rm L}_{\alpha\beta}(\Omega) \equiv \frac{1}{2} \bigl[ \sigma_{\alpha\beta}(\Omega) + \sigma_{\beta\alpha}(\Omega) \bigr] \,.
\end{equation}
Then, \Eq{eq:sigma_chiR} can be rewritten using the fluctuation-dissipation relations \eqref{eq:fdr_boson} as
\begin{align} \label{eq:sigma_chi}
    \Re \sigma^{\rm L}_{\alpha\beta}(\Omega)
    &= -\frac{e^2}{2 \Omega V^{\rm uc}} \Im \bigl[ \Lambda^{\rm R}_{\alpha\beta}(\Omega) - (\Lambda^{\rm R}_{\beta\alpha}(\Omega))^* \bigr]
    \\
    &= -\frac{e^2}{2 \Omega V^{\rm uc}} \Im \bigl[ \Lambda^{\rm R}_{\alpha\beta}(\Omega) - \Lambda^{\rm A}_{\alpha\beta}(\Omega) \bigr]
    \nnnl
    &= -\frac{e^2}{2 \Omega[1 + 2n(\Omega)] V^{\rm uc}} \Im [\Lambda^{>}_{\alpha\beta}(\Omega) + \Lambda^{<}_{\alpha\beta}(\Omega)] \,. \nonumber
\end{align}
The prefactor in Eq.~\eqref{eq:sigma_chi} remains finite as $\Omega \to 0$,
\begin{equation}
    \lim_{\Omega \to 0} \frac{1}{2\Omega[1 + 2n(\Omega)]} = \frac{1}{4T} \,,
\end{equation}
enabling one to write the longitudinal dc conductivity as
\begin{equation} \label{eq:sigma_chi_dc}
    \Re \sigma^{\rm L}_{\alpha\beta}(0)
    = -\frac{e^2}{4T V^{\rm uc}} \Im [\Lambda^{>}_{\alpha\beta}(0) + \Lambda^{<}_{\alpha\beta}(0)] \,.
\end{equation}
This expression allows for a direct evaluation of the dc conductivity at $\Omega = 0$, avoiding the need for a finite-frequency extrapolation for the limit $\Omega \to 0$.

Since the current has both electronic [$\opmbJ^{\rm e}$, \Eq{eq:current_Je}] and phonon-assisted [$\opmbJ^{\rm p}$, \Eq{eq:current_Jp}] components, the current-current susceptibility splits into four contributions:
\begin{subequations} \label{eq:Lambda_def}
\begin{alignat}{2}
    \Lambda^{c\cpr}_{\alpha\beta}(\Omega)
    \label{eq:Lambda_def_ee}
    &= \quad \chi^{c\cpr}_{J_{\alpha}^{\rm e} J_{\beta}^{\rm e}}(\Omega, \bQ = \mb{0})
    &&\  \bigl( \ = \Lambda^{{\rm (ee)}c\cpr}_{\alpha\beta}(\Omega) \ \bigr)
    \\
    \label{eq:Lambda_def_pe}
    &\quad + \chi^{c\cpr}_{J_{\alpha}^{\rm p} J_{\beta}^{\rm e}}(\Omega, \bQ = \mb{0})
    &&\  \bigl( \ = \Lambda^{{\rm (pe)}c\cpr}_{\alpha\beta}(\Omega) \ \bigr)
    \\
    \label{eq:Lambda_def_ep}
    &\quad + \chi^{c\cpr}_{J_{\alpha}^{\rm e} J_{\beta}^{\rm p}}(\Omega, \bQ = \mb{0})
    &&\  \bigl( \ = \Lambda^{{\rm (ep)}c\cpr}_{\alpha\beta}(\Omega) \ \bigr)
    \\
    \label{eq:Lambda_def_pp}
    &\quad + \chi^{c\cpr}_{J_{\alpha}^{\rm p} J_{\beta}^{\rm p}}(\Omega, \bQ = \mb{0})
    &&\  \bigl( \ = \Lambda^{{\rm (pp)}c\cpr}_{\alpha\beta}(\Omega) \ \bigr)\,.
\end{alignat}
\end{subequations}
The corresponding Feynman diagrams appear on the left column of \Fig{fig:feynman_conductivity}.
Now, let us evaluate each term within the self-consistent ladder approximation.

\subsection{Purely electronic current-current susceptibility} \label{sec:transport_el}

\begin{figure}[tb]
\centering
\includegraphics[width=0.99\linewidth]{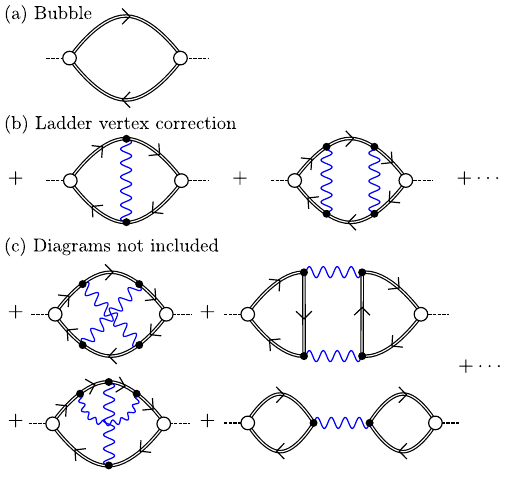}
\caption{
    Feynman diagrams for the purely electronic current susceptibility $\Lambda^{\rm (ee)}$ [\Eq{eq:Lambda_def_ee}, \Fig{fig:feynman_conductivity}(a)]:
    (a) the bubble diagram [\Eq{eq:sigma_bubble}],
    (b) the ladder vertex correction, and
    (c) diagrams which are not included in the ladder approximation (see also \Fig{fig:feynman_not_included}).
}
\label{fig:feynman_conductivity_expansion}
\end{figure}

We start with the purely electronic term [\Eq{eq:Lambda_def_ee}], which is written using \Eq{eq:sigma_XY_R_Sigma} as
\begin{equation} \label{eq:sigma_chi_from_F}
    \Lambda^{{\rm (ee)}c\cpr}_{\alpha\beta}(\Omega)
    = \sum_{12} \intkk \delta^c_{J^{\rm e}_\alpha} \Sigma_{0, 21}(k+\Omega, -\Omega)
    \delta^{\cpr}_{J^{\rm e}_\beta} G_{12}(k, \Omega) \,.
\end{equation}
Here, $k + \Omega = (\veps + \Omega, \bk)$, and the bare vertex for the electronic current [\Eq{eq:current_Je}] is
\begin{equation} \label{eq:sigma_V0_el}
    \delta^c_{J^{\rm e}_\alpha} \Sigma_{0, 12}(k, \Omega)
    = \mb{v}_{n_1 n_2 \bk} \, \delta_{c_1 c} \delta_{c_2 c} \,.
\end{equation}

The simplest approach to evaluate this susceptibility is to use the bubble approximation, which ignores all vertex corrections and uses the bare response function [\Eq{eq:lr_R0_def}] instead of the full one.
Then, the susceptibility becomes
\begin{align} \label{eq:sigma_Lambda_ee_bubble}
    &\Lambda^{{\rm (Bubble)}c\cpr}_{\alpha\beta}(\Omega)   = \sum_{n_1 n_2} \intkk v_{n_2 n_1\bk}^{\alpha} \,
    \delta^{\cpr}_{J^{\rm e}_\beta} G^{cc}_{0, n_1 n_2}(k, \Omega) \nonumber \\
    &= \sum_{\substack{n_1 n_2 \\ m_1 m_2}} \intkk v_{n_1 n_2\bk}^{\alpha *} \, v^\beta_{m_1 m_2 \bk} \,
    G_{n_1 m_1}^{c\cpr}(k+\Omega) G_{m_2 n_2}^{\cpr c}(k) \,.
\end{align}
The corresponding Feynman diagram is shown in \Fig{fig:feynman_conductivity_expansion}(a).
Within the diagonal Green's function approximation, \Eq{eq:sc_G_diag_approx}, the longitudinal bubble conductivity reads
\begin{multline} \label{eq:sigma_bubble}
    \Re \sigma^{\rm L,\, (Bubble)}_{\alpha\beta}(\Omega)
    = \frac{\pi e^2}{V^{\rm uc}} \sum_{n_1 n_2} \intbk \Re \bigl( v^{\alpha*}_{n_1 n_2 \bk} v^\beta_{n_1 n_2 \bk} \bigr)
    \\
    \times \nint \dd \veps \, \frac{f^+(\veps) - f^+(\veps + \Omega)}{\Omega} A_{n_1 \bk}(\veps + \Omega) A_{n_2 \bk}(\veps) \,,
\end{multline}
where we provide the derivation in \Sec{sec:sigma_bubble_derivation} of the SM~\cite{supplemental}.
This expression is commonly used to compute the dc and ac conductivity from spectral functions~\cite{Basov2011RMP, Zhou2019STO, Chang2022, Pickem2022, Abramovitch2023, Abramovitch2024}.

To go beyond the bubble approximation, we solve the self-consistent ladder equation for the current vertex.
Explicitly writing out \Eqs{eq:lr_R0_def} and \eqref{eq:lr_ladder} for the electron current vertex, we obtain
\begin{equation} \label{eq:sigma_V_to_R}
    \delta^c_{J^{\rm e}_\beta} G_{12}(k, \Omega)
    = \sum_{34} \Pi_{1234}(k, \Omega) \,
    \delta^c_{J^{\rm e}_\beta} \Sigma_{34}(k, \Omega)
\end{equation}
and
\begin{multline} \label{eq:sigma_BSE}
    \delta^c_{J^{\rm e}_\beta} \Sigma_{12}(k, \Omega)
    =
    \delta^c_{J^{\rm e}_\beta} \Sigma_{0,12}(k, \Omega)
    \\
    + \! \!  \sum_{n_3 n_4} \intqq
    W^{c_1 c_2 c_1 c_2}_{n_1 n_2 n_3 n_4}(\bk, \bk, q)
    \delta^c_{J^{\rm e}_\beta} G^{c_1 c_2}_{n_3 n_4}(k + q, \Omega)
    \,.
\end{multline}
The self-consistent solution of these equations gives the renormalized current response function including vertex corrections.
The current-current susceptibility and conductivity can then be computed using \Eqs{eq:sigma_chi} and \eqref{eq:sigma_chi_from_F}.
The ladder vertex correction to the susceptibility is depicted in \Fig{fig:feynman_conductivity_expansion}(b).
The non-ladder diagrams shown in \Fig{fig:feynman_conductivity_expansion}(c) are not included in this approximation.

We note that applying a weak-coupling approximation to the full ladder equation yields simplified expressions only involving the fully retarded current vertex~\cite{Holstein1964}.
With the additional assumption that the renormalization depends solely on energy and not on momentum, the conductivity can be expressed in terms of a transport-corrected Eliashberg function~\cite{Scher1970, Allen1971, Kupcic2014, Kupcic2015}, analogous to the momentum relaxation-time approximation.
A systematic comparison between these simple yet approximate expressions and the full ladder-\scGD is a subject of further study.

\subsection{Phonon-assisted current-current susceptibility} \label{sec:transport_ph}

Next, we consider the phonon-assisted current.
Unlike the electronic case, the phonon-assisted current operator [\Eq{eq:current_Jp}] involves a phonon operator alongside electronic operators, preventing direct treatment with the ladder equation.
To define the bare \eph current vertex as an electronic vertex function, we eliminate the phonon operator using the noncrossing approximation.
This attaches the phonon propagator to one of the two electronic lines, as shown in \Fig{fig:feynman_current}(b), leaving only electronic external lines that can be treated with the ladder equation.
The phonon-assisted current vertex is then given by
\begin{align} \label{eq:lr_Vbare_ep}
    &\delta^c_{J^{\rm p}_\beta} \Sigma_{0,12}(k, \Omega)
    \nnnl
    &= - Z^{c_1} \delta_{c_2 c} \!\! \sum_{n_3 n_4 \nu} \intqq G^{c_1 c_2}_{n_3 n_4}(k+q + \Omega)
    \nnnl
    &\qquad \times g^*_{n_3 n_1 \nu}(\bk, \bq) (\mb{\mc{D}} g)_{n_4 n_2 \nu}(\bk, \bq)
    D^{c_2 c_1}_{0, \nu}(q)
    \nnnl
    &\quad - Z^{c_2} \delta_{c_1 c} \!\! \sum_{n_3 n_4 \nu} \intqq G^{c_1 c_2}_{n_3 n_4}(k+q)
    \nnnl
    &\qquad \times (\mb{\mc{D}} g)^*_{n_3 n_1 \nu}(\bk, \bq) g_{n_4 n_2 \nu}(\bk, \bq)
    D^{c_2 c_1}_{0, \nu}(q)
    \,,
\end{align}
which corresponds to the two diagrams on the right in \Fig{fig:feynman_current}(b).
Comparing \Eq{eq:lr_Vbare_ep} with \Eq{eq:sc_Sigma_c}, we observe that the phonon-assisted current vertex has a structure similar to the \scGD self-energy, except that one of the two \eph vertices is replaced with its covariant derivative.

Importantly, this step is not an ad hoc approximation.
The phonon-assisted current vertex in this form can be derived by first writing the ladder vertex correction in the length gauge, where the electric field is introduced through a scalar potential, and then performing a transformation to the velocity gauge.
The noncrossing approximation is required to ensure the equivalence between the length- and velocity-gauge formulations.
Technically, \Eq{eq:lr_Vbare_ep} is derived from the Ward--Takahashi identity; see \App{app:Ward} for details.

Using the bare phonon-assisted current vertex, the $\Lambda^{\rm (pe)}$ susceptibility [\Eq{eq:Lambda_def_pe}] can be written as
\begin{equation} \label{eq:sigma_Lambda_pe}
    \Lambda^{{\rm (pe)}c\cpr}_{\alpha\beta}(\Omega)
    = \sum_{12} \intkk
    \delta^c_{J^{\rm p}_\alpha} \Sigma_{0, 21}(k+\Omega, -\Omega) \,
    \delta^{\cpr}_{J^{\rm e}_\beta} G_{12}(k, \Omega) \,.
\end{equation}
Similarly, by solving the self-consistent ladder equations, analogous to \Eqs{eq:sigma_V_to_R} and \eqref{eq:sigma_BSE}, but replacing $\delta^c_{J^{\rm e}_\beta} \Sigma_{0, 12}$ with $\delta^c_{J^{\rm p}_\beta} \Sigma_{0, 12}$, we obtain the $\Lambda^{\rm (ep)}$ susceptibility:
\begin{equation} \label{eq:sigma_Lambda_ep}
    \Lambda^{{\rm (ep)}c\cpr}_{\alpha\beta}(\Omega)
    = \sum_{n_1 n_2} \intkk v^{\alpha*}_{n_1 n_2 \bk} \,
    \delta^{\cpr}_{J^{\rm p}_\beta} G^{cc}_{n_1 n_2}(k, \Omega) \,.
\end{equation}
These two cross terms are related by symmetry [\Eq{eq:chi_symmetry}]:
\begin{equation}
    \Lambda^{{\rm (ep)}c\cpr}_{\alpha\beta}(\Omega)
    = \Lambda^{{\rm (pe)}\cpr c}_{\beta\alpha}(-\Omega) \,.
\end{equation}

For the $\Lambda^{\rm (pp)}$ susceptibility, which involves two phonon-assisted current operators, there is an additional contribution beyond the ladder diagrams.
This arises because the two phonon operators, generated by the external field, can be directly connected as illustrated in \Fig{fig:feynman_conductivity}(e).
We call this the phonon-assisted bubble diagram, which represents an electron-hole pair and a phonon created by the external field propagating independently.
The phonon-phonon current susceptibility is given by
\begin{equation} \label{eq:sigma_Lambda_pp}
     \Lambda^{{\rm (pp)}c\cpr}_{\alpha\beta}(\Omega)
    = \Lambda^{{\rm (pp\tbar Ladder)}c\cpr}_{\alpha\beta}(\Omega)
    + \Lambda^{{\rm (pp\tbar Bubble)}c\cpr}_{\alpha\beta}(\Omega) \,,
\end{equation}
where
\begin{multline} \label{eq:sigma_Lambda_pp_Ladder}
    \Lambda^{{\rm (pp\tbar Ladder)}c\cpr}_{\alpha\beta}(\Omega)
    \\
    = \sum_{12} \intkk
    \delta^c_{J^{\rm p}_\alpha} \Sigma_{0, 21}(k+\Omega, -\Omega) \,
    \delta^{\cpr}_{J^{\rm p}_\beta} G^{c}_{12}(k, \Omega) \,,
\end{multline}
and
\begin{multline} \label{eq:sigma_Lambda_pp_Bubble}
    \Lambda_{\alpha\beta}^{{\rm (pp\tbar Bubble)} c\cpr}(\Omega)
    = \!\! \sum_{\substack{n_1 n_2 \\ m_1 m_2 \nu}} \nsint{kq}
    G^{c\cpr}_{n_1 m_1}(k+q+\Omega) G^{\cpr c}_{m_2 n_2}(k)
    \\
    \times (\mc{D}_\alpha g)^*_{n_1 n_2 \nu}(\bk, \bq)
    (\mc{D}_\beta g)_{m_1 m_2 \nu}(\bk, \bq)
    D^{\cpr c}_{0, \nu}(q) \,.
\end{multline}
Interestingly, the phonon-assisted bubble conductivity naturally appears when deriving the conductivity using the continuity equation and Ward identity, as discussed in \Sec{sec:dielectric} and \App{app:Ward}.

Within the diagonal Green's function approximation [\Eq{eq:sc_G_diag_approx}], the contribution of the phonon-assisted bubble to the longitudinal conductivity is given by
\begin{align} \label{eq:sigma_pp_bubble}
    & \sigma_{\alpha\beta}^{{\rm L,\,(pp\tbar Bubble)}}(\Omega)
    \nnnl
    &= \frac{\pi e^2}{V^{\rm uc}}
    \! \sum_{n_1 n_2 \nu} \nsint{\bk \bq}
    (\mc{D}_\alpha g)^*_{n_1 n_2\nu}(\bk, \bq) (\mc{D}_\beta g)_{n_1 n_2 \nu}(\bk, \bq)
    \nnnl
    &\ \times \sum_\pm \nint \dd \veps \, \Bigl\{
    \bigl[ n_\nuq \!+\! f^\pm(\veps) \bigr]
    \frac{f^+(\veps \!\mp\! \omega_\nuq) - f^+(\veps \!\mp\! \omega_\nuq \!+\! \Omega)}{\Omega}
    \nnnl
    &\qquad \times
    A_{n_1 \bkq}(\veps \!\mp\! \omega_\nuq \!+\! \Omega) A_{n_2 \bk}(\veps)
    \Bigr\} \,,
\end{align}
with the derivation in \Sec{sec:sigma_bubble_derivation} of SM~\cite{supplemental}.
This equation describes transport due to an electron-hole pair at states $n_1\bkq$ and $n_2\bk$, with energies $\veps \mp \omega_\nuq + \Omega$ and $\veps$, created while emitting a phonon at $\nu\!-\!\bq$ (or absorbing a phonon at $\nuq$).

The phonon-assisted bubble introduces the possibility of creating an electron-hole pair with a nonzero total momentum, similar to indirect optical absorption processes~\cite{Cheeseman1952Indabs, Hall1954Indabs, BassaniBook}.
However, we emphasize that the physical origin of the phonon-assisted bubble is distinct.
The phonon-assisted bubble is induced by the phonon-induced perturbation of the optical matrix elements, which is absent in the semiclassical theory of indirect optical absorption~\cite{Kioupakis2010, Noffsinger2012, Tiwari2024}.

\subsection{Ladder-\texorpdfstring{\scGD}{scGD0} equations under the diagonal approximation} \label{sec:transport_diagonal}

In practice, we use the diagonal approximation for Green's functions [\Eq{eq:sc_G_diag_approx}] and neglect interband current matrix elements.
Hence, we write $\mb{v}_\nk = \mb{v}_{nn\bk}$ for short.
The diagonal approximation is justified when interband contributions to the conductivity are negligible, as in transport at DC and THz frequencies.
However, it should be relaxed when studying conductivities in the infrared or optical regimes where interband effects become significant.

Below, we summarize the central equations for ladder-\scGD transport calculations within this diagonal approximation.
The bare vertices [\Eqs{eq:sigma_V0_el} and \eqref{eq:lr_Vbare_ep}] read
\begin{equation} \label{eq:sigma_V0_el_diag}
    \delta^c_{\mb{J}^{\rm e}} \Sigma_{0, \nk}^{c_1 c_2}(\veps, \Omega)
    = \mb{v}_\nk \delta_{c_1 c} \delta_{c_2 c} \,,
\end{equation}
and
\begin{widetext}
\begin{align} \label{eq:lr_Vbare_ep_diag}
    \delta^c_{\mb{J}^{\rm p}} \Sigma^{c_1 c_2}_{0, \nk}(\veps, \Omega)
    &= - Z^{c_1} \delta_{c_2 c} \sum_{m\nu} \intqq D^{c_2 c_1}_{0, \nu}(q)
    g^*_{mn\nu}(\bk, \bq) (\mb{\mc{D}} g)_{mn\nu}(\bk, \bq)
    G^{c_1 c_2}_{\mkq}(\veps + \omega + \Omega)
    \nnnl
    &\quad - Z^{c_2}\delta_{c_1 c}  \sum_{m\nu} \intqq D^{c_2 c_1}_{0, \nu}(q)
    (\mb{\mc{D}} g)^*_{mn\nu}(\bk, \bq) g_{mn\nu}(\bk, \bq)
    G^{c_1 c_2}_{\mkq}(\veps + \omega)
    \,.
\end{align}
Note that here $G$ denotes the \scGD Green's functions computed based on the theory of \Sec{sec:selfen}.
The self-consistent ladder equations [\Eqs{eq:sigma_V_to_R} and \eqref{eq:sigma_BSE}] have the same form for the electronic and phonon-assisted responses and read
\begin{equation} \label{eq:sigma_V_to_R_diag}
    \delta^c_{J^{\rm e,p}_\beta} G^{c_1 c_2}_{\nk}(\veps, \Omega)
    = \sum_{c_3 c_4} G^{c_1 c_3}_{\nk}(\veps + \Omega) \,
    \delta^c_{J^{\rm e,p}_\beta} \Sigma^{c_3 c_4}_\nk(\veps, \Omega) \,
    G^{c_4 c_2}_\nk(\veps) \,,
\end{equation}
and
\begin{equation} \label{eq:sigma_BSE_diag}
    \delta^c_{J^{\rm e,p}_\beta} \Sigma^{c_1 c_2}_{\nk}(\veps, \Omega)
    = \delta^c_{J^{\rm e,p}_\beta} \Sigma^{c_1 c_2}_{0, \nk}(\veps, \Omega)
    - Z^{c_1} Z^{c_2} \sum_{m \nu}
    \intqq
    D^{c_2 c_1}_{0, \nu}(q)
    \abs{g_{m n \nu}(\bk, \bq)}^2
    \delta^c_{J^{\rm e,p}_\beta} G^{c_1 c_2}_{\mkq}(\veps + \omega, \Omega)
    \,.
\end{equation}
\end{widetext}
The current-current susceptibility [\Eq{eq:Lambda_def}] then reads

\begin{align} \label{eq:sigma_Lambda_diag}
    \Lambda^{c\cpr}_{\alpha\beta}(\Omega)
    &= \sum_{n} \intkk
    v^\alpha_\nk \,  \Bigl[
          \delta^{\cpr}_{J^{\rm e}_\beta} G^{cc}_{\nk}(\veps, \Omega)
        + \delta^{\cpr}_{J^{\rm p}_\beta} G^{cc}_{\nk}(\veps, \Omega)
    \Bigr]
    \nnnl
    &+ \! \sum_{n c_1 c_2} \intkk
    \delta^c_{J^{\rm p}_\alpha} \Sigma^{c_2 c_1}_{0, \nk}(\veps + \Omega, -\Omega)
    \nnnl
    &\hspace{1.5em} \times \Bigl[
          \delta^{\cpr}_{J^{\rm e}_\beta} G^{c_1 c_2}_{\nk}(\veps, \Omega)
        + \delta^{\cpr}_{J^{\rm p}_\beta} G^{c_1 c_2}_{\nk}(\veps, \Omega)
    \Bigr]
    \,.
\end{align}

Finally, we obtain the real part of the longitudinal ac and dc conductivities [\Eqs{eq:sigma_chi} and \eqref{eq:sigma_chi_dc}] using
\begin{multline} \label{eq:sigma_chiR_diag}
    \Re \sigma^{{\rm L,\, Ladder\tbar sc}GD_0}_{\alpha\beta}(\Omega)
    \\
    = -\frac{e^2}{2 \Omega[1 + 2n(\Omega)] V^{\rm uc}} \Im [\Lambda^{>}_{\alpha\beta}(\Omega) + \Lambda^{<}_{\alpha\beta}(\Omega)] \,,
\end{multline}
where the $\Omega \to 0$ limit of the prefactor is well defined: $1 / 2 \Omega[1 + 2n(\Omega)] = 1 / 4T$.
Neglecting the vertex correction, we obtain the bubble-\scGD conductivity [\Eq{eq:sigma_bubble}]
\begin{multline} \label{eq:sigma_bubble_scGD_diag}
    \Re \sigma^{{\rm L,\, Bubble\tbar sc}GD_0}_{\alpha\beta}(\Omega)
    = \frac{\pi e^2}{V^{\rm uc}} \sum_{n} \intbk \Re \bigl( v^{\alpha*}_\nk v^\beta_\nk \bigr) \\
   \! \! \times \! \nint \dd \veps \, \frac{f^+(\veps) \!-\! f^+(\veps \!+\! \Omega)}{\Omega} A^{\mathrm{sc}GD_0}_{\nk}(\veps \!+\! \Omega) A^{\mathrm{sc}GD_0}_{\nk}(\veps) \,,
\end{multline}
where we again neglected the interband velocity.
By using the \GD spectral function instead of the \scGD one, we obtain the bubble-\GD conductivity
\begin{multline} \label{eq:sigma_bubble_G0D0_diag}
    \Re \sigma^{{\rm L,\, Bubble\tbar}G_0 D_0}_{\alpha\beta}(\Omega)
    = \frac{\pi e^2}{V^{\rm uc}} \sum_{n} \intbk \Re \bigl( v^{\alpha*}_\nk v^\beta_\nk \bigr)    \\
   \! \times \! \nint \dd \veps \, \frac{f^+(\veps) \!-\! f^+(\veps \!+\! \Omega)}{\Omega} A^{G_0 D_0}_{\nk}(\veps \!+\! \Omega) A^{G_0 D_0}_{\nk}(\veps) \,.
\end{multline}
The complex conductivity for \Eqs{eq:sigma_chiR_diag}--\eqref{eq:sigma_bubble_G0D0_diag} follows from the Kramers--Kronig relation:
\begin{equation} \label{eq:sigma_KK}
    \sigma^{\rm L}_{\alpha\beta}(\Omega)
    = \Re \sigma^{\rm L}_{\alpha\beta}(\Omega)
    - \frac{i}{\pi} \mcP \! \nint \dd\Omega' \, \frac{
        \Re \sigma^{\rm L}_{\alpha\beta}(\Omega')
    }{\Omega' - \Omega} \,.
\end{equation}

\begin{figure}[b]
\centering
\begin{tikzpicture}[node distance = 1.4cm and 2cm]
    \node (start0) [startstop, align=center,] {
        Input: $\veps_\nk$, $\omega_\nuq$, $g_{mn\nu}(\bk, \bq)$
    };
    \node (bef0) [wrapout, below of=start0, align=center,] {};
    \node (bef1) [wrap, below of=start0, align=center,] {
        \scGD (\Fig{fig:selfen_flowchart})\\$G_\nk(\veps)$
    };
    \node (bef11) [io, align=center, right of=bef1, xshift=2.5cm] {
        Output (Bubble):\\
        $\sigma^{{\rm L,\, Bubble\tbar sc}GD_0}_{\alpha\beta}(\Omega)$\\
        {[}Eqs.~(\ref{eq:sigma_bubble_scGD_diag}, \ref{eq:sigma_KK}){]}
    };
    \node (bef3) [process, below of=bef1] {
        $\delta_{\mb{J}^{\rm e}} \Sigma_0$ [\Eq{eq:sigma_V0_el}]
        $\delta_{\mb{J}^{\rm p}} \Sigma_0$ [\Eq{eq:lr_Vbare_ep}]
    };
    \node (start) [process, below of=bef3] {
        Initial guess: $\delta_{\mb{J}} \Sigma = \delta_{\mb{J}} \Sigma_0$\\
        {[}Eqs.~(\ref{eq:sigma_V0_el}, \ref{eq:lr_Vbare_ep}){]}
    };
    \node (pro2) [process, below of=start] {
        $\delta_\mb{J} G$ [\Eq{eq:sigma_V_to_R}]
    };
    \node (pro3) [process, below of=pro2, yshift=0.2cm] {
        $\Lambda_{\alpha\beta}(\Omega)$\\
        {[}Eqs.~(\ref{eq:sigma_chi_from_F}, \ref{eq:sigma_Lambda_ep}-\ref{eq:sigma_Lambda_pp}){]}
    };
    \node (pro4) [process, below of=pro3, yshift=0.2cm] {
        $\sigma_{\alpha\beta}(\Omega)$ [\Eq{eq:sigma_chi}]
        };
    \node (dec) [decision, below of=pro4, aspect=3, align=center, inner sep=-0.1ex, yshift=-0.1cm] {
        $\textrm{max} \abs{\sigma^{\rm in} - \sigma^{\rm out}} < s$?
    };
    \node (mix2) [process, right of=pro2, xshift=2.5cm, yshift=-0.5cm] {
        $\delta_\mb{J} \Sigma$ [\Eq{eq:sigma_BSE}]
    };
    \node (mix1) [process, right of=pro4, xshift=2.5cm, yshift=+0.5cm] {
        Mix $\delta_\mb{J} G$ [\Eq{eq:sigma_mixing}]
    };
    \node (out) [io, align=center, below of=dec, yshift=-0.3cm] {
        Output (Ladder):\\
        $\sigma^{{\rm L,\, Ladder\tbar sc}GD_0}_{\alpha\beta}(\Omega)$\\
        {[}Eqs.~(\ref{eq:sigma_chiR_diag}, \ref{eq:sigma_KK}){]}
    };
    \draw [arrow] (start0) -- (bef1);
    \draw [arrow] (bef0) -- (bef11);
    \draw [arrow] (bef1) -- (bef3);
    \draw [arrow] (bef3) -- (start);
    \draw [arrow] (start) -- (pro2);
    \draw [arrow] (pro2) -- (pro3);
    \draw [arrow] (pro3) -- (pro4);
    \draw [arrow] (pro4) -- (dec);
    \draw [arrow] (dec) -| node[anchor=west] {no} (mix1);
    \draw [arrow] (mix1) -- (mix2);
    \draw [arrow] (mix2) |- (start);
    \draw [arrow] (dec) -- node[anchor=west] {yes} (out);
\end{tikzpicture}
\caption{
    Flowchart for the ladder-\scGD calculation of electronic conductivity.
}
\label{fig:sigma_flowchart}
\end{figure}
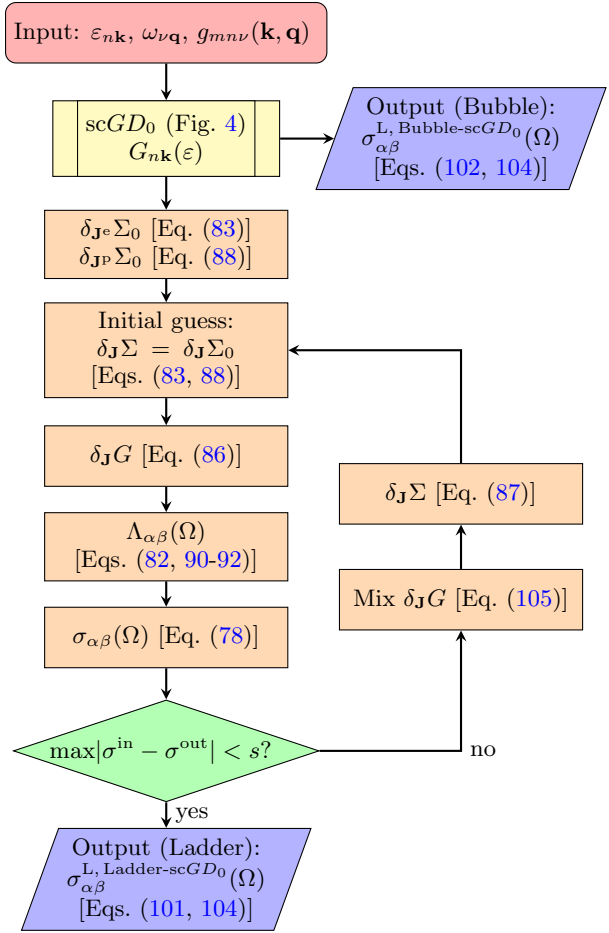

The workflow for solving the self-consistent ladder equations is summarized in \Fig{fig:sigma_flowchart}.
After each iteration, we use linear mixing to update the current response function:
\begin{equation} \label{eq:sigma_mixing}
    \delta^c_{J^{\rm e}_\beta} G^{\rm next}
    = \lambda^{\rm mix} \, \delta^c_{J^{\rm e}_\beta} G^{\rm out}
    + (1 - \lambda^{\rm mix}) \delta^c_{J^{\rm e}_\beta} G^{\rm in} \,.
\end{equation}
We use $\lambda^{\rm mix} = 1$ unless convergence is not achieved, in which case we set $\lambda^{\rm mix} = 0.5$.
The iteration continues until the relative change in conductivity is below 1\%.
Since the ladder equation does not couple the response functions for different external frequencies $\Omega$, the equations can be solved separately for each $\Omega$.

\begin{figure*}[tb]
\centering
\includegraphics[width=0.99\linewidth]{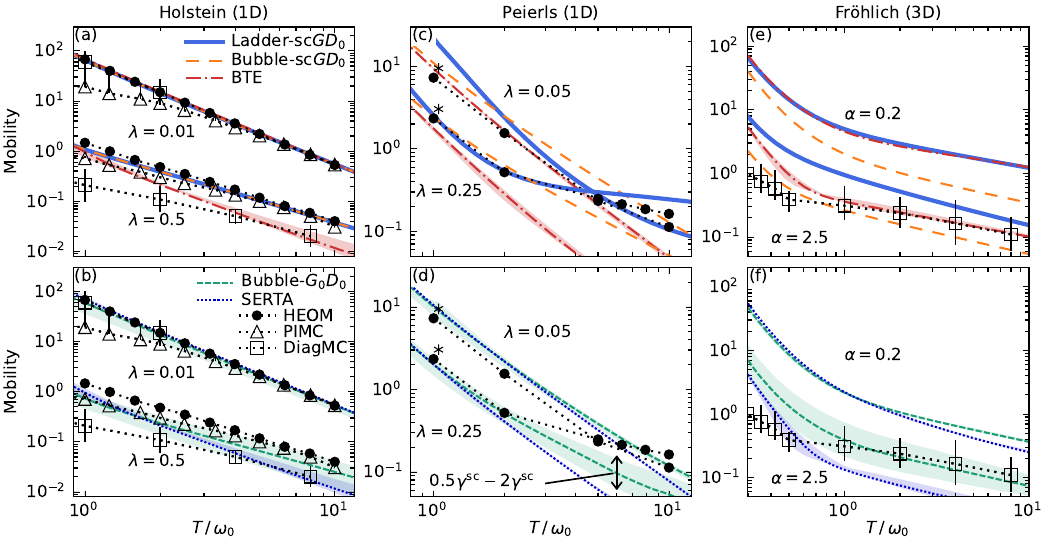}
\caption{
    Mobility of (a, b) the 1D Holstein model with $t = \omega_0 = 1$ and $\lambda = 0.01$ or $0.5$,
    (c, d) the 1D Peierls model with $t = \omega_0 = 1$, and $\lambda = 0.05$ or $0.25$, and
    (e, f) the 3D Fr\"ohlich model with $m_0 = \omega_0 = 1$ and $\alpha = 0.2$ or $2.5$.
    In the first row, we compare the ladder-\scGD method, featuring the fewest approximations, with the bubble-\scGD method, which neglects the vertex corrections, and the BTE, which adopts the quasiparticle approximation and neglects the phonon-assisted current (see \Fig{fig:sigma_methods}).
    In the second row, we show the bubble-\GD and SERTA results.
    Shaded regions represent the variation in mobility when the artificial broadening is varied from $0.5\gamma^{\rm sc}$ to $2\gamma^{\rm sc}$.
    The mobility is compared against the hierarchical equations of motion (HEOM) results of Ref.~\cite{Jankovic2023} (Holstein) and Ref.~\cite{Jankovic2025a} (Peierls),
    the path integral quantum Monte Carlo (PIMC) results of Ref.~\cite{Miladic2023} (Holstein),
    and the diagrammatic Monte Carlo (DiagMC) results of Ref.~\cite{Mishchenko2015DMC} (Holstein) and  Ref.~\cite{Mischenko2019DMC} (Fr\"ohlich).
    Asterisks in (c, d) indicate that the HEOM results may not be fully converged~\cite{Jankovic2025a}.
}
\label{fig:conductivity_models}
\end{figure*}

By imposing the quasiparticle approximation, which assumes that the spectral function is a sharp delta function, one can derive the BTE from the ladder formalism~\cite{MahanBook, Butler1985, Kim2019Vertex}.
The SERTA can be derived either from the BTE by neglecting the vertex correction from the scattering-in processes, or from the bubble conductivity by using the quasiparticle approximation.
Explicit expressions for the dc and ac conductivities within BTE and SERTA are given in \App{sec:transport_bte}.

We implement and solve \Eqs{eq:sigma_V0_el_diag}--\eqref{eq:sigma_bubble_G0D0_diag} using our in-house developed \EPjl package~\cite{Lihm2024NLHE, EPjl}.
Solving the ladder equations requires significant computational effort, which we mitigate by optimizing frequency integration and convolution, as detailed in \App{app:implementation}.
We also parallelize the calculation over the bands and $\bk$ points using \textsc{Julia}'s native multithreading.
As an example, a converged calculation for SrVO$_3$ using a 80$\times$80$\times$80 $\bk$ and $\bq$ grids filtered with the active-space window of width 0.3~eV considers 126,916 electronic states (3,010 inside the irreducible Brillouin zone) and takes 0.7 hours per iteration with 128 cores.
In most cases, the self-consistent ladder equations converged after 5 to 10 iterations.

\subsection{Application to model Hamiltonians}  \label{sec:transport_model}

We now apply the ladder-\scGD method, beginning with model Hamiltonians.
As in \Sec{sec:spectral_models}, we consider the Holstein, Fr\"ohlich, and Peierls models, defined in \Eqs{eq:holstein_def}--\eqref{eq:frohlich_def}.
We compare the resulting mobility with the bubble approximation and the BTE, along with those from three numerically exact methods: hierarchical equations of motion (HEOM)~\cite{Jankovic2023, Jankovic2025b}, path-integral quantum Monte Carlo (PIMC)~\cite{Miladic2023}, and diagrammatic Monte Carlo (DiagMC)~\cite{Mishchenko2015DMC, Mischenko2019DMC}.
For the \GD, BTE, and SERTA methods, which require a finite artificial broadening, we use the self-consistent broadening [\Eq{eq:gamma_sc}] and report results in the range $0.5\gamma^{\rm sc}$ to $2\gamma^{\rm sc}$ to assess the sensitivity to this parameter.
We simulate the dilute polaronic limit with a carrier concentration of $10^{-4}$ per unit cell.
Further decreasing the carrier density does not change the computed mobility.

\Figus{fig:conductivity_models}(a, b) show the mobility of the Holstein model at weak ($\lambda = 0.01$) and intermediate ($\lambda = 0.5$) coupling.
At weak coupling, all methods produce similar mobilities, in excellent agreement with the HEOM~\cite{Jankovic2023} and DiagMC~\cite{Mishchenko2015DMC} references, as well as with the PIMC~\cite{Miladic2023} result within the error bar.
This indicates that the quasiparticle approximation is valid and the vertex correction is small~\cite{Mitric2023Thesis, Jankovic2024Holstein}.
At intermediate coupling, the \scGD mobility agrees well with the HEOM result, while the BTE underestimates the mobility at $T/t \geq 2$.
As the temperature exceeds the bandwidth $4t$, the quasiparticle approximation breaks down, rendering the BTE invalid.
The DiagMC result underestimates the mobility by nearly an order of magnitude compared to the HEOM result, which could be related to the use of numerical analytic continuation~\cite{Miladic2023, Jankovic2024Holstein}.
The ladder-\scGD and bubble-\scGD mobilities are identical since the \eph coupling is local and the ladder vertex correction is zero~\cite{Basov2011RMP, Mitric2023Thesis}.
From \Fig{fig:conductivity_models}(b), we find that the bubble-\GD and SERTA are both valid at weak coupling but yield larger errors at intermediate coupling.
Notably, at intermediate coupling, they show significant variability depending on the value of the artificial broadening, limiting their predictive power.

\Figus{fig:conductivity_models}(c, d) shows the mobility of the Peierls model at weak ($\lambda = 0.05$) and intermediate ($\lambda = 0.25$) coupling.
Compared with the HEOM benchmark~\cite{Jankovic2025a}, we observe a quantitative discrepancy between ladder-\scGD and HEOM at $\lambda = 0.05$ and $T \leq 2 \omega_0$, and at $\lambda = 0.25$ and $T \geq 5 \omega_0$.
Nevertheless, ladder-\scGD captures the trend that the mobility flattens as temperature increases, a behavior not observed in any of the other approximate methods.
In \Fig{fig:conductivity_Peierls_assisted} in SM~\cite{supplemental}, we show that this temperature dependence is due to the increasing contribution of the phonon-assisted current at high temperatures, consistent with the finding from HEOM calculations~\cite{Jankovic2025a, Jankovic2025b}.

\Figus{fig:conductivity_models}(e, f) show a similar trend for the Fr\"ohlich model at weak ($\alpha = 0.2$) and intermediate ($\alpha = 2.5$) coupling.
At weak coupling, the quasiparticle approximation holds, and both ladder-\scGD and BTE yield similar mobilities.
Unlike the Holstein model, the current vertex correction remains significant even at weak coupling, evident from the difference between the bubble- and ladder-\scGD mobilities.
This results from the strong momentum dependence of the Fr\"ohlich \eph coupling, which leads to strong forward scattering and weak backscattering.
Forward scattering does not dissipate current~\cite{ZimanBook}, a feature captured only with vertex correction.
We note that the correction to the \textit{current} vertex appears already at the lowest order in the \eph coupling, $O(g^2)$ [first figure in \Fig{fig:feynman_conductivity_expansion}(b)].
In contrast, the corrections to the \textit{\eph} vertex, which affect the self-energy [first figure in \Fig{fig:feynman_selfen}(d)] and the conductivity [first and third figures in \Fig{fig:feynman_conductivity_expansion}(c)] through crossing diagrams, arise only at $O(g^4)$.
This result is consistent with a recent study of transport in the Hubbard model~\cite{Kovacevic2025}, which also reported a finite vertex correction to the conductivity in the weak-coupling limit.

At intermediate coupling, we find a noticeable difference between the ladder-\scGD and BTE mobilities.
However, neither method agrees well with DiagMC~\cite{Mischenko2019DMC} at $T \leq 0.5$.
It remains unclear whether the discrepancy between ladder-\scGD and DiagMC is due to the approximations made in the ladder-\scGD method or due to the numerical analytic continuation used in DiagMC (which led to an underestimation of mobility in the Holstein model~\cite{Miladic2023, Jankovic2024Holstein}).

\begin{figure}[tb]
\centering
\includegraphics[width=0.99\linewidth]{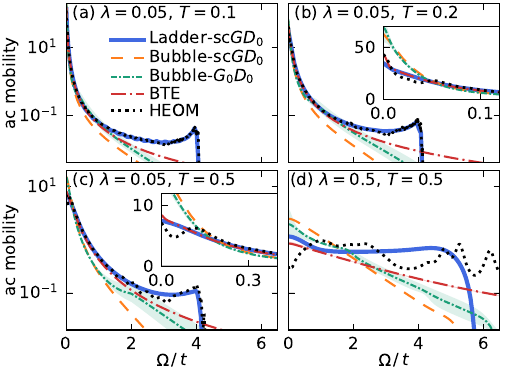}
\caption{
    Ac mobility of the Peierls model with $t = 1$ and $\omega_0 = 0.044$, compared with the HEOM results from Ref.~\cite{Jankovic2025b}.
}
\label{fig:Peierls_ac}
\end{figure}

We also study the ac mobility of the Peierls model, as shown in \Fig{fig:Peierls_ac}.
Here, we focus on the case with a low phonon frequency $\omega_0 = 0.044$, which is representative of phonons in molecular crystals~\cite{Jankovic2025b}.
Results for $\omega_0 = 1.0$ is shown in \Fig{fig:Peierls_ac_omega1.0} of SM~\cite{supplemental}.
At weak coupling [\Fig{fig:Peierls_ac}(a)], ladder-\scGD quantitatively reproduces the HEOM result, except for a low-energy dip and peak around $\Omega \approx \omega_0$.
The kink at $\Omega = 4t$ is particularly well reproduced.
By decomposing the conductivity into different contributions, we find that the kink arises from the phonon-assisted bubble: see \Fig{fig:Peierls_ac_decomp} of SM~\cite{supplemental}.
At intermediate coupling [\Fig{fig:Peierls_ac}(b)], ladder-\scGD fails to capture the displaced Drude peak seen in the HEOM result, characterized by a zero-frequency dip and a finite-frequency peak.
Higher-order vertex corrections are expected to be essential in this regime.

\begin{figure}[tb]
\centering
\includegraphics[width=0.99\linewidth]{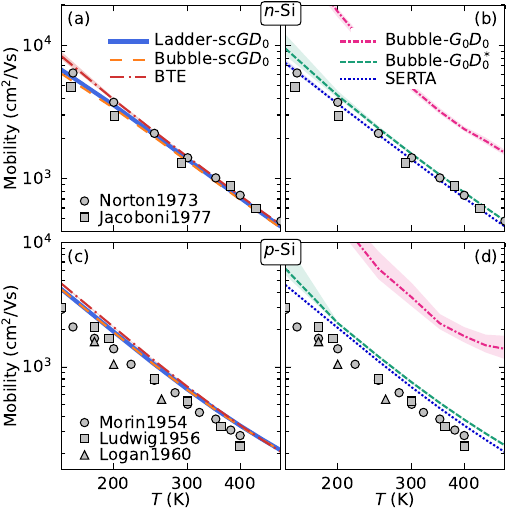}
\caption{
    Temperature dependence of the intrinsic mobility of (a, b) $n$-doped Si and (c, d) $p$-doped Si.
    Markers indicate experimental Hall mobilities from Refs.~\cite{Morin1954Si, Ludwig1956Si, Logan1960Si, Norton1973Si, Jacoboni1977Si}.
    The left column compares the ladder-\scGD, bubble-\scGD, and BTE results.
    The right column shows bubble-\GD calculations with and without (indicated by an asterisk) the static correction [\Eqs{eq:scGD0_static}, \eqref{eq:sigma_G0D0}], along with the SERTA result.
    Including the static correction in the \GD calculation leads to a significant overestimation of the mobility.
    Shaded regions represent the variation in mobility when the artificial broadening is varied from $0.5\gamma^{\rm sc}$ to $2\gamma^{\rm sc}$.
}
\label{fig:Si_dc}
\end{figure}

\begin{figure}[tb]
\centering
\includegraphics[width=0.99\linewidth]{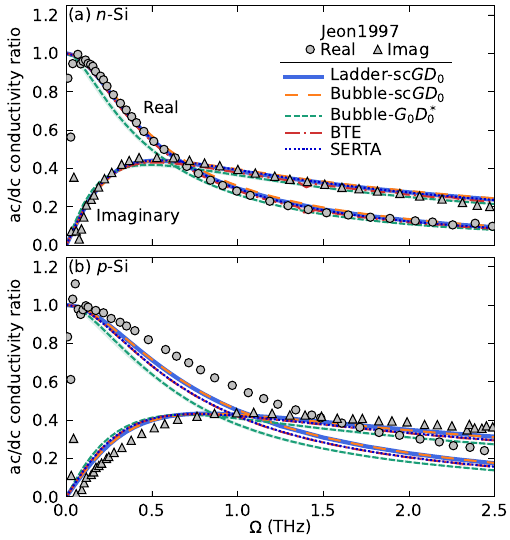}
\caption{
    Normalized ac conductivity of (a) $n$-doped Si and (b) $p$-doped Si at $T = 300$~K.
    Markers show experimental data from Ref.~\cite{Jeon1997Si}.
}
\label{fig:Si_ac}
\end{figure}

\subsection{Application to real materials: Si, ZnO, and \texorpdfstring{SrVO$_3$}{SrVO3}}  \label{sec:transport_materials}

We now turn to transport in real materials, beginning with semiconductors and then addressing metals.
In nondegenerate semiconductors, the chemical potential lies within the band gap, and mobility is independent of carrier density.
In our calculations, we fix the carrier density to $10^{15}~\mathrm{cm}^{-3}$ and iteratively adjust the chemical potential to match this target.
For the \GD spectral functions, we apply the truncation scheme outlined in Ref.~\cite{Lihm2025} to mitigate problems arising from the long tail of the spectral functions.

\Figu{fig:Si_dc} shows the temperature dependence of the dc mobility in $n$- and $p$-doped Si.
All methods except bubble-\GD yield similar mobilities and agree well with the experimental data.
This is consistent with the fact that Si has weak and short-ranged \eph coupling where quasiparticle approximation holds and the vertex correction is small~\cite{Ponce2018, Ponce2021}.
For the $p$-doped case, the calculated value is slightly overestimated, likely because of an underestimated effective mass~\cite{Ponce2018}, as elaborated at the end of this section.
The pure electronic contribution dominates the mobility, and the phonon-assisted and cross contributions are less than 0.1\% for the full temperature range.
The bubble-\GD method is the only outlier that significantly overestimates mobility.
This overestimation arises from the band structure renormalization, which shifts the bands toward the gap, resulting in a smaller imaginary part of the one-shot self-energy and thus an artificially high mobility.
When the static band structure correction is neglected (bubble-$G_0 D_0^*$, green dashed lines), the overestimation is reduced.
We note that in this work we compute the drift mobility, but the Hall factor ranges from 0.9--1.5 in Si~\cite{Ponce2021}.

\Figu{fig:Si_ac} shows the ac mobility normalized to the dc value.
All methods show excellent agreement with the experimental data from Ref.~\cite{Jeon1997Si}.
The width of the real part of the conductivity reflects the scattering rate, which is higher for $p$-doped Si than for $n$-doped Si~\cite{Ponce2021}.

\begin{figure}[tb]
\centering
\includegraphics[width=0.99\linewidth]{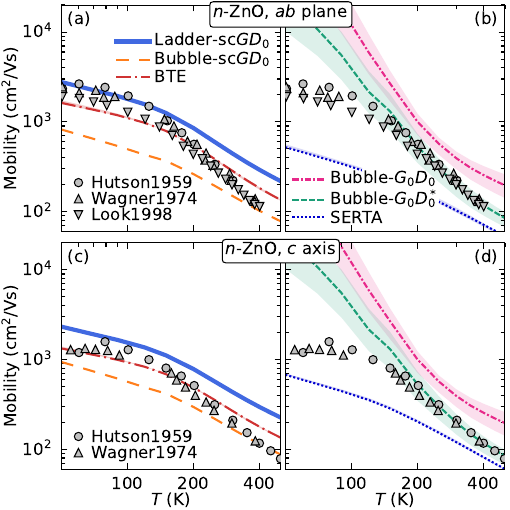}
\caption{
    Temperature dependence of the intrinsic mobility of $n$-doped ZnO (a, b) along the $ab$ plane, and (c, d) along the hexagonal $c$ axis.
    Markers show experimental Hall mobilities from Refs.~\cite{Hutson1959, Wagner1974ZnO, Look1998ZnO}.
    See \Fig{fig:Si_dc} for details on the lines.
}
\label{fig:ZnO_dc}
\end{figure}

\begin{figure}[tb]
\centering
\includegraphics[width=0.99\linewidth]{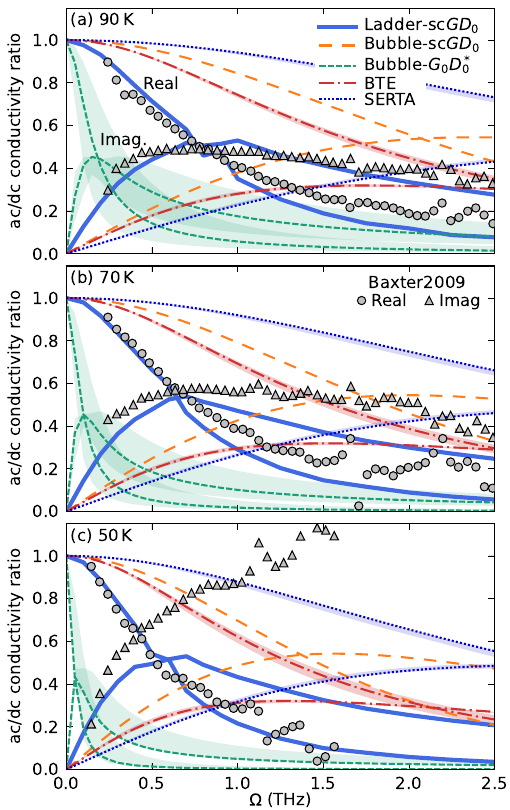}
\caption{
    Normalized ac conductivity of ZnO along the $ab$ plane at (a) 90~K, (b) 70~K, and (c) 50~K.
    Markers show experimental data from Ref.~\cite{Baxter2009}.
}
\label{fig:ZnO_ac}
\end{figure}

\Figu{fig:ZnO_dc} shows the temperature dependence of the dc mobility in $n$-doped ZnO.
Since ZnO is a polar semiconductor, it exhibits strong long-range \eph coupling of the Fr\"ohlich type~\cite{Verdi2015, Sjakste2015}.
This interaction causes significant spectral function renormalization as well as large vertex corrections.
Among the methods, ladder-\scGD yields the highest mobility, indicating that both the quasiparticle approximation and the neglect of vertex corrections lead to an underestimation of mobility.
The overestimation of mobility compared to experiments may stem from the neglect of other scattering mechanisms.
In addition, a recent calculation of the BTE mobility of ZnO using DFPT+U~\cite{Yang2025} reported that the Hubbard correction can further reduce the mobility.
We note that our calculation gives the drift mobility, while the experiments report Hall mobility, which the BTE predicts to be 1.1--1.2 times larger~\cite{Rode1975}.
The change in the slope around 150~K is attributed to the activation of the longitudinal optical phonon scattering~\cite{Makino2005}.
As in Si, the mobility is dominated by the pure electronic contribution, and the phonon-assisted and cross contributions are less than 0.1\% for the full temperature range.
Also, the bubble-\GD method, with or without static correction, strongly overestimates mobility due to renormalization of the band structure.

\Figu{fig:ZnO_ac} presents the ac mobility of $n$-doped ZnO, normalized by the dc value.
The ladder-\scGD results show excellent agreement with the experimental data~\cite{Baxter2009} for the real part at all three temperatures.
For the imaginary part, the agreement is good at 90~K but worsens at lower temperatures.
We note that the experimental data in the displayed range do not satisfy the Kramers--Kronig relation, suggesting that the rise in the imaginary part likely includes contributions beyond the free-carrier response, possibly from higher-frequency resonances.
This discrepancy is more pronounced at lower temperatures as the free carrier density is significantly reduced~\cite{Baxter2009}.
Overall, the close quantitative match between ladder-\scGD and experiment highlights its strength in capturing transport in polar semiconductors.

Compared to ladder-\scGD, both bubble-\scGD and BTE yield ac conductivities that decay slowly at higher frequencies, indicating higher effective scattering rates.
For the bubble-\scGD, this discrepancy reflects the absence of vertex correction.
For the BTE, it shows that the quasiparticle approximation overestimates the scattering rates, an effect also seen in monolayer InSe~\cite{Lihm2025}.
SERTA incorporates both approximations, resulting in the slowest decay.
For the bubble-\GD case, the characteristic frequency is significantly underestimated due to the underestimation of the quasiparticle linewidth [see \Fig{fig:spectral_all}(m)].

\begin{figure}[tb]
\centering
\includegraphics[width=0.99\linewidth]{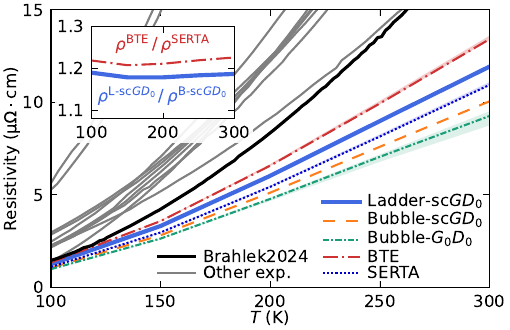}
\caption{
    Temperature dependence of the resistivity of SrVO$_3$.
    Experimental data are from the compilation of Ref.~\cite{LeeHand2024SVO} (see Fig.~1(a) therein), with the residual $T=0$~K resistivity subtracted.
    The latest experimental result with the lowest resistivity~\cite{Brahlek2024SVO} is shown in black, while others~\cite{Inoue1998SVO, Reyes2000SVO, Brahlek2015SVO, Fouchet2016SVO, Zhang2016SVO, Xu2019SVO, Shoham2020SVO, Mirjolet2021SVO, Roth2021SVO, Ahn2022SVO} appear in gray.
    See \Fig{fig:SrVO3_dc_logT} in SM~\cite{supplemental} for the same data on a logarithmic scale.
    The inset shows the ratio of the resistivity with and without vertex corrections for many-body and quasiparticle calculations.
}
\label{fig:SrVO3_dc}
\end{figure}

\begin{figure}[tb]
\centering
\includegraphics[width=0.99\linewidth]{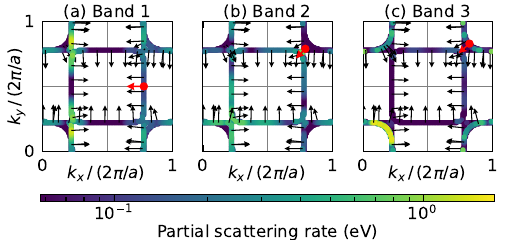}
\caption{
    BTE partial scattering rates [\Eq{eq:sigma_bte_partial_decay}] of SrVO$_3$ at 300~K for states on the Fermi surface in the $k_z = 0$ plane.
    Scattering rates are shown for states marked by the red circles: (a) $\bk = (0.78, 0.5, 0.0)$ (in units of $2\pi/a$) and band 1, (b) $\bk = (0.79, 0.79, 0.0)$ and band 2, and (c) $\bk = (0.82, 0.83, 0.0)$ and band 3, to other states.
    Arrows indicate the in-plane band velocity.
    Bright colors indicate strong scattering.
    We observe a strong backscattering: scattering is stronger between states with anti-parallel velocities than between states with parallel velocities.
}
\label{fig:SrVO3_scattering}
\end{figure}

We now turn to SrVO$_3$.
A previous study has shown that the resistivity of SrVO$_3$ above 100~K is dominated by phonon scattering, but examined vertex corrections only at the level of BTE~\cite{Abramovitch2024}.
Here, we go beyond and analyze the vertex corrections using the ladder-\scGD approach at the many-body level.
\Figu{fig:SrVO3_dc} shows the temperature dependence of the resistivity of SrVO$_3$.
Comparing the ladder- and bubble-\scGD results, we find a $\sim$20\% increase in resistivity due to vertex correction, as highlighted in the inset.
A similar increase is seen between the BTE and SERTA calculations, consistent with the results of Ref.~\cite{Abramovitch2024} (see Fig.~S2 therein).

The increase in resistivity with vertex correction contrasts with the behavior in polar semiconductors, where vertex corrections typically reduce resistivity~\cite{Ponce2021}.
To understand this difference, we analyze scattering processes at the microscopic level.
As shown in \Fig{fig:SrVO3_scattering}, we find strong backscatterings (scattering between states with anti-parallel velocities), while forward scatterings (scattering between states with parallel velocities) are weak.
This finding is consistent with the observation of large \eph matrix elements at the M and R points of the Brillouin zone~\cite{Abramovitch2024, Abramovitch2025}.
This behavior is markedly different from that of polar semiconductors, which exhibit strong forward scattering.
The key difference is that metallic screening suppresses the long-wavelength (small $\bq$) \eph coupling.
The ladder vertex correction captures the velocity flip due to backscattering, which is not accounted for in the bubble approximation, thus leading to an increase in resistivity.

Comparing the ladder-\scGD and BTE results, we find that going beyond the quasiparticle approximation has a minor effect, reducing the resistivity by about 10\%.
As in Si and ZnO, the phonon-assisted and cross contributions are less than 0.1\% for the full temperature range.
We also note that our bubble-\GD, BTE, and SERTA resistivity match those of Ref.~\cite{Abramovitch2024} (see \Fig{fig:SrVO3_dc_comparison}~\cite{supplemental}).

Overall, our calculation including vertex correction and beyond-quasiparticle effects underestimates the experimental resistivity of SrVO$_3$~\cite{Fouchet2016SVO, Ahn2022SVO} (with the residual $T=0$~K resistivity subtracted) by around 40\%.
This discrepancy is in part due to the neglect of electron-electron scattering, which is accounted for in Ref.~\cite{Abramovitch2024} using dynamical mean-field theory.
Another factor is the use of the \eph coupling calculated using DFT.
References~\cite{Abramovitch2024, Coulter2025} demonstrated that including electronic correlations through DFPT+U~\cite{Zhou2021} leads to a sizable increase in the resistivity.
Since the ladder-\scGD method is based on the Green's function formalism, which is also the standard approach for treating electronic correlation, it could be extended to include electronic correlation as well.
This would likely improve the accuracy of the resistivity calculation for SrVO$_3$.
Furthermore, incorporating the off-diagonal velocity would enable a detailed study of the temperature dependence of the infrared optical conductivity~\cite{Ahn2022SVO}.

A more accurate description of the electronic structure, such as with hybrid functionals or the $GW$ approximation, could improve the quantitative agreement with experimental DC mobility, mainly through a better estimation of the effective mass.
In hole-doped Si, the DC mobility [Fig.~17(c, d)] and the inverse lifetime from the AC conductivity [Fig.~18(b)] are overestimated primarily due to an underestimated valence-band effective mass~\cite{Ponce2018}.
Improved electronic-structure methods that yield more accurate effective masses would therefore yield better agreement with experiment.
It has been found that the correction to the \eph coupling strength remains small ($\sim$10\%)~\cite{Poliukhin2025}.
Similarly, for ZnO, DFT+U reduces the DC BTE mobility mainly by increasing the effective mass~\cite{Yang2025}; hence, the ladder-\scGD mobility would also decrease when computed with DFT+U.

In contrast, for SrVO$_3$, electron-electron interactions treated at the DFT+U or DMFT level significantly modify the \eph coupling strength~\cite{Abramovitch2024, Coulter2025, Abramovitch2025}.
At the DFT+U level, the coupling is enhanced~\cite{Abramovitch2024, Coulter2025}, leading to a reduction in the ladder-\scGD mobility.
Within DMFT, the \eph vertex acquires a nontrivial frequency dependence~\cite{Abramovitch2025}, making its effect on the spectral functions and mobility difficult to predict.

\section{Connecting dielectric and optical properties via charge conservation} \label{sec:dielectric}

The ladder-\scGD formalism is a generic linear response theory applicable to a wide range of response properties.
In this section, we use it to study the dielectric function, which characterizes the response of electric polarization to an external electric field.

The dielectric function is defined as a linear response to a scalar potential with a sinusoidal spatial variation in the long-wavelength limit.
The scalar potential is described by the finite-$\bQ$ density operator
\begin{equation}
    \hat{N}(\bQ)
    = \nint \dd\br \, e^{i\bQ \cdot \br} \hat{\psi}^\dagger(\br) \hat{\psi}(\br) \,,
\end{equation}
where $\hat{\psi}(\br)$ is the field operator.
In the eigenstate basis, this operator becomes
\begin{equation}
    \hat{N}(\bQ)
    = \sum_\mnk N_{\mnk}(\bQ) \opcd_\mkQ \opc_\nk \,,
\end{equation}
where
\begin{equation} \label{eq:N_mel_def}
    N_{\mnk}(\bQ) = \braket{u_\mkQ | u_\nk} \,.
\end{equation}
The dielectric function is given by the second derivative of the density-density correlation function with respect to $\bQ$:
\begin{equation} \label{eq:ward_epsilon_def}
    \epsilon_{\alpha\beta}(\Omega)
    = 1 - \frac{4\pi e^2}{V^{\rm uc}} \frac{1}{2} \frac{\partial^2 \chi^{\rm R}_{N\!N}(Q)}{\partial Q_\alpha \partial Q_\beta} \biggr\rvert_{\bQ = \mb{0}} \,.
\end{equation}

The evaluation of \Eq{eq:ward_epsilon_def} is numerically challenging, as the calculation of the second derivatives not only increases computational cost but also introduces numerical noise.
Moreover, for \textit{ab initio} Hamiltonians, applying a finite-$\bQ$ perturbation raises the issue of phase consistency.
The bare eigenstates are defined up to an arbitrary phase factor $e^{i\phi_\nk}$.
Although physical observables are independent of this phase, intermediate quantities, such as the response and vertex functions, depend on the choice of phase.
Maintaining a consistent phase across all quantities is difficult, particularly when symmetries are used to reduce the computational cost.
In contrast, optical conductivities only involve the $\bQ = \mb{0}$ response, where the phase choice is irrelevant under the diagonal Green's function approximation.

Therefore, it is desirable to compute the dielectric function from the optical conductivity, using the continuity equation~\cite{FoxBook, Vucicevic2023}
\begin{equation} \label{eq:epsilon_sigma_relation}
    \epsilon_{\alpha\beta}(\Omega)
    = 1 + i \frac{4\pi}{\Omega} \sigma^{\rm L}_{\alpha\beta}(\Omega) \,.
\end{equation}
However, not all approximations preserve this relation: only ``charge-conserving approximations,'' in which the self-energy and correlation functions are derived from the same class of diagrams~\cite{Baym1962}, satisfy \Eq{eq:epsilon_sigma_relation}.
Notably, ladder-\scGD is a charge-conserving approximation~\cite{Engelsberg1963}.
In \App{app:Ward}, we explicitly prove this by showing that the ladder-\scGD method satisfies the Ward--Takahashi identity~\cite{Ward1950, Takahashi1957} and the continuity equation.
With the ladder-\scGD ac conductivities, one can then use \Eq{eq:epsilon_sigma_relation} to compute the dielectric function.

When the \eph coupling is nonlocal, the phonon-assisted current term is essential to satisfy the continuity equation.
The reason is that a nonlocal \eph coupling does not commute with the electronic polarization operator and thus generates an additional term in the dynamics of the polarization.
A related situation has been extensively discussed in the context of electron-electron interactions in Hubbard and extended Hubbard models~\cite{Krien2017, Krien2018Thesis}.

The other approximations shown in \Fig{fig:sigma_methods}, namely bubble-\scGD, bubble-\GD, BTE, and SERTA, violate charge conservation by neglecting the vertex correction or the phonon-assisted current.
Therefore, these approximations do not satisfy the continuity equation.
One may still use \Eq{eq:epsilon_sigma_relation} to obtain dielectric functions from the conductivities computed using these methods.
But, those results must be interpreted with caution as they will not be equal to those obtained from the density-density susceptibility computed within the same approximation.

\begin{figure}[tb]
\centering
\includegraphics[width=0.99\linewidth]{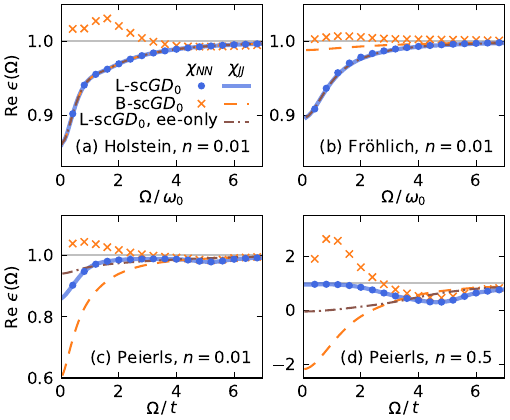}
\caption{
    Real part of the dielectric functions for (a) the Holstein model ($T=1$, $\lambda=0.5$), (b) the Fr\"ohlich model ($T=1$, $\alpha=2.5$), and (c, d) the Peierls model ($T=0.5$, $\omega_0=0.044$, $\lambda=0.5$) calculated using the ladder-\scGD (L-\scGD) and bubble-\scGD (B-\scGD) methods.
    We use different carrier concentrations, $n = 0.01$ in (a, b, c), and $n = 0.5$ (half-filling) in (d).
    Other model parameters are the same as in \Fig{fig:conductivity_models}.
    Circles and crosses represent direct calculations using the density-density susceptibility, while lines show results obtained from the ac conductivity using the continuity equation \eqref{eq:epsilon_sigma_relation}.
    The two results agree for ladder-\scGD (blue) but not for bubble-\scGD (orange).
    In the Peierls model, neglecting the phonon-assisted current in the conductivity (brown) also causes a large discrepancy.
    The imaginary part is shown in \Fig{fig:dielectric_models_imag} of SM~\cite{supplemental}.
}
\label{fig:dielectric_models}
\end{figure}

\subsection{Dielectric function in model Hamiltonians}

We now apply the ladder-\scGD method to compute dielectric functions.
We begin by numerically verifying the consistency between the dielectric function and optical conductivity within the ladder-\scGD formalism, using model Hamiltonians where a direct finite-$\bQ$  calculation is feasible.
\Figu{fig:dielectric_models} presents the real part of the dielectric function for the Holstein, Fr\"ohlich, and Peierls models.
The blue circles show the results of a direct calculation via the density-density susceptibility [\Eq{eq:ward_epsilon_def}], while the solid blue curves represent the dielectric function derived from the optical conductivity using \Eq{eq:epsilon_sigma_relation}.
The close agreement between the two confirms that ladder-\scGD respects the continuity equation.
In contrast, for the bubble-\scGD approximation, the dielectric functions computed from the two approaches (orange crosses and dashed curves) differ significantly, indicating the violation of the charge conservation and continuity equation.
Notably, the discrepancy occurs even in the Holstein model where the ladder vertex correction to the conductivity is zero, but the vertex correction to the dielectric function is nonzero.

For the Peierls model [\Figs{fig:dielectric_models}(c, d)], neglecting the phonon-assisted current contribution to the optical conductivity leads to a significant violation of the continuity equation (dot-dashed brown curves).
This result highlights the necessity of including the phonon-assisted current for an accurate calculation of the dielectric properties from the optical conductivity.

\subsection{Dielectric function in Si and ZnO}

\begin{figure}[tb]
\centering
\includegraphics[width=0.99\linewidth]{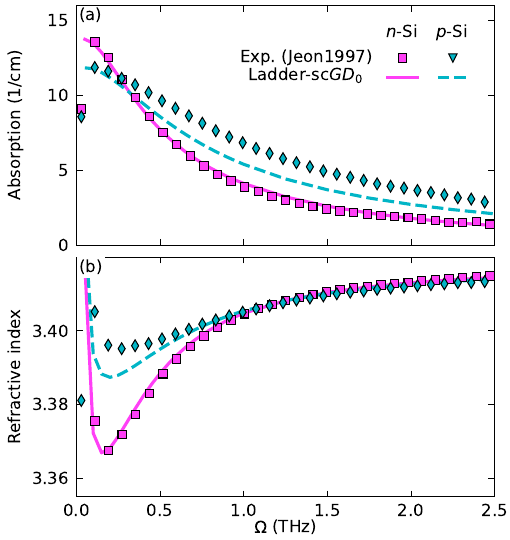}
\caption{
    (a) Absorption coefficient and (b) refractive index of nondegenerate $n$- and $p$-doped Si at 300~K calculated using the ladder-\scGD method.
    Results are compared with experimental data from Ref.~\cite{Jeon1997Si}.
    The absorption coefficient and refractive index were obtained from the calculated ac conductivity (\Fig{fig:Si_ac}) using \Eqs{eq:epsilon_sigma_relation} and \eqref{eq:epsil_to_alpha_n}, with $\epsilon^{\rm undoped} = 11.68$~\cite{Jeon1997Si}.
}
\label{fig:dielectric_Si}
\end{figure}

\begin{figure}[tb]
\centering
\includegraphics[width=0.99\linewidth]{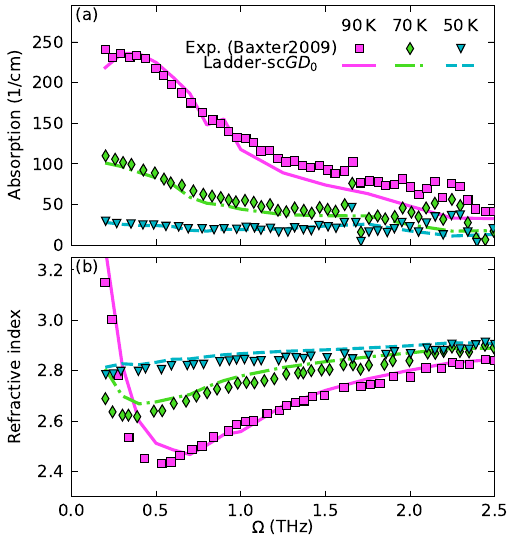}
\caption{
    (a) Absorption coefficient and (b) refractive index of nondegenerate $n$-doped ZnO calculated using ladder-\scGD as in \Fig{fig:dielectric_Si}, compared with experimental data from Ref.~\cite{Baxter2009}.
    We extract $\epsilon^{\rm undoped}(\Omega)$ from the experimental data at $T=10~\mathrm{K}$, where the carrier concentration is negligible~\cite{Baxter2009}.
    We only compute results for frequencies between 0.2 and 2.5~THz, where the experimental $\epsilon^{\rm undoped}(\Omega)$ is available.
}
\label{fig:dielectric_ZnO}
\end{figure}

\begin{figure}[tb]
\centering
\includegraphics[width=0.99\linewidth]{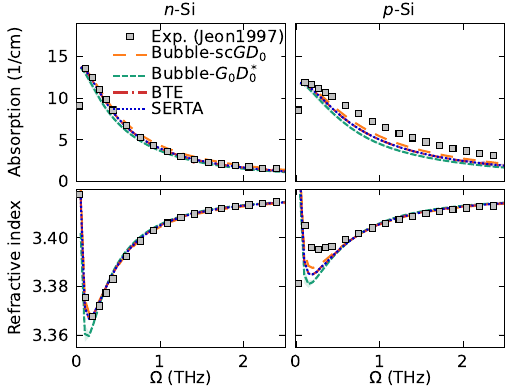}
\caption{
    Absorption coefficient and refractive index of nondegenerate $n$- and $p$-doped Si at 300~K, as in \Fig{fig:dielectric_Si}, but using the more approximate methods.
    The results are computed from the optical conductivities of \Fig{fig:Si_ac}, using \Eqs{eq:epsilon_sigma_relation} and \eqref{eq:epsil_to_alpha_n}.
    We note that the continuity equation \eqref{eq:epsilon_sigma_relation} is not valid for these approximations; a direct calculation using the density-density susceptibility [\Eq{eq:ward_epsilon_def}] would yield different results than those shown here.
}
\label{fig:dielectric_Si_no_ladder}
\end{figure}

\begin{figure}[t]
\centering
\includegraphics[width=0.99\linewidth]{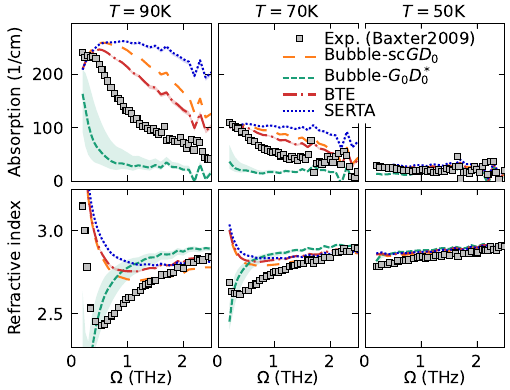}
\caption{
    Absorption coefficient and refractive index of nondegenerate $n$-doped ZnO as in \Fig{fig:dielectric_ZnO}, but using the more approximate methods.
    See \Fig{fig:dielectric_Si_no_ladder} for a detailed description.
    The large noise above $2~\mathrm{THz}$ is due to noise in the experimental data for $\epsilon^{\rm undoped}(\Omega)$.
}
\label{fig:dielectric_ZnO_no_ladder}
\end{figure}

Having confirmed that the ladder-\scGD method satisfies the continuity equation \eqref{eq:epsilon_sigma_relation}, we now apply it to compute the dielectric function of Si and ZnO.
In THz experiments, the measured quantities are the optical absorption and the refractive index.
These are related to the dielectric function through~\cite{Baxter2009}
\begin{subequations} \label{eq:epsil_to_alpha_n}
\begin{align}
    \alpha(\Omega) &= \frac{2\Omega}{c} \Im \sqrt{\epsilon^{\rm undoped}(\Omega) + \epsilon(\Omega) - 1} \,,
    \\
    n(\Omega) &= \Re \sqrt{\epsilon^{\rm undoped}(\Omega) + \epsilon(\Omega) - 1} \,.
\end{align}
\end{subequations}
Here, $c$ is the speed of light, $\epsilon^{\rm undoped}(\Omega)$ is the dielectric function of the undoped system, and $\epsilon(\Omega)$ is the contribution of free carriers.
We compute $\epsilon(\Omega)$ from the ac conductivity using \Eq{eq:epsilon_sigma_relation}.
Because the experimental carrier concentrations are not known directly, we determine them such that the calculated dc conductivity at each temperature corresponds to the experimental value.
For the undoped dielectric function, we use the experimental data: $\epsilon^{\rm undoped}(\Omega) = 11.68$ for Si~\cite{Jeon1997Si}, and the low-temperature ($T=10~\mathrm{K}$) measurements for ZnO~\cite{Baxter2009}, for which the free carrier contribution is negligible.

\Figu{fig:dielectric_Si} shows the absorption coefficient and refractive index of $n$- and $p$-doped Si.
The ladder-\scGD results agree well with the experimental data from Ref.~\cite{Jeon1997Si}.
In particular, the smaller width of the absorption peak and sharper dip in the refractive index with $n$-doping, attributable to the longer lifetime of the conduction band states compared to the valence band states, are well captured.
The result for ZnO is shown in \Fig{fig:dielectric_ZnO}, where again the ladder-\scGD calculations quantitatively agree with the experimental data from Ref.~\cite{Baxter2009}.
We note that these agreements are expected from the good agreement between the ac conductivities in \Figs{fig:Si_ac} and \ref{fig:ZnO_ac}.

Figures~\ref{fig:dielectric_Si_no_ladder} and \ref{fig:dielectric_ZnO_no_ladder} show the dielectric functions computed from the ac conductivities using the more approximate methods, bubble-\scGD, bubble-\GD, BTE, and SERTA.
We used the continuity equation \eqref{eq:epsilon_sigma_relation}, although this step introduces a formal inconsistency as the approximations violate charge conservation.
For Si, all methods yield similar results because the \eph coupling is weak and the vertex correction is small.
In contrast, for ZnO, all methods other than ladder-\scGD either significantly overestimate (bubble-\scGD, BTE, and SERTA) or underestimate (bubble-\GD) the optical absorption at $\Omega > 0.5~\mathrm{THz}$.
Furthermore, the dip in refractive index for the $T=90~\mathrm{K}$ and $70~\mathrm{K}$ results is not captured.
These findings indicate that both beyond-quasiparticle effects and vertex corrections are essential for accurately capturing the THz dielectric properties of ZnO.

\section{Conclusion and outlook} \label{sec:conclusion}

In this work, we demonstrated that the ladder-\scGD method unifies and improves on the two most widely used approaches for first-principles phonon-limited transport: the BTE and the bubble approximation.
Other approaches to phonon-limited transport also exist, and one example is the nonequilibrium Green's function (NEGF)~\cite{Haug2008Book, Luisier2009NEGF, Szabo2015NEGF, Backman2024NEGF} method.
When the electron self-energy is computed self-consistently~\cite{Luisier2009NEGF, Szabo2015NEGF, Backman2024NEGF}, NEGF in the small bias limit is equivalent to ladder-\scGD.
The difference lies in their focus: ladder-\scGD targets linear response, while NEGF is a real-time method capable of handling both linear and nonlinear responses.
Due to its narrower scope, ladder-\scGD is simpler and more efficient to implement, delivering fully converged results without massive parallelization or extra approximations~\cite{Backman2024NEGF}.
Furthermore, the momentum-space formulation enables the inclusion of the long-range Fr\"ohlich interaction, which is crucial in polar semiconductors.
In contrast, the NEGF method excels at modeling realistic device geometries and handling large bias voltages.

Another approach recently proposed for \eph transport is the semiconductor electron-phonon equations (SEPE)~\cite{Stefanucci2024SEPE}, which generalizes the BTE to include coherence effects.
Although SEPE has not yet been implemented, its computational cost scales in the same way as the BTE~\cite{Stefanucci2024SEPE}.
Its main approximation is the use of the (mirrored) generalized Kadanoff--Baym ansatz~\cite{Lipavsky1986GKBA, Karlsson2021GKBA, Pavlyukh2022GKBA, Schafer2013GKBA}, which evaluates the retarded and advanced self-energies under the quasiparticle approximation.
In contrast, the ladder-\scGD method does not rely on the quasiparticle approximation and captures full spectral features, such as significant broadening and satellite structures.
Extending ladder-\scGD to use renormalized instead of bare-phonon Green's functions would allow for the inclusion of nuclear displacements and the electron-induced phonon self-energy, which are accounted for in SEPE.

Before concluding, let us discuss possible extensions of the ladder-\scGD method.
First, one can improve the Green's functions by extending the \scGD method.
Phonon renormalization can be added to develop a fully self-consistent $GD$ method, which has been achieved in model systems~\cite{Sakkinen2015A, Sakkinen2015B}.
The phonon self-energy may be computed using a one-loop diagram~\cite{Calandra2010, Berges2023}, or using the random phase approximation including free-carrier screening~\cite{Macheda2024, Lihm2024PlPh, Krsnik2024}.
One may also allow for spontaneous symmetry breaking of the self-consistent Green's functions, capturing the polaronic nuclear displacements via the tadpole (or Ehrenfest) self-energy.
Such an extension would allow the study of self-trapped polarons~\cite{Sio2019, Sio2019a, Lafuente2022, Lafuente2022a}, as well as the vibrational absorption process [\Fig{fig:feynman_not_included}(b)].
While symmetry-broken self-consistent Green's functions have been studied in the two-site Holstein model~\cite{Sakkinen2015A, Sakkinen2015B}, first-principles calculations~\cite{Lafuente2022, Lafuente2022a} have focused on a single, lowest-lying eigenstate rather than the full, frequency-dependent Green's function.
To handle the increased complexity due to the symmetry breaking, further numerical approximations and optimizations would be needed.

Second, the ladder-\scGD method could be applied to a broader class of response functions.
One example is optical properties in the infrared and visible ranges arising from interband transitions.
This would provide a quantum many-body description of indirect optical absorption in semiconductors, complementing the semiclassical theories~\cite{Cheeseman1952Indabs, Hall1954Indabs, BassaniBook, Kioupakis2010, Noffsinger2012, Tiwari2024}.
It could also be used to investigate phonon-assisted bubble processes in the infrared absorption.
Although the phonon-assisted current was negligible (below 1\%) for all materials studied in this work, systems with flat bands or molecular semiconductors exhibiting Peierls-like \eph coupling could display much stronger effects.
Other promising applications include the optical spatial dispersion~\cite{LandauBook, MelroseBook}, as well as photovoltaic and Hall effects.
The ladder formalism could also, in principle, be used to evaluate vertex corrections to the \eph coupling itself.

Finally, incorporating electron-electron interactions and electronic correlations is another important direction.
Since ladder-\scGD is based on the Green's function formalism, it can be naturally combined with methods for electronic correlations, such as the dynamical mean-field theory~\cite{Georges1996, Kotliar2006DMFT} and its diagrammatic extensions~\cite{Rohringer2018}.
This would allow for a unified treatment of electronic~\cite{Kauch2020, Krsnik2024PiTon, Gleis2025, Kovacevic2025} and lattice contributions to transport with vertex corrections.

To conclude, we have introduced the many-body ladder-\scGD formalism for studying \eph transport in solids.
This method unifies the bubble approximation and the BTE, capturing both beyond-quasiparticle dynamics and vertex corrections.
With its balance of formal accuracy and computational efficiency, the ladder-\scGD method is well-suited for studying \eph transport in real materials.
We have demonstrated the applicability of ladder-\scGD in model Hamiltonians and applied it to Si, ZnO, and SrVO$_3$, finding good agreement with numerically exact benchmarks and experimental data.
In addition, we applied ladder-\scGD to calculate the dielectric function of Si and ZnO, finding a significant improvement in the calculated optical absorption and refractive index compared to the state-of-the-art methods.
This work lays the groundwork for using many-body Green's function methods to study \eph interactions and transport in real materials.

\bigskip

The code and data that support the findings of this article are openly available~\cite{EPjl, MaterialsCloudArchive}.

\begin{acknowledgments}
We thank David Abramovitch, Antoine Georges, Nina Girotto, Raveena Gupta, Matthew Houtput, Sunghoon Kim, Francesco Macheda, Hongki Min, Dino Novko, Matthieu Verstraete, and Wooil Yang for helpful discussions, and Fabian Kugler for sharing the data of Ref.~\cite{LeeHand2024SVO}.
S.P. is a Research Associate of the Fonds de la Recherche Scientifique - FNRS.
This work was supported by the Fonds de la Recherche Scientifique - FNRS under Grants number T.0183.23 (PDR) and T.W011.23 (PDR-WEAVE).
This publication was supported by the Walloon Region in the strategic axe FRFS-WEL-T.
Computational resources have been provided
by the EuroHPC JU award granting access to MareNostrum5 at Barcelona Supercomputing Center (BSC), Spain (Project ID: EHPC-EXT-2023E02-050),
by the Consortium des \'Equipements de Calcul Intensif (C\'ECI), funded by the FRS-FNRS under Grant No.~2.5020.11,
and by Lucia, the Tier-1 supercomputer of the Walloon Region with infrastructure funded by the Walloon Region under the grant agreement n°1910247.
\end{acknowledgments}

\appendix

\section{Details of the Keldysh Green's functions} \label{sec:green_details}

\begin{figure*}[tb]
\centering
\includegraphics[width=0.99\linewidth]{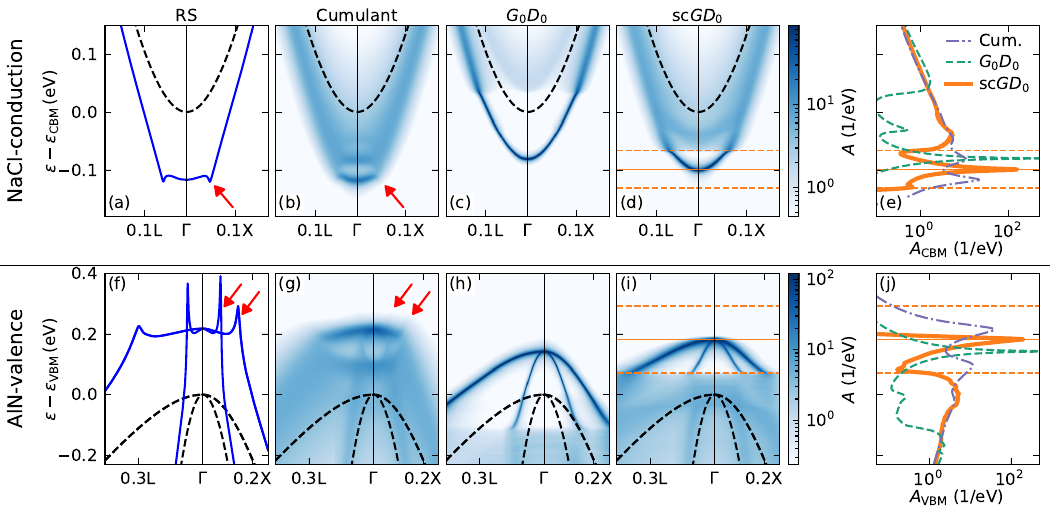}
\caption{
    Spectral functions as in \Fig{fig:spectral_models}, for (a)--(e) the conduction band of NaCl at $100~\mathrm{K}$ and (f)--(j) the valence bands of cubic AlN at $100~\mathrm{K}$.
    Red arrows mark the unphysical features in the RS dispersion and the cumulant spectral function.
}
\label{fig:spectral_NaCl_AlN}
\end{figure*}

The electron Green's function in the Keldysh formalism can be written in terms of the spectral function using the fermionic fluctuation-dissipation relations~\cite{StefanucciBook}:
\begin{subequations} \label{eq:sc_fdt_G}
\begin{align}
    \label{eq:sc_fdt_G_A}
    G^{\rm A}_{n_1 n_2 \bk}(\veps)
    &= \bigl[ G^{\rm R}_{n_2 n_1\bk}(\veps) \bigr]^* \,,
    \\
    \label{eq:sc_fdt_G_gtrless}
    G^{\gtrless}_{n_1 n_2 \bk}(\veps)
    &= \mp 2\pi i f^\mp(\veps) A_{n_1 n_2 \bk}(\veps) \,,
    \\
    \label{eq:sc_fdt_G_T}
    G^{\rm T/\bar{T}}_{n_1 n_2 \bk}(\veps)
    &= \pm G^{\rm R}_{n_1 n_2 \bk}(\veps) + G^{\lessgtr}_{n_1 n_2 \bk}(\veps) \,.
\end{align}
\end{subequations}
Electronic self-energies satisfy the same fluctuation-dissipation relations.
With the single-particle density matrix [\Eq{eq:sc_rho_def}], the integral of the Green's function is written as
\begin{equation} \label{eq:sc_G_int}
    \nint \frac{\dd\veps}{2\pi i} \, G_{12}(k)
    = \begin{pmatrix}
        \rho_{n_1 n_2 \bk} - \frac{1}{2} & \rho_{n_1 n_2 \bk} \\
        \rho_{n_1 n_2 \bk} - 1 & \rho_{n_1 n_2 \bk} - \frac{1}{2}
    \end{pmatrix} \,.
\end{equation}
Applying Eqs.~\eqref{eq:sc_fdt_G} for the bare spectral function [\Eq{eq:Gbare}], we find the bare electron Green's functions:
\begin{align} \label{eq:Gbare_contour}
    G^{\pm\pm}_{0,\nk}(\veps) &= \frac{\mp f_\nk^\pm}{\veps - \veps_\nk + i0^+} \mp \frac{f_\nk^\mp}{\veps - \veps_\nk - i0^+} \,,
    \nnnl
    G^{\pm\mp}_{0,\nk}(\veps) &= \mp 2\pi i f_\nk^\mp \delta(\veps - \veps_\nk) \,.
\end{align}
The bare phonon Green's functions are given by
\begin{align} \label{eq:D0}
    D^\RA_{0, \nuq}(\omega)
    &= \frac{1}{\omega - \omega_\nuq \pm i0^+} - \frac{1}{\omega + \omega_\nuq \pm i0^+} \nnnl
    &= \frac{2\omega_\nuq}{(\omega \pm i0^+)^2 - \omega_\nuq^2} \,,
    \nnnl
    D^{\pm\pm}_{0, \nuq}(\omega)
    &=
    \mp \mcP \frac{1}{\omega - \omega_\nuq}
    \pm \mcP \frac{1}{\omega + \omega_\nuq}
    \nnnl
    &\quad  - \!\pi i (2n_\nuq \!+\! 1) [\delta(\omega \!-\! \omega_\nuq) + \delta(\omega \!+\! \omega_\nuq)] \,,
    \nnnl
    D^{\pm\mp}_{0, \nuq}(\omega)
    &= - 2\pi i \bigl[ (n_\nuq + 1) \delta(\omega \mp \omega_\nuq)
    \nnnl
    &\hspace{4em} + n_\nuq \delta(\omega \pm \omega_\nuq) \bigr] \,.
\end{align}

\section{Spectral functions of NaCl and AlN} \label{app:additional_spectral}

In the main text, we show the spectral functions for Si, ZnO, and SrVO$_3$.
Here, we present additional results for NaCl and cubic AlN, where the polar \eph interaction is stronger.
\Figus{fig:spectral_NaCl_AlN}(a)--(e) display the conduction band spectral functions of NaCl at $100~\mathrm{K}$.
Similarly to the Fr\"ohlich model, both the RS dispersion and the cumulant spectral function exhibit kinks at $\bk$ points where $\veps_\nk \approx \veps_{\rm CBM} + \omega_{\rm LO}$.
The \scGD spectral function does not show this artifact and instead captures the onset of the phonon emission continuum at $E_{\rm CBM} + \omega_{\rm LO}$.
Moreover, the cumulant spectral function lacks a well-defined quasiparticle peak, as the \scGD spectral function does.
The CBM energy renormalization of NaCl at 100~K computed using the cumulant, \GD, and \scGD methods was $-117$, $-80$, and $-99$~meV, respectively.

\Figus{fig:spectral_NaCl_AlN}(f)--(j) show the valence band spectral functions of cubic AlN at $100~\mathrm{K}$.
Similarly to NaCl, an anomalous kink appears in the RS dispersion.
This behavior is also seen in Figure 16 of Ref.~\cite{Ponce2015}, although it is less pronounced there due to the limited $\bk$-point resolution.
The \scGD spectral function again displays relatively well-defined quasiparticle peaks above $E_{\rm VBM} - \omega_{\rm LO}$.
The renormalization of the VBM energy computed using the RS, cumulant, \GD, and \scGD methods was 218, 214, 142, and 182~meV, respectively.

\section{Details of linear response theory} \label{sec:lr_detail}

In this Appendix, we augment \Sec{sec:lr_keldysh} by providing more details on the derivation of the linear response theory within the Keldysh formalism~\cite{StefanucciBook}.
The linear response of $\hat{X}^\dagger_\bQ$ to the perturbation $\hat{Y}_\bQ$ in the time domain is given by~\cite{StefanucciBook}
\begin{align} \label{eq:lr_chi_response}
    \delta_Y X_{\bQ}(t^{c})
    &= \sum_{\cpr} Z^{\cpr} \!\nint\! \dd t' \, \chi_{XY}^{c\cpr}(t - t'; \bQ) \, e^{-i\Omega t'} Z^{\cpr} \delta Y^{\cpr}(Q)
    \nnnl
    &= e^{-i\Omega t} \sum_{\cpr} \chi_{XY}^{c\cpr}(Q) \, \delta Y^{\cpr}(Q) \,,
\end{align}
where
\begin{equation} \label{eq:lr_chi_def}
    \chi^{c\cpr}_{XY}(t - t'; \bQ)
    \equiv \frac{-i}{N_\bk} \expval{\mathcal{T} \hat{X}^\dagger_{\bQ}(t^{c}) \hat{Y}_{\bQ}(t^{\prime\cpr})}
    + \frac{i}{N_\bk^2} \expval{\hat{X}^\dagger_{\bQ}} \expval{\hat{Y}_{\bQ}}
\end{equation}
is the susceptibility between $\hat{X}^\dagger_\bQ$ and $\hat{Y}_\bQ$, and
\begin{align} \label{eq:lr_chi_FT}
    \chi^{c\cpr}_{XY}(Q)
    &= \nint \dd t \, \chi^{c\cpr}_{XY}(t; \bQ) \, e^{i\Omega t}
\end{align}
its Fourier transform.
By Fourier transforming \Eq{eq:lr_chi_response}, we obtain \Eq{eq:lr_chi_Q}.
The susceptibility is a bosonic correlation function and thus satisfies the bosonic fluctuation-dissipation relations~\cite{Kubo1957, Martin1959, StefanucciBook}
\begin{subequations} \label{eq:fdr_boson}
\begin{align}
    \chi^{\rm A}_{XY}(Q)
    &= \bigl[ \chi^{\rm R}_{YX}(Q) \bigr]^* \,,
    \\
    \chi^{>}_{XY}(Q)
    &= \bigl[ 1 + n(\Omega) \bigr] \bigl[ \chi^{\rm R}_{XY}(Q) - \chi^{\rm A}_{XY}(Q) \bigr] \,,
    \\
    \chi^{<}_{XY}(Q)
    &= n(\Omega) \bigl[ \chi^{\rm R}_{XY}(Q) - \chi^{\rm A}_{XY}(Q) \bigr] \,,
    \\
    \chi^{\rm T/\bar{T}}_{XY}(Q)
    &= \pm \chi^{\rm R}_{XY}(Q) + \chi^{\lessgtr}_{XY}(Q) \,.
\end{align}
\end{subequations}

In an experimentally relevant scenario, the external perturbation depends only on time and not on the contour variable, i.e.,
\begin{equation} \label{eq:lr_Y_physical}
    \delta Y^-(Q) = -\delta Y^+(Q) = \delta Y(Q) \,.
\end{equation}
The minus sign arises from the $Z^c$ factor in \Eq{eq:lr_H_pert}.
The physical response is then given by the retarded susceptibility~\cite{StefanucciBook}:
\begin{equation} \label{eq:lr_retarded}
    \delta_Y X(Q) = \delta_Y X^{\pm}(Q) = \chi^{\rm R}_{XY}(Q) \, \delta Y(Q) \,.
\end{equation}

The response function $\delta_Y G$ can be expressed in terms of a three-point correlation function between the bosonic perturbation operator $\hat{Y}_\bQ$, a creation operator $\opc_{n_1\bkQ}$, and an annihilation operator $\opcd_{n_2\bk}$~\cite{StefanucciBook}:
\begin{multline}
    \delta_Y G_{n_1 n_2 \bk}(t_1^{c_1}, t_2^{c_2}, t^{c}; \bQ) \\
    = (-i)^2 \expval{\mcT \, \opc_{n_1\bkQ}(t_1^{c_1}) \, \opcd_{n_2\bk}(t_2^{c_2}) \, \hat{Y}_{\bQ}(t^{c})}
    \\
    + \frac{1}{N_\bk}  \expval{\mcT \, \opc_{n_1 \bkQ}(t_1^{c_1}) \, \opcd_{n_2\bk}(t_2^{c_2})} \,
    \expval{\hat{Y}_{\bQ}} \,.
\end{multline}
Its Fourier transform yields the response function:
\begin{multline}
    \delta_Y^c G_{12}(k, Q)
    = \nint \dd t_1 \, \dd t \, \delta_Y G_{n_1 n_2 \bk}(t_1^{c_1}, t_2^{c_2}, t^{c}; \bQ)
    \\
    \times e^{i (\veps + \Omega - \mu) t_1} \, e^{-i (\veps - \mu) t_2} \, e^{-i \Omega t}
    \,.
\end{multline}

Given the response function, one can write the $XY$ susceptibility in the time domain as
\begin{align} \label{eq:chi_XY_from_R_time}
    &\chi^{c c'}_{XY}(t; \bQ)
    \nnnl
    &= -i \sum_{mn} \intbk X_\mnk^*(\bQ) \Bigl[ \expval{\mathcal{T} \opcd_\nk(t^{c}_+) \opc_\mkQ(t^{c}) \hat{Y}_{\bQ}(0^{c'})}
    \nnnl
    &\hspace{10em} - \frac{1}{N_\bk} \expval{\opcd_\nk (t^{c_1}_+) \opc_\mkQ (t^{c}) } \expval{\hat{Y}_{\bQ}}
    \Bigr]
    \nnnl
    &= -i \sum_{mn} \intbk X_\mnk^*(\bQ) \, \delta_Y G_{\mnk}(t^{c}, t^{c}_+, 0^{c'};\bQ) \,.
\end{align}
Here, $t^{c}_+$ is a contour argument infinitesimally after $t^{c}$ such that the creation operator is ordered after the annihilation operator.
By Fourier transforming \Eq{eq:chi_XY_from_R_time} while using time translational invariance, we find \Eq{eq:sigma_XY_from_R}:
\begin{align}
    &\chi^{c\cpr}_{XY}(Q)
    \nnnl
    &= -i \nint \dd t \sum_{mn} \intbk X_{\mnk}^*(\bQ) \, \delta_Y G_{\mnk}(t^{c}, t^{c}_+, 0^{\cpr}; \bQ) \, e^{i\Omega t}
    \nnnl
    &= \sum_{mn} \intkk X_{\mnk}^*(\bQ) \, \delta_Y^{\cpr} G^{cc}_{mn}(k, Q) \,.
\end{align}
In the second equality, the $-i$ factor is absorbed into $\intkk$, see \Eq{eq:intkk_def}.

\section{Velocity gauge Hamiltonian and current operators} \label{app:velocity_gauge}

One way to include the effect of an external electric field on the electron dynamics is to use the Hamiltonian with a scalar potential of the form
\begin{equation} \label{eq:gauge_H_E}
    \hat{H}^{\rm E}(t) = \hat{H} + e \, \mb{E}(t) \cdot \hat{\br} \,,
\end{equation}
where
\begin{equation}
    \hat{\br} = \nint \dd\br\,  \hat{\psi}^\dagger(\br) \hat{\psi}(\br) \, \br
\end{equation}
is the electron polarization operator.
However, using this length-gauge Hamiltonian~\cite{Aversa1995} introduces numerical challenges due to the breaking of translational invariance or due to the evaluation of momentum space derivative~\cite{Blount1962} which represents the polarization operator~\cite{Ventura2017, Passos2018}.

To avoid these problems, we use the velocity gauge Hamiltonian shown in \Eq{eq:gauge_H_A} of the main text.
Here, we derive the velocity-gauge Hamiltonian and the corresponding current operators [\Eqs{eq:current_Je} and \eqref{eq:current_Jp}].
The vector potential $\mb{A}(t)$ is related to the electric field $\mb{E}(t)$ via the relation
\begin{equation} \label{eq:time_dep}
    \mb{E}(t) = -\frac{\dd\mb{A}(t)}{\dd t} \,.
\end{equation}
Following Ref.~\cite{Ventura2017}, we consider a time-dependent unitary transformation
\begin{equation}
    \hat{U}(t) = e^{i e \mb{A}(t) \cdot \hat{\br}} \,.
\end{equation}
By applying this unitary transformation to the length-gauge Hamiltonian [\Eq{eq:gauge_H_E}], we find that the velocity-gauge Hamiltonian reads
\begin{equation} \label{eq:gauge_H_transf}
    \hat{H}^{\rm A}(t) = \hat{U}^\dagger(t) \hat{H}^{\rm E} \hat{U}(t) - i \hat{U}^\dagger(t) \frac{\dd \hat{U}(t)}{\dd t} \,.
\end{equation}

To evaluate this expression, we calculate the commutator $i [\hat{\br}, \hat{O}]$, where $\hat{O}$ is a generic electronic operator defined as
\begin{equation} \label{eq:gauge_O_r_def}
    \hat{O} = \nint \dd\br \dd\br' O(\br, \br') \hat{\psi}^\dagger(\br) \hat{\psi}(\br') \,,
\end{equation}
where $\hat{\psi}(\br)$ is the electron field operator, with which the electron annihilation operator $\opc_\nk$ can be written as
\begin{equation}
    \opc_\nk = \int \dd\br \, \hat{\psi}(\br) \psi_\nk^*(\br) \,,
\end{equation}
where $\psi_\nk(\br) = u_\nk(\br) e^{i\bk\cdot\br}$.
In the Bloch eigenstate basis, the operator $\hat{O}$ becomes
\begin{equation} \label{eq:gauge_O_k_def}
    \hat{O} = \sum_\mnkq O_\mnkq \, \opcd_\mkq \, \opc_\nk \,,
\end{equation}
with the matrix elements
\begin{multline} \label{eq:gauge_O_mnkq}
    O_\mnkq
    = \nint \dd\br \dd\br' \,
    e^{-i \bq \cdot \br} e^{-i \bk \cdot (\br - \br')}
    \\
    \times u_\mkq^*(\br) O(\br, \br') u_\nk(\br') \,.
\end{multline}
The commutator between the polarization operator and $\hat{O}$ is then
\begin{equation} \label{eq:gauge_rO_der1}
    -i \bigl[ \hat{\br}, \hat{O} \bigr]
    = -\nint \dd\br \dd\br'\, i(\br - \br') O(\br, \br') \, \hat{\psi}^\dagger(\br) \, \hat{\psi}(\br') \,.
\end{equation}
Using integration by parts, we find
\begin{align} \label{eq:gauge_DO_in_r}
    &-i(\br - \br') O(\br, \br')
    \nnnl
    &= \sum_{mn\bk \bq} e^{i \bq \cdot \br} e^{i\bk \cdot (\br - \br')}
    \nabla_\bk \bigl[u_\mkq(\br) O_\mnkq u^*_\nk(\br') \bigr]
    \nnnl
    &= \sum_{mn\bk \bq} e^{i \bq \cdot \br} e^{i\bk \cdot (\br - \br')} u_\mkq(\br) (\mb{\mc{D}} O)_\mnkq u^*_\nk(\br') \,,
\end{align}
where
\begin{multline} \label{eq:gauge_cov_der_def}
    (\mb{\mc{D}} O)_\mnkq
    = \nabla_\bk O_\mnkq \\
    -i (\xi_\bkq O_{\bk\bq})_{mn} +i (O_{\bk\bq} \xi_\bk)_{mn}
\end{multline}
is the covariant derivative of $O$, with $\xi_\bk$ the electron Berry connection [\Eq{eq:current_xi_def}].
Substituting this result into \Eq{eq:gauge_rO_der1} and using \Eq{eq:gauge_O_mnkq}, we obtain
\begin{align} \label{eq:gauge_rO}
    -i \bigl[ \hat{\br}, \hat{O} \bigr]
    = \mb{\mc{D}} \hat{O}
    \equiv \sum_\mnkq (\mb{\mc{D}} O)_\mnkq \, \opcd_\mkq \, \opc_\nk \,.
\end{align}

We now evaluate the two terms in \Eq{eq:gauge_H_transf}.
For the first term, applying the Baker--Hausdorff formula and using \Eq{eq:gauge_rO}, we obtain
\begin{align} \label{eq:gauge_H_term1}
    &\hat{U}^\dagger(t) \hat{H}^E \hat{U}(t)
    \\
    &= \hat{H}^E
    - i e \bigl[ \mb{A}(t) \cdot \hat{\br}, \hat{H} \bigr]
    \nnnl
    &\quad + \frac{(i e)^2}{2} \Bigl[ \mb{A}(t) \cdot \hat{\br}, \bigl[ \mb{A}(t) \cdot \hat{\br}, \hat{H} \bigr] \Bigr]
    + O(A^3)
    \nnnl
    &= \hat{H}^E
    + e \mb{A}(t) \cdot \mb{\mc{D}} \hat{H}
    + \frac{e^2}{2} \bigl[ \mb{A}(t) \cdot \mb{\mc{D}} \bigr]^2 \hat{H}
    + O(A^3) \,.
    \nonumber
\end{align}
The first term is the original length-gauge Hamiltonian.
The second and third terms correspond to the paramagnetic and diamagnetic currents, respectively:
\begin{equation}
    \mb{\mc{D}} \hat{H} = \opmbJ \,,\quad
    \mb{\mc{D}}^2 \hat{H} = \opmbJ^{\rm D} \,.
\end{equation}
We note that the commutator between $\hat{\br}$ and $\mb{E}(t) \cdot \hat{\br}$ is zero.
Next, for the second term of \Eq{eq:gauge_H_transf}, we find
\begin{equation} \label{eq:gauge_H_term2}
    -i \hat{U}^\dagger(t) \frac{\dd \hat{U}(t)}{\dd t}
    = e \frac{\dd \mb{A}(t)}{\dd t} \cdot \hat{\br}
    = -e\, \mb{E} \cdot \hat{\br} \,.
\end{equation}
This term cancels exactly the electronic potential term in the length-gauge Hamiltonian in \Eq{eq:gauge_H_E}.
By adding \Eqs{eq:gauge_H_term1} and \eqref{eq:gauge_H_term2}, we obtain the velocity-gauge Hamiltonian \Eq{eq:gauge_H_A}.

The paramagnetic current operator is given in \Eqs{eq:current_J_def}, \eqref{eq:current_Je}, and \eqref{eq:current_Jp}.
Here, we present an explicit expression for the diamagnetic current operator.
Similarly to the paramagnetic current, the diamagnetic current operator consists of two contributions:
\begin{equation} \label{eq:current_JD}
    \opmbJ^{\rm D}
    = \opmbJ^{\rm D,\,e} + \opmbJ^{\rm D,\,p} \,.
\end{equation}
The electronic contribution is given by
\begin{equation} \label{eq:current_JDe}
    \opmbJ^{\rm D,\,e}
    = \sum_{mn\bk} (\mb{\mc{D}} \mb{v})_\mnk \, \opcd_\mk \, \opc_\nk \,,
\end{equation}
while the phonon-assisted contribution is given by
\begin{equation} \label{eq:current_JDp}
    \opmbJ^{\rm D,\,p}
    = \frac{1}{\sqrt{N_\bk}} \sum_{\substack{\bk\bq \\ mn\nu}} (\mb{\mc{D}}^2 g)_{mn\nu}(\bk, \bq) \opcd_\mkq \, \opc_\nk \, \hat{x}_\nuq \,.
\end{equation}

Finally, we comment on the evaluation of the covariant derivative of the \eph coupling in \textit{ab initio} calculations.
Using the real-space representation [\Eq{eq:gauge_DO_in_r}], the covariant derivative of $g$ can be written as
\begin{align} \label{eq:gauge_Dg_real_space}
    (\mb{\mc{D}} & g)_{mn\nu}(\bk, \bq)
    \nnnl
    &= i \nint \dd\br \dd \br' (\br' - \br) \delta V_\nuq(\br, \br') \psi^*_\mkq(\br) \psi_\nk(\br')
    \nnnl
    &= i \mel{\psi_\mkq}{[\delta \hat{V}_\nuq, \hat{\mathbf{r}}]}{\psi_\nk} \,.
\end{align}
This result shows that only nonlocal potentials, such as nonlocal pseudopotentials or Hubbard corrections, contribute to the covariant derivative of $g$; local potentials do not.
This situation is analogous to the velocity matrix, which includes contributions from the nonlocal potential $\hat{V}^{\rm nl}$ but not from the local one~\cite{Adolph1996}:
\begin{equation}
    \mb{v}_\mnk
    = \mel{\psi_\mk}{\hat{\mathbf{p}}}{\psi_\nk}
    + i\mel{\psi_\mk}{[\hat{V}^{\rm nl}, \hat{\br}]}{\psi_\nk} \,.
\end{equation}
Since the momentum operator $\hat{\mathbf{p}}$ is independent of atomic displacements, only the variation of the nonlocal potential contributes to the covariant derivative of the \eph coupling.

In practice, evaluating the \textit{ab initio} $\mb{\mc{D}} g$ using \Eq{eq:gauge_Dg_real_space} requires substantial implementation.
Using \Eq{eq:current_Dg} is even more demanding, as it involves summing over all electronic bands.
Therefore, in this work, we adopt the tight-binding approximation, where the Berry connection is assumed to be diagonal in the Wannier function basis~\cite{Graf1995}.
The Wannier interpolation formula for \eph coupling reads~\cite{Giustino2007EPW}
\begin{equation}
    g_{mn\nu}(\bk, \bq)
    = \sum_{ij\bR} e^{i\bk\cdot\bR} U_{mi\bkq}^\dagger \, g_{ij\nu}(\bR, \bq) U_{jn\bk} \,,
\end{equation}
where $U_{in\bk}$ is the eigenvector in the Wannier basis, and $g_{ij\nu}(\bR, \bq)$ is the \eph coupling in the electronic Wannier and phonon Bloch basis.
Within the tight-binding approximation, the covariant derivative of the \eph coupling is given by
\begin{multline} \label{eq:current_Dg_tight-binding}
    (\mb{\mc{D}} g)_{mn\nu}(\bk, \bq)
    \approx i \sum_{ij\bR} (\bR + \br_j - \br_i) e^{i\bk\cdot\bR}
    \\
    \times U_{mi\bkq}^\dagger \, g_{ij\nu}(\bR, \bq) U_{jn\bk} \,,
\end{multline}
where $\br_i$ denotes the center of the $i$-th Wannier function.
This approximation can be systematically improved by including inter-orbital and inter-atomic position matrix elements~\cite{Bennetto1996, Ibanez2022}.

\section{Boltzmann transport equations} \label{sec:transport_bte}

It has been shown~\cite{MahanBook, Butler1985, Kim2019Vertex} that the self-consistent ladder equations reduce to the BTE~\cite{MahanBook, ZimanBook, Ponce2020Review} under the quasiparticle approximation.
This approximation assumes that the spectral function is a sharp delta function, satisfying
\begin{subequations}
\begin{align}
    A_\nk(\veps) &\approx \delta(\veps - \veps_\nk) \,,
    \\
    [A_\nk(\veps)]^2 &\approx \frac{\tau_\nk}{\pi} \delta(\veps - \veps_\nk) \,,
\end{align}
\end{subequations}
where $\tau_\nk = 1 / 2\abs{\Im \Sigma_\nk(\veps_\nk)}$ is the quasiparticle lifetime.
In addition, we neglect the phonon-assisted current and interband velocity.
With these approximations, the self-consistent ladder equation for the electronic current reduces to the BTE~\cite{MahanBook, Butler1985, Kim2019Vertex}
\begin{multline} \label{eq:sigma_bte_equation}
    -e\, v_\nk^\alpha \, f'_\nk
    = - \tau^{-1}_\nk \partial_{E_\alpha} f_\nk
    \\
    + \sum_m \intbq \tau^{-1}_{\nk \leftarrow \mkq} \partial_{E_\alpha} f_\mkq \,,
\end{multline}
where $\partial_{E_\alpha} f_\nk$ is the change in electron occupation at state $\nk$ due to an electric field along the $\alpha$ direction, and $f'_\nk =  \dd f^+(\veps) / \dd\veps |_{\veps = \veps_\nk}$.
The partial decay rate is given by
\begin{multline} \label{eq:sigma_bte_partial_decay}
    \tau^{-1}_{\mkq \leftarrow \nk} = 2\pi \sum_\nu \abs{g_{mn\nu}(\bk, \bq)}^2
    \\
    \times \sum_{\pm} \delta(\veps_\nk - \veps_\mkq \pm \omega_\nuq) (n_\nuq + f^\pm_\mkq) \,,
\end{multline}
and the total decay rate is
\begin{equation}
    \tau^{-1}_\nk = \sum_m \intbq \tau^{-1}_{\mkq \leftarrow \nk} \,.
\end{equation}
After solving \Eq{eq:sigma_bte_equation}, the dc BTE conductivity is computed as
\begin{equation} \label{eq:sigma_bte}
    \sigma^{\rm BTE}_{\alpha\beta}(0) = -\frac{e}{V^{\rm uc}} \sum_n \intbk v_\nk^\alpha \, \partial_{E_\beta} f_\nk \,.
\end{equation}

With the quasiparticle approximation, the dc bubble conductivity [\Eq{eq:sigma_bubble}] reduces to the self-energy relaxation time approximation (SERTA)~\cite{Ponce2020Review} conductivity:
\begin{equation} \label{eq:sigma_serta}
    \sigma^{\rm SERTA}_{\alpha\beta}(0)
    = -\frac{e^2}{V^{\rm uc}} \sum_{n} \intbk v_\nk^\alpha v_\nk^\beta f'_\nk \tau_\nk \,.
\end{equation}
The SERTA conductivity is also obtained by neglecting the second scattering term in the BTE \eqref{eq:sigma_bte_equation}, which corresponds to disregarding the vertex correction.

The ac conductivity can be calculated using the BTE by considering a time-dependent electric field.
For a monochromatic electric field, the BTE reads~\cite{Lihm2024NLHE}:
\begin{multline} \label{eq:sigma_bte_ac_equation}
    -e\, v_\nk^\alpha \, f'_\nk
    = - (i\Omega + \tau^{-1}_\nk) \partial_{E_\alpha} f_\nk(\Omega)
    \\
    + \sum_m \intbq \tau^{-1}_{\nk \leftarrow \mkq} \partial_{E_\alpha} f_\mkq(\Omega) \,,
\end{multline}
where the only difference from the dc case is the $i \Omega$ term on the right side.
The corresponding ac BTE conductivities are given by
\begin{align}
    \label{eq:sigma_bte_ac}
    \sigma^{\rm BTE}_{\alpha\beta}(\Omega)
    &= -\frac{e}{V^{\rm uc}} \sum_n \intbk
    v_\nk^\alpha \, \partial_{E_\beta} f_\nk(\Omega) \,,
    \\
    \label{eq:sigma_serta_ac}
    \sigma^{\rm SERTA}_{\alpha\beta}(\Omega)
    &= -\frac{e^2}{V^{\rm uc}} \sum_{n} \intbk
    v_\nk^\alpha v_\nk^\beta f'_\nk
    \frac{\tau_\nk}{1 + i\Omega \tau_\nk} \,.
\end{align}

\section{Implementation details} \label{app:implementation}

\subsection{Hilbert transform with linear interpolation} \label{app:implementation_spline}

We represent frequency-dependent functions on a frequency grid defined by points $\veps_0, \cdots, \veps_{N+1}$, which may be nonuniform.
Function values are linearly interpolated between grid points and are assumed to vanish at the endpoints and outside the grid range:
\begin{equation} \label{eq:itp_f}
    f(\veps) = \sum_{i=1}^{N} f_i b_i(\veps)\,,
\end{equation}
where $b_i(\veps)$ is a linear spline basis function that linearly interpolates $(\veps_{i-1}, 0)$, $(\veps_{i}, 1)$, and $(\veps_{i+1}, 0)$:
\begin{equation}
    b_i(\veps) = \begin{cases}
        \frac{\veps - \veps_{i-1}}{\veps_i - \veps_{i-1}} & \text{if $\veps_{i-1} \leq \veps \leq \veps_i$} \\
        \frac{\veps_{i+1} - \veps}{\veps_{i+1} - \veps_i} & \text{if $\veps_{i} \leq \veps \leq \veps_{i+1}$} \\
        0 & \text{otherwise}\,.
    \end{cases}
\end{equation}
The coefficients $f_i$ are the function values at the grid points $\veps_i$:
\begin{equation}
    f_i = f(\veps_i) \,.
\end{equation}

By expressing functions in terms of the linear spline basis coefficients, frequency integrals can be recast as matrix multiplications involving these coefficients.
The matrix representation of the integral operator can be computed once and reused, reducing the computational cost.
This approach has previously been applied to the Hilbert transform~\cite{Brambilla2009, Bilato2014}.
For completeness, we summarize the relevant details in the following.

We aim to evaluate the Kramers--Kronig transformation of a function $f(\veps)$, given by
\begin{equation} \label{eq:spline_KK}
    \mathrm{KK}[f](\veps) = \frac{1}{\pi} \mcP\! \nint \dd\veps' \, \frac{f(\veps')}{\veps' - \veps} \,,
\end{equation}
where $\mc{P}$ denotes the Cauchy principal value.
We compute this integral at the grid points $\veps = \veps_1, \cdots, \veps_N$ by analytically integrating over the linear spline segments.
This yields
\begin{equation}
    \mathrm{KK}[f](\veps_i)
    = \frac{1}{\pi} \mcP\! \nint \dd\veps' \, \frac{f(\veps')}{\veps' - \veps_i}
    = \sum_{j = 1}^N h_j(\veps_i) f_j \,,
\end{equation}
where the kernel $h_j(\veps)$ is the Kramers--Kronig transformation of the linear spline basis function $b_j(\veps)$ and is given by
\begin{align}
    h_j(x)
    &= \frac{1}{\pi} \mcP\! \nbint{\veps_{j-1}}{\veps_{j+1}} \dd\veps' \, \frac{b_j(\veps')}{\veps' - x}
    \nnnl
    &=\quad \frac{(\veps_{j-1} - x) \log \abs{\veps_{j-1} - x}}{\pi (\veps_{j} - \veps_{j-1})}
    \nnnl
    &\quad + \frac{(\veps_{j+1} - x) \log \abs{\veps_{j+1} - x}}{\pi (\veps_{j+1} - \veps_{j})}
    \nnnl
    &\quad - \frac{(\veps_{j+1} - \veps_{j-1}) (\veps_{j} - x) \log \abs{\veps_{j} - x}}{\pi (\veps_{j} - \veps_{j-1}) (\veps_{j+1} - \veps_{j})}\,.
\end{align}

\subsection{Optimization of convolutions}  \label{app:implementation_convolution}

The main computational bottleneck in a ladder-\scGD calculation is the convolution between $D$ and $\delta G$ that appears in \Eq{eq:sigma_BSE}.
This convolution is expressed as
\begin{equation} \label{eq:conv_opt_main}
    H^{c c'\nsp}(\veps) = \frac{1}{N_q} \sum_{\nuq} \nint \frac{\dd\omega}{2\pi i} \, \abs{g_\nuq}^2 D^{c'\nsp c}_{0, \nuq}(\omega) F^{c c'\nsp}(\veps + \omega) \,,
\end{equation}
where $D_{0,\nuq}$ is defined in \Eq{eq:D0}, and $F$ is an arbitrary function of frequency.
In the following, we outline the efficient strategy used to evaluate this convolution.

As a first step, for each phonon mode $\nuq$, we linearly interpolate the function $F$ and construct shifted functions
\begin{equation} \label{eq:conv_opt_step1}
    F^{c c'\nsp}_{\pm \nuq}(\veps) = F^{c c'\nsp}(\veps \pm \omega_\nuq) \,.
\end{equation}
Since this interpolation is performed most frequently and constitutes a major computational bottleneck, we precompute and reuse the interpolation weights for all inequivalent phonon frequencies.

Next, we compute six intermediate functions by summing the shifted versions of $F$ over all phonon modes:
\begin{subequations} \label{eq:conv_opt_step2}
\begin{align}
    H^{(1)} &= \frac{1}{N_q} \sum_\nuq \tfrac{1}{2} \bigl[ F^{--}_{+ \nuq} - F^{--}_{- \nuq} \bigr] \,,
    \\
    H^{(2)} &= \frac{1}{N_q} \sum_\nuq -(n_\nuq + \tfrac{1}{2}) \bigl[ F^{--}_{+ \nuq} + F^{--}_{- \nuq} \bigr] \,,
    \\
    H^{(3)} &= \frac{1}{N_q} \sum_\nuq \tfrac{1}{2} \bigl[ F^{++}_{+ \nuq} - F^{++}_{- \nuq} \bigr] \,,
    \\
    H^{(4)} &= \frac{1}{N_q} \sum_\nuq -(n_\nuq + \tfrac{1}{2}) \bigl[ F^{++}_{+ \nuq} + F^{++}_{- \nuq} \bigr] \,,
    \\
    H^{(5)} &= \frac{1}{N_q} \sum_\nuq -\bigl[ n_\nuq F^{+-}_{+ \nuq} + (n_\nuq + 1) F^{+-}_{- \nuq} \bigr] \,,
    \\
    H^{(6)} &= \frac{1}{N_q} \sum_\nuq -\bigl[ (n_\nuq+1) F^{-+}_{+ \nuq} + n_\nuq F^{-+}_{- \nuq} \bigr] \,.
\end{align}
\end{subequations}
We omitted the argument $\veps$ for brevity.
To proceed, we use the identity
\begin{equation}
    \frac{1}{\pi} \,\mcP\! \nint \dd\omega \, \frac{f(\veps + \omega)}{\omega \mp \omega_\nuq}
    =
    \frac{1}{\pi} \,\mcP\! \nint \dd\omega \, \frac{f(\omega \pm \omega_\nuq)}{\omega - \veps} \,,
\end{equation}
which allows the convolution in \Eq{eq:conv_opt_main} to be expressed as
\begin{subequations} \label{eq:conv_opt_step3}
\begin{align}
    H^{--}(\veps) &= -i\,\mathrm{KK}\bigl[ H^{(1)} \bigr](\veps)
    + H^{(2)}(\veps) \,,
    \\
    H^{++}(\veps) &= +i\,\mathrm{KK}\bigl[ H^{(3)} \bigr](\veps)
    + H^{(4)}(\veps) \,,
    \\
    H^{+-}(\veps) &= H^{(5)}(\veps) \,,
    \\
    H^{-+}(\veps) &= H^{(6)}(\veps) \,,
\end{align}
\end{subequations}
The Kramers--Kronig transformations are evaluated using the spline-based method described in \Eq{eq:spline_KK}.

This scheme is numerically efficient because it avoids performing any frequency integrals inside the loop over $\bq$.
Instead, only linear interpolation is performed inside the $\bq$-loop.
The frequency integration is performed only \textit{after} completing the summation over $\bq$.

\subsection{Cumulant approximation} \label{app:cumulant}

Here, we describe our implementation of the retarded cumulant approximation, which we find to be robust for all the systems studied in this work.
Our goal is to evaluate the cumulant Green's function [\Eq{eq:cum_G}]
\begin{equation} \label{eq:cumulant_G}
    G^{\rm R,\,Cum.}_\nk(\veps) =-i \nint \dd t \, \Theta(t) e^{i (\veps - \veps_\nk - \Sigma^{\rm static}_{nn\bk}) t} e^{C_\nk(t)} \,,
\end{equation}
where the cumulant function $C_\nk(t)$ is defined in \Eq{eq:cum_C}.
Defining
\begin{equation} \label{eq:cumulant_beta}
    \beta(\veps) = \frac{1}{\pi} \abs{\Im \Sigma_\nk^{{\rm R}, G_0 D_0}(\veps)} \,,
\end{equation}
we can rewrite \Eq{eq:cum_C} as
\begin{equation} \label{eq:cumulant_C_beta}
    C_\nk(t) = \nbint{-\infty}{\infty} \dd\veps \, \beta(\veps + \veps_\nk) \frac{e^{-i\omega t} + i\veps t - 1}{\veps^2} \,.
\end{equation}
The real part of the retarded self-energy is given by the Kramers--Kronig relation
\begin{equation} \label{eq:cumulant_KK}
    \Re \Sigma_\nk^{{\rm R}, G_0 D_0}(\veps) = - \mc{P} \nbint{-\infty}{\infty} \dd\veps' \, \frac{\beta(\veps')}{\veps' - \veps} \,.
\end{equation}

In principle, one could compute $C_\nk(t)$ by directly evaluating \Eq{eq:cumulant_C_beta} on a dense and extended time grid.
For efficiency, however, we make use of its analytic long-time limit, given by
\begin{equation} \label{eq:cumulant_C_limit}
    \lim_{t \to \infty} C_\nk(t)
    = -i \Sigma_\nk^{{\rm R}}(\veps_\nk) \, t
    + \left. \frac{\dd \Sigma_\nk^{{\rm R}}(\veps)}{\dd \veps} \right|_{\veps = \veps_\nk} \,,
\end{equation}
as shown in \Sec{sec:supp_cumulant_proof} of the SM~\cite{supplemental} (we omitted the $G_0 D_0$ superscript on $\Sigma$).
We explicitly evaluate the integral in \Eq{eq:cumulant_C_beta} on a short, dense time grid.
Then, we extrapolate to arbitrarily long times using the asymptotic form, \Eq{eq:cumulant_C_limit}.
Using a sufficiently large time cutoff ensures good energy resolution after the Fourier transformation.

We also find that numerical stability critically depends on evaluating the exponential $e^{C_\nk(t)}$ in the time domain before performing the Fourier transformation.
In contrast, evaluating \Eq{eq:cumulant_beta} through repeated convolutions in the frequency domain leads to numerical instabilities and convergence issues.
Since $C_\nk(t)$ diverges in time, its Fourier transform has sharp nonanalytic features which requires extremely dense frequency sampling to resolve.
By contrast, the exponential form $e^{C_\nk(t)}$ decays exponentially to zero when $\Im \Sigma^{\rm R}(\veps_\nk) \neq 0$ and, otherwise remains a bounded oscillatory function.
Its Fourier transform is therefore much better behaved than that of $C_\nk(t)$.

Our workflow for the cumulant approximation is as follows.
(1) Calculate $C_\nk(t)$ on a dense, short time grid using \Eq{eq:cumulant_C_beta}. We use the analytic form of $\Sigma^{\rm R}$ if known; otherwise, we compute $\Sigma^{\rm R}$ on a grid of frequencies and then linearly interpolate it.
(2) Extrapolate to a long-time grid by using the linear asymptotic form \Eq{eq:cumulant_C_limit}.
(3) Evaluate $e^{C_\nk(t)}$ for the extrapolated cumulant function.
(4) Evaluate the Green's function with \Eq{eq:cumulant_G}, using fast Fourier transformation.

We note that the same idea can be applied to the time-ordered cumulant approximation, which defines the cumulant function as \Eq{eq:cumulant_C_beta}, with the $\beta$ function replaced by
\begin{equation}
    \beta^\pm(\omega) = \beta(\omega)\Theta[\pm(\mu - \omega)] \,.
\end{equation}
The workflow described above applies with a minimal modification.

\section{Derivation of the Ward--Takahashi identity and the continuity equation} \label{app:Ward}

In this Appendix, we derive the Ward--Takahashi identity, the continuity equation, and \Eq{eq:epsilon_sigma_relation}, which relates the conductivity to the dielectric function for the ladder-\scGD method.
Here, we consider the single-band case, where the derivation is simplified by the trivial overlap matrix [\Eq{eq:N_mel_def}], $N_{\bk}(\bQ) = 1$.
We omit band indices, e.g., we write the \eph coupling as $g_\nu(\bk, \bq)$.
For multiband systems, the same derivation can be applied by fixing the wavefunction gauge: see \Sec{sec:ward_identity_multiband} of SM for details~\cite{supplemental}.

Before presenting the full derivation, we summarize the key steps.
First, we define a finite-$\bQ$ generalization of the current vertex $\delta^{c}_{\Delta^{\rm e, p}} \Sigma_0$ [\Eq{eq:ward_1band_current_vertex}], where the covariant derivatives are replaced by finite differences.
The usual current vertex, defined in \Eqs{eq:sigma_V0_el} and \eqref{eq:lr_Vbare_ep}, is recovered in the long-wavelength limit $\bQ \to \mb{0}$:
\begin{equation} \label{eq:ward_Delta_long_range}
    \delta^{c}_{\Delta^{\rm e, p}} \Sigma_{0, 12}(k, Q)
    = i \bQ \cdot \delta^{c}_{\mb{J}^{\rm e, p}} \Sigma_{0, 12}(k, \Omega)
    + O(Q^2) \,.
\end{equation}
Next, we prove the Ward--Takahashi identity relating density and current response functions as
\begin{equation} \label{eq:ward_1band_continuity}
    Z^{c} P^c_{12}(k ,Q)
    = i \delta^{c}_{\Delta} G_{12}(k, Q)
    - \Omega \, \delta^{c}_N G_{12}(k, Q) \,,
\end{equation}
where
\begin{equation} \label{eq:ward_1band_P_def}
    P^c_{12}(k, Q)
    \equiv G_{12}(k+Q) \delta_{c_2 c} - G_{12}(k) \delta_{c_1 c} \,.
\end{equation}
Using this identity, we represent the current-current response function in terms of the density-density response function
\begin{equation} \label{eq:ward_1band_chi_final}
    \chi^{c\cpr}_{\Delta \Delta}(Q)
    = \Omega^2 \chi^{c\cpr}_{N N}(Q)
    -i I^{c\cpr}_{\Delta}(Q)
    \,,
\end{equation}
where the additional term $I^{c\cpr}_{\Delta}(Q)$ yields the diamagnetic and phonon-assisted bubble contributions in the long-wavelength limit:
\begin{multline} \label{eq:ward_I_term}
    I^{c\cpr}_{\Delta}(Q)
    = i \sum_{\alpha\beta} \Bigl[ Z^{c\cpr} \expval{\hat{J}^{\rm D}_{\alpha\beta}}
    - \Lambda_{\alpha\beta}^{{\rm (pp\tbar Bubble)} c\cpr}(\Omega)
    \Bigr] Q^\alpha Q^\beta
    \\
    + O(\bQ^3) \,.
\end{multline}
Taking the long-wavelength limit of $\chi_{\Delta\Delta}(Q)$ yields the usual current-current susceptibility [\Eq{eq:chi_def}] minus the phonon-assisted bubble term:
\begin{multline} \label{eq:ward_chi_DeltaDelta}
    \chi^{c\cpr}_{\Delta \Delta}(Q)
    = \sum_{\alpha \beta} \Bigl[ \Lambda^{c\cpr}_{\alpha \beta}(\Omega) - \Lambda_{\alpha\beta}^{{\rm (pp\tbar Bubble)} c\cpr}(\Omega) \Bigr] Q^\alpha Q^\beta
    \\
    + O(\bQ^3) \,.
\end{multline}
Finally, from the second derivative of \Eq{eq:ward_1band_chi_final} with respect to $\bQ$, we obtain the relation between conductivity and dielectric function:
\begin{multline} \label{eq:ward_1band_result1}
    \Lambda^{c\cpr}_{\alpha \beta}(\Omega, \mb{0})
    + \Lambda^{c\cpr}_{\beta \alpha}(\Omega, \mb{0})
    \\
    = \Omega^2 \frac{\partial^2 \chi^{c\cpr}_{N N}(Q)}{\partial Q^\alpha \partial Q^\beta} \biggr\rvert_{\bQ = \mb{0}}
    + Z^{c\cpr} \bigl(
        \expval{\hat{J}^{\rm D}_{\alpha\beta}}
        + \expval{\hat{J}^{\rm D}_{\beta\alpha}}
    \bigr) \,.
\end{multline}
Using the definition of the dielectric function \Eq{eq:ward_epsilon_def}, we find
\begin{align} \label{eq:ward_1band_result2}
    \Re \sigma^{\rm L}_{\alpha\beta}(\Omega)
    &= -\frac{e^2}{2 \Omega V^{\rm uc}} \Im \Lambda^{\rm R}_{\alpha \beta}(\Omega, \mb{0})
    + (\alpha \leftrightarrow \beta)
    \nnnl
    &= -\frac{\Omega}{V^{\rm uc}} \frac{e^2}{2} \Im \frac{\partial^2 \chi^{\rm R}_{N N}(Q)}{\partial Q^\alpha \partial Q^\beta} \biggr\rvert_{\bQ = \mb{0}}
    \nnnl
    &= \frac{\Omega}{4\pi} \Im \epsilon_{\alpha\beta}(\Omega) \,.
\end{align}
This equation is the imaginary part of the relation between the longitudinal conductivity and dielectric function \Eq{eq:epsilon_sigma_relation}, which we copy here for convenience:
\begin{equation} \label{eq:ward_1band_result3}
    \epsilon_{\alpha\beta}(\Omega)
    = 1 + i \frac{4\pi}{\Omega} \sigma^{\rm L}_{\alpha\beta}(\Omega) \,.
\end{equation}
The equality for the real part follows from the Kramers--Kronig relation.

We now present the detailed derivation.
The finite-$\bQ$ bare current vertex is defined as
\begin{subequations} \label{eq:ward_1band_current_vertex}
\begin{equation}
    \delta^{c}_{\Delta} \Sigma_{0, 12}(k, Q)
    = \delta^{c}_{\Delta^{\rm e}} \Sigma_{0, 12}(k, Q)
    + \delta^{c}_{\Delta^{\rm p}} \Sigma_{0, 12}(k, Q) \,,
\end{equation}
where the electronic and phonon-assisted contributions are
\begin{equation} \label{eq:ward_1band_Sigma_Delta_e_def}
    \delta^{c}_{\Delta^{\rm e}} \Sigma_{0, 12}(k, Q)
    = i (\veps_\bkQ - \veps_\bk) \delta_{c_1 c} \delta_{c_2 c} \,,
\end{equation}
and
\begin{align} \label{eq:ward_1band_Sigma_Delta_p_def}
    &\delta^{c}_{\Delta^{\rm p}} \Sigma_{0, 12}(k, Q)
    = -i Z^{c_1} Z^{c_2} \intqq D_{0,\nu}^{c_2 c_1}(q) \, \Bigl[
    \nnnl
    &\qquad g^*_\nu(\bkQ, \bq) (\Delta_\bQ g_\nu)(\bk, \bq)
    G_{12}(k+q+Q) Z^{c_2 c}
    \nnnl
    &\qquad  + (\Delta_\bQ g_\nu)^*(\bk, \bq) g_\nu(\bk, \bq)
    G_{12}(k+q) Z^{c_1 c}
    \Bigr]
    \,,
\end{align}
\end{subequations}
where
\begin{align} \label{eq:ward_1band_Delta_def}
    (\Delta_\bQ g_\nu)(\bk, \bq)
    &\equiv g_\nu(\bkQ, \bq) - g_\nu(\bk, \bq)
    \nnnl
    &= \bQ \cdot \nabla_\bk g_\nu(\bk, \bq) + O(\bQ^2)
    \,.
\end{align}
In the long-wavelength limit $\bQ \to \mb{0}$, the leading order term is the usual current vertex, see \Eq{eq:ward_Delta_long_range}.
We also define the bare finite-$\bQ$ density vertex as
\begin{equation} \label{eq:ward_1band_Sigma_N_def}
    \delta^{c}_{N} \Sigma_{0, 12}(k, Q)
    \equiv N_\bk(\bQ) \delta_{c_1 c} \delta_{c_2 c}
    = \delta_{c_1 c} \delta_{c_2 c} \,,
\end{equation}
where we used $N_\bk(\bQ) = 1$ for a single-band system.

We now prove the Ward--Takahashi identity, \Eq{eq:ward_1band_continuity}.
It suffices to show that the $P^c_{12}(k, Q)$ defined in \Eq{eq:ward_1band_P_def} satisfies the ladder equation
\begin{multline} \label{eq:ward_1band_bse}
    \quad P^c_{12}(k ,Q)
    - (\Pi \circ W \circ P^c)_{12}(k, Q)
    \\
    = i Z^{c} \bigl[ \Pi \circ (\delta^{c}_{\Delta} \Sigma_{0}
    - \Omega \, \delta^{c}_N \Sigma_{0} ) \bigr]_{12}(k, Q)
    \,.
\end{multline}
Substituting \Eq{eq:ward_1band_Delta_def} into \Eq{eq:ward_1band_Sigma_Delta_p_def}, we obtain
\begin{align} \label{eq:ward_1band_der2}
    &-i \delta^{c}_{\Delta^{\rm p}} \Sigma_{0, 12}(k, Q)
    = - Z^{c_1} Z^{c_2} \intqq D_{0,\nu}^{c_2 c_1}(q)
    \nnnl
    &\hspace{0.5em} \times \Bigl[ g^*_\nu(\bkQ, \bq) g_\nu(\bk, \bq)
    \nnnl
    &\hspace{4em} \times \bigl( G_{12}(k+q+Q) Z^{c_2 c} + G_{12}(k+q) Z^{c_1 c} \bigr)
    \nnnl
    &\hspace{2em} - g^*_\nu(\bkQ, \bq) g_\nu(\bkQ, \bq) G_{12}(k+q+Q) Z^{c_2 c}
    \nnnl
    &\hspace{2em} - g^*_\nu(\bk, \bq) g_\nu(\bk, \bq) G_{12}(k+q) Z^{c_1 c}
    \Bigr]
    \,.
\end{align}
The first term inside the square bracket contains the factor $g^*_\nu(\bkQ, \bq) g_\nu(\bk, \bq)$, which matches the phonon-mediated electron-electron interaction at finite $\bQ$ [\Eq{eq:lr_W_def}].
Also, the sum of the Green's functions multiplying it can be expressed as
\begin{multline}
    G_{12}(k+q+Q) Z^{c_2 c} + G_{12}(k+q) Z^{c_1 c}
    \\
    = P^c_{12}(k+q, Q) Z^{c} \,,
\end{multline}
with $P$ defined in \Eq{eq:ward_1band_P_def}.
Hence, the first term of \Eq{eq:ward_1band_der2} simplifies to $Z^{c} (W \circ P^c)_{12}(k+q, Q)$.
The second and third terms each involve two \eph couplings at momenta $\bkQ$ or $\bk$, respectively.
These momenta also appear in the corresponding Green's function $G_{12}(k+q+Q)$ or $G_{12}(k+q)$.
Thus, these terms correspond to the \scGD self-energy equation \eqref{eq:sc_Sigma_c}.
Combining these results, we get
\begin{multline} \label{eq:ward_1band_der1}
    -i \delta^{c}_{\Delta^{\rm p}} \Sigma_{0, 12}(k, Q)
    = Z^{c} (W \circ P^c)_{12}(k+q, Q)
    \\
    - Z^{c} \bigl[ \Sigma_{12}(k+Q) \delta_{c_2 c} - \Sigma_{12}(k) \delta_{c_1 c} \bigr]
    \,.
\end{multline}

Next, starting from the Dyson equation \eqref{eq:sc_Dyson_inv}, we matrix-multiply $G_\bk$ from the left or right to find
\begin{subequations} \label{eq:ward_dyson2}
\begin{align}
    \label{eq:ward_dyson2a}
    G_{12}(k) (\veps - \veps_\bk) Z^{c_2}
    &= \delta_{c_1 c_2} + \sum_{3} G_{13}(k) \Sigma_{32}(k) \,, \\
    \label{eq:ward_dyson2b}
    (\veps - \veps_\bk) Z^{c_1} G_{12}(k)
    &= \delta_{c_1 c_2} + \sum_{3} \Sigma_{13}(k) G_{32}(k) \,.
\end{align}
\end{subequations}
We apply these identities to the bubble function $\Pi_{1234}(k, Q) = G_{13}(k+Q) G_{42}(k)$.
Applying \Eq{eq:ward_dyson2a} to the first Green's function gives
\begin{multline} \label{eq:ward_bubble1}
    (\veps + \Omega - \veps_\bkQ) Z^{c_3} \Pi_{1234}(k, Q)
    \\
    = \delta_{c_1 c_3} G_{42}(k)
    + \sum_5 \Pi_{1254}(k, Q) \Sigma_{53}(k+Q) \,.
\end{multline}
Similarly, applying \Eq{eq:ward_dyson2b} to the second Green's function yields
\begin{multline} \label{eq:ward_bubble2}
    (\veps - \veps_\bk) Z^{c_4} \Pi_{1234}(k, Q)
    \\
    = G_{13}(k+Q) \delta_{c_4 c_2}
    + \sum_5 \Pi_{1235}(k, Q) \Sigma_{45}(k) \,.
\end{multline}
Setting $c_3 = c_4$ and subtracting these equations, we find
\begin{multline} \label{eq:ward_bubble_final}
    (\veps_\bkQ - \veps_\bk - \Omega) Z^{c_3} \Pi_{1233}(k, Q)
    \\
    = P_{123}(k ,Q)
    - \sum_{1' 2'} \Pi_{121'2'}(k, Q) \bigl[ \Sigma_{1'3}(k+Q) \delta_{c_2' c_3}
    \\
    - \Sigma_{32'}(k) \delta_{c_1' c_3} \bigr]
    \,.
\end{multline}
Using the definitions \Eqs{eq:ward_1band_Sigma_Delta_e_def} and \eqref{eq:ward_1band_Sigma_N_def}, we rewrite \Eq{eq:ward_bubble_final} as
\begin{multline} \label{eq:ward_bubble_final2}
    P^c_{12}(k ,Q)
    = i Z^{c} \bigl[ \Pi \circ (\delta^{c}_{\Delta^{\rm e}} \Sigma_{0}
    - \Omega \, \delta^{c}_N \Sigma_{0} ) \bigr]_{12}(k, Q)
    \\
    + \sum_{1' 2'} \Pi_{121'2'}(k, Q) \bigl[ \Sigma_{1'2'}(k+Q) \delta_{c_2' c}
    - \Sigma_{1'2'}(k) \delta_{c_1' c} \bigr].
\end{multline}
Substituting \Eq{eq:ward_1band_der1} into \Eq{eq:ward_bubble_final2} yields \Eq{eq:ward_1band_bse}, completing the proof of the Ward--Takahashi identity \eqref{eq:ward_1band_continuity}.

As a side note, the single-particle Ward identity follows directly from \Eq{eq:ward_1band_continuity}.
Setting $\Omega = 0$ and summing over $c$, we obtain the Ward--Takahashi identity for the single-particle self-energy:
\begin{multline}
    G_{12}(k + \bQ) - G_{12}(k)
    \\
    = i \sum_{c} Z^{c} \, \delta^{c}_{\Delta} G_{12}(k, \bQ, \Omega=0) \,.
\end{multline}
Taking the $O(\bQ)$ term yields the single-particle Ward identity for the Green's functions
\begin{equation} \label{eq:ward_1band_G_ward}
    \nabla_\bk G^{c_1 c_2}_\bk(\veps)
    = - \sum_{c} Z^{c} \, \delta^{c}_{\mb{J}} G^{c_1 c_2}_\bk(\veps, \Omega=0) \,,
\end{equation}
and that for the self-energy
\begin{equation}
    \nabla_\bk \Sigma^{c_1 c_2}_\bk(\veps)
    = - \sum_{c} Z^{c} \, \delta^{c}_{\mb{J}} \Sigma^{c_1 c_2}_\bk(\veps, \Omega=0) \,.
\end{equation}
When the \eph coupling is $\bk$ dependent, the phonon-assisted contribution must be included in these Ward identities.
If the \eph coupling is $\bk$-independent, these single-band expressions reduce to the Ward identities in Refs.~\cite{Engelsberg1963, MahanBook, Kim2019Vertex} (see, for example, Eq.~(C8) of Ref.~\cite{Kim2019Vertex}).
When there are multiple electron bands, the change in the wavefunction character (Berry connection) also contributes to the phonon-assisted current: see \Sec{sec:ward_identity_multiband} of SM for details~\cite{supplemental}.

\begin{figure}[tb]
    \centering
    \includegraphics[width=0.99\linewidth]{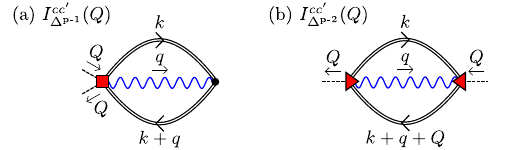}
    \caption{
    Feynman diagrams for
    (a) $I^{c\cpr}_{\Delta^{{\rm p}\tbar1}}(Q)$ [\Eq{eq:ward_1band_I_p1}]
    (b) and $I^{c\cpr}_{\Delta^{{\rm p}\tbar2}}(Q)$ [\Eq{eq:ward_1band_I_p2}].
    The red square represents the second-order finite difference of the \eph coupling: $g_\nu(\bkQ, \bq) + g_\nu(\bkmQ, \bq) - 2g_\nu(\bk, \bq)$, while the red triangles represent $(\Delta_{\bQ} g_\nu)(\bk, \bq)$.
    In the long-wavelength limit $\bQ \to \mb{0}$, (b) reduces to the phonon-assisted bubble shown in \Fig{fig:feynman_conductivity}(e).
    }
    \label{fig:feynman_ward_derivation}
\end{figure}

We now use the Ward--Takahashi identity to relate the $\Delta\Delta$ and $NN$ susceptibilities and derive \Eq{eq:ward_1band_chi_final}.
By contracting the response function with any perturbation $X$, as in \Eq{eq:sigma_XY_from_R}, we find a relation between the $X\Delta$ and $XN$ susceptibilities:
\begin{equation} \label{eq:ward_1band_chi_1}
    i \chi^{c\cpr}_{X \Delta}(Q)
    = \Omega \, \chi^{c\cpr}_{X N}(Q)
    + I^{c\cpr}_X(Q) \,,
\end{equation}
where
\begin{equation} \label{eq:ward_1band_chi_2}
    I^{c\cpr}_X(Q)
    \equiv
    Z^{\cpr} \sum_{c_1 c_2} \intkk \delta^c_{X} \Sigma_{0}^{c_2 c_1}(k+Q, -Q)
    P^{c_1 c_2\cpr}(k, Q) \,.
\end{equation}
Setting $X = \Delta$ in Eq.~\eqref{eq:ward_1band_chi_1} gives
\begin{equation} \label{eq:ward_1band_chi_3}
    i \chi^{c\cpr}_{\Delta \Delta}(Q)
    = \Omega \chi^{c\cpr}_{\Delta N}(Q)
    + I^{c\cpr}_{\Delta}(Q)
    \,.
\end{equation}
Similarly, with $X=N$, we get
\begin{align} \label{eq:ward_1band_chi_4}
    \chi^{c\cpr}_{\Delta N}(Q)
    &= \chi^{\cpr c}_{N \Delta}(-Q) \nnnl
    &= i \Omega \chi^{\cpr c}_{N N}(-Q) + I^{\cpr c}_N(-Q) \nnnl
    &= i \Omega \chi^{c\cpr}_{N N}(Q) + I^{\cpr c}_N(-Q) \,.
\end{align}
Here, we used the symmetry relation \Eq{eq:chi_symmetry} in the first and last equalities.
Substituting \Eq{eq:ward_1band_chi_4} into \Eq{eq:ward_1band_chi_3} yields
\begin{equation} \label{eq:ward_1band_chi_5}
    \chi^{c\cpr}_{\Delta \Delta}(Q)
    = \Omega^2\chi^{c\cpr}_{N N}(Q)
    -i \Omega I^{\cpr c}_N(-Q)
    -i I^{c\cpr}_{\Delta}(Q)
    \,.
\end{equation}

Next, we evaluate $I^{c\cpr}_X(Q)$ for $X = N$ and $X = \Delta = \Delta^{\rm e} + \Delta^{\rm p}$.
For $X = N$, the integral vanishes:
\begin{align}
    I^{c\cpr}_N(Q)
    &= Z^{\cpr} \intkk (N_\bk - N_\bkQ) P^{cc\cpr}(k, Q)
    \nnnl
    &= Z^{c\cpr} \intkk (\rho_\bkQ - \rho_\bk)
    \nnnl
    &= 0 \,.
\end{align}
Substituting this into \Eq{eq:ward_1band_chi_5} recovers \Eq{eq:ward_1band_chi_final}.
For $X = \Delta^{\rm e}$, we have
\begin{align}
    I^{c\cpr}_{\Delta^{\rm e}}(Q)
    &= i Z^{\cpr} \intkk (\veps_\bk - \veps_\bkQ) P^{cc\cpr}(k, Q)
    \nnnl
    &= -i Z^{c\cpr} \intkk (\veps_\bkQ - \veps_\bk) (\rho_\bkQ - \rho_\bk)
    \nnnl
    &= -i Z^{c\cpr} \intkk (2\veps_\bk - \veps_\bkQ - \veps_\bkmQ) \rho_\bk
    \,,
\end{align}
where we used \Eqs{eq:sc_G_int} and \eqref{eq:ward_1band_P_def} in the first equality, and a change of variables $\bk \to \bkmQ$ for the $(\veps_\bkQ - \veps_\bk) \rho_\bkQ$ term in the second equality.
In the long-wavelength limit, this reduces to the electronic diamagnetic current [\Eq{eq:current_JDe}]:
\begin{align} \label{eq:ward_I_e}
    I^{c\cpr}_{\Delta^{\rm e}}(Q)
    &= i Z^{c\cpr} \intkk \sum_{\alpha\beta} Q^\alpha Q^\beta \frac{\partial^2 \veps_\bk}{\partial Q^\alpha \partial Q^\beta} \rho_\bk
    + O(\bQ^3)
    \nnnl
    &= i Z^{c\cpr} \bQ \cdot \expval{\opmbJ^{\rm D,\,e}} \cdot \bQ
    + O(\bQ^3)
    \,.
\end{align}

\begin{widetext}
For $X = \Delta^{\rm p}$, we find
\begin{align}
    I^{c\cpr}_{\Delta^{\rm p}}(Q)
    &= -i Z^{\cpr} \sum_{c_1 c_2} Z^{c_1} Z^{c_2} \nsint{kq}
    D_{0,\nu}^{c_1 c_2}(q) P^{c_1 c_2\cpr}(k, Q) \Bigl[
        g^*_\nu(\bk, \bq) (\Delta_{-\bQ} g_\nu)(\bkQ, \bq) G_{21}(k+q) Z^{c_1 c}
    \\
    &\hspace{20em}
        + (\Delta_{-\bQ} g_\nu)^*(\bkQ, \bq) g_\nu(\bkQ, \bq) G_{21}(k+q+Q) Z^{c_2 c}
    \Bigr]
    \nnnl
    &= -i Z^{\cpr} \sum_{c_1 c_2} Z^{c_1} Z^{c_2} \nsint{kq}
    D_{0,\nu}^{c_1 c_2}(q)
    \Bigl[ G_{12}(k+Q) \delta_{c_2 \cpr} - G_{12}(k) \delta_{c_1 \cpr} \Bigr]
    \nnnl
    &\quad \times \Bigl[
        g^*_\nu(\bk, \bq) (\Delta_{-\bQ} g_\nu)(\bkQ, \bq) G_{21}(k+q) Z^{c_1 c}
        + (\Delta_{-\bQ} g_\nu)^*(\bkQ, \bq) g_\nu(\bkQ, \bq) G_{21}(k+q+Q) Z^{c_2 c}
    \Bigr]
    \nonumber
    \,.
\end{align}
We split this into two parts,
\begin{equation}
    I^{c\cpr}_{\Delta^{\rm p}}(Q)
    = I^{c\cpr}_{\Delta^{{{\rm p}\tbar1}}}(Q)
    + I^{c\cpr}_{\Delta^{{{\rm p}\tbar2}}}(Q) \,.
\end{equation}
The first part is
\begin{align} \label{eq:ward_1band_I_p1}
    I^{c\cpr}_{\Delta^{{\rm p}\tbar1}}(Q)
    &= -i Z^{\cpr} \sum_{c_1 c_2} Z^{c_1} Z^{c_2} \nsint{kq}
    D_{0,\nu}^{c_1 c_2}(q) \Bigl[
        G_{12}(k+Q) \delta_{c_2 \cpr} (\Delta_{-\bQ} g_\nu)^*(\bkQ, \bq) g_\nu(\bkQ, \bq) G_{21}(k+q+Q) Z^{c_2 c}
    \nnnl
    &\hspace{15em}
        - G_{12}(k) \delta_{c_1 \cpr} g^*_\nu(\bk, \bq) (\Delta_{-\bQ} g_\nu)(\bkQ, \bq) G_{21}(k+q) Z^{c_1 c}
    \Bigr]
    \nnnl
    &= -i Z^{\cpr} \sum_{c_1 c_2} Z^{c_1} Z^{c_2} \nsint{kq}
    D_{0,\nu}^{c_1 c_2}(q) G_{21}(k+q) G_{12}(k)
    \Bigl[
         \delta_{c_2 \cpr} (\Delta_{-\bQ} g_\nu)^*(\bk, \bq) g_\nu(\bk, \bq) Z^{c_2 c}
    \nnnl
    &\hspace{15em}
        - \delta_{c_1 \cpr} g^*_\nu(\bk, \bq) (\Delta_{-\bQ} g_\nu)(\bkQ, \bq) Z^{c_1 c}
    \Bigr]
    \nnnl
    &= -i Z^{\cpr} \sum_{c_1 c_2} Z^{c_1} Z^{c_2} \delta_{c_2 \cpr} Z^{c_2 c} \nsint{kq}
    D_{0,\nu}^{c_1 c_2}(q) G_{21}(k+q) G_{12}(k)
    \bigl[ (\Delta_{-\bQ} g_\nu)(\bk, \bq) + (\Delta_{\bQ} g_\nu)(\bk, \bq) \bigr]^* g_\nu(\bk, \bq)
    \nnnl
    &= -i Z^{c\cpr} \sum_{c_1} Z^{c_1} \nsint{kq}
    D_{0,\nu}^{c_1 c}(q) G^{c c_1}(k+q) G^{c_1 c}(k)
    \bigl[ g_\nu(\bkQ, \bq) + g_\nu(\bkmQ, \bq) - 2g_\nu(\bk, \bq) \bigr]^* g_\nu(\bk, \bq)
    \,.
\end{align}
In the second equality, we changed the integration variable from  $k$ to $k-Q$ in the first term.
In the third equality, we replaced $1 \leftrightarrow 2$, $q \to -q$, and $k \to k+q$ in the second term, and used
\begin{equation} \label{eq:ward_g_sym_1}
    g^*_\nu(\bkq, -\bq) = g_\nu(\bk, \bq) \,,
\end{equation}
and
\begin{equation} \label{eq:ward_g_sym_2}
    (\Delta_{-\bQ} g_\nu)(\bkqQ, -\bq)
    = g_\nu(\bkq, -\bq) - g_\nu(\bkqQ, -\bq)
    = g^*_\nu(\bk, \bq) - g^*_\nu(\bkQ, \bq)
    = -(\Delta_{\bQ} g_\nu)^*(\bk, \bq) \,.
\end{equation}
The last line of \Eq{eq:ward_1band_I_p1} corresponds to the Feynman diagram in \Fig{fig:feynman_ward_derivation}(a).
In the long-wavelength limit, this term becomes the diamagnetic phonon-assisted current  [\Eq{eq:current_JDp}]:
\begin{align} \label{eq:ward_I_p1}
    I^{c\cpr}_{\Delta^{{\rm p}\tbar1}}(Q)
    &= -i Z^{c\cpr} \sum_{c_1} Z^{c_1} \nsint{kq}
    D_{0,\nu}^{c_1 c}(q) G^{c c_1}(k+q) G^{c_1 c}(k)
    (\mb{\mc{D}}^2 g_\nu)^*(\bk, \bq) g_\nu(\bk, \bq)
    + O(\bQ^3)
    \nnnl
    &= i Z^{c\cpr} \, \bQ \cdot \expval{\opmbJ^{\rm D,\,p}} \cdot \bQ
    + O(\bQ^3)
    \,.
\end{align}

The second part is
\begin{align} \label{eq:ward_1band_I_p2}
    I^{c\cpr}_{\Delta^{{\rm p}\tbar2}}(Q)
    &= -i Z^{\cpr} \sum_{c_1 c_2} Z^{c_1} Z^{c_2} \nsint{kq}
    D_{0,\nu}^{c_1 c_2}(q) \Bigl[
        G_{12}(k+Q) \delta_{c_2 \cpr} g^*_\nu(\bk, \bq) (\Delta_{-\bQ} g_\nu)(\bkQ, \bq) G_{21}(k+q) Z^{c_1 c}
    \nnnl
    &\hspace{15em}
        - G_{12}(k) \delta_{c_1 \cpr} (\Delta_{-\bQ} g_\nu)^*(\bkQ, \bq) g_\nu(\bkQ, \bq) G_{21}(k+q+Q) Z^{c_2 c}
    \Bigr]
    \nnnl
    &= -i \nsint{kq}
    \Bigl[
        D_{0,\nu}^{c \cpr}(q) G^{c \cpr}(k+Q)
        g^*_\nu(\bk, \bq) (\Delta_{-\bQ} g_\nu)(\bkQ, \bq) G^{\cpr c}(k+q)
    \nnnl
    &\hspace{5em}
        - D_{0,\nu}^{\cpr c}(q) G^{\cpr c}(k)
        (\Delta_{-\bQ} g_\nu)^*(\bkQ, \bq) g_\nu(\bkQ, \bq) G^{c \cpr}(k+q+Q)
    \Bigr]
    \nnnl
    &= -i \nsint{kq} D_{0,\nu}^{\cpr c}(q) G^{c \cpr}(k+q+Q) G^{\cpr c}(k)
    \Bigl[
        - (\Delta_{\bQ} g_\nu)^*(\bk, \bq) g_\nu(\bk, \bq)
        - (\Delta_{-\bQ} g_\nu)^*(\bkQ, \bq) g_\nu(\bkQ, \bq)
    \Bigr]
    \nnnl
    &= -i \nsint{kq} D_{0,\nu}^{\cpr c}(q) G^{c \cpr}(k+q+Q) G^{\cpr c}(k)
    (\Delta_{\bQ} g_\nu)^*(\bk, \bq) (\Delta_{\bQ} g_\nu)(\bk, \bq)
    \,.
\end{align}
\end{widetext}
In the third equality, we replaced $q \to -q$ and $k \to k+q$ in the first term and used \Eqs{eq:ward_g_sym_1} and \eqref{eq:ward_g_sym_2}.
\Equ{eq:ward_1band_I_p2} corresponds to the Feynman diagram shown in \Fig{fig:feynman_ward_derivation}(b).
In the long-wavelength limit, it becomes the phonon-assisted bubble contribution to the current-current susceptibility [\Eq{eq:sigma_Lambda_pp_Bubble}]:
\begin{equation} \label{eq:ward_I_p2}
    I^{c\cpr}_{\Delta^{{\rm p}\tbar2}}(Q)
    = -i \sum_{\alpha\beta} Q^\alpha Q^\beta \Lambda_{\alpha\beta}^{{\rm (pp\tbar Bubble)} c\cpr}(\Omega)
    + O(\bQ^3) \,.
\end{equation}
Adding the three contributions from \Eqs{eq:ward_I_e}, \eqref{eq:ward_I_p1}, and \eqref{eq:ward_I_p2}, we recover \Eq{eq:ward_I_term}.
This completes the proof of the continuity equation, as taking the second derivative with respect to $\bQ$ of \Eq{eq:ward_1band_chi_final} yields \Eqs{eq:ward_1band_result1}--\eqref{eq:ward_1band_result3}.

We emphasize that it is crucial to use the \scGD self-energy and Green's functions to ensure that the Ward--Takahashi identity and the continuity equation are satisfied.
We used this property to derive the second line of \Eq{eq:ward_1band_der1} from \Eq{eq:ward_1band_der2}.
In addition, unless the \eph coupling is $\bk$-independent, including the phonon-assisted contribution is essential for the Ward--Takahashi identity to hold.

\FloatBarrier 

\clearpage
\newpage
\def\myroot{main}
\ifdefined\myroot\else  
\documentclass[
    reprint, twocolumn,
    aps,prx,10pt,amsmath,amssymb,
    longbibliography,superscriptaddress,
    showpacs,preprintnumbers,
]{revtex4-2}

\usepackage{jmlstyle}

\myexternaldocument{main}{}

\newcommand{\RA}{{\mathrm{R\!/\! A}}}
\newcommand{\nsp}[0]{\mspace{-2mu}}  
\newcommand{\Cpr}[0]{C'\nsp}  
\newcommand{\cpr}[0]{c'\nsp}

\newcommand{\eph}[0]{e-ph\xspace}
\newcommand{\EPjl}[0]{\textsc{ElectronPhonon.jl}\xspace}
\newcommand{\scGD}[0]{sc$GD_0$\xspace}
\newcommand{\GD}[0]{$G_0 D_0$\xspace}
\newcommand{\opmbJ}[0]{\hat{\mathbf{J}}}

\begin{document}
\fi

\setcounter{equation}{0}
\setcounter{figure}{0}
\setcounter{table}{0}
\setcounter{page}{1}
\setcounter{section}{0}

\renewcommand{\theequation}{S\arabic{equation}}
\renewcommand{\thefigure}{S\arabic{figure}}
\renewcommand{\thetable}{S\arabic{table}}
\renewcommand{\thepage}{S\arabic{page}}
\renewcommand{\thesection}{S-\Roman{section}}

\makeatletter
\renewcommand{\theHsection}{supp.\Roman{section}}
\renewcommand{\theHequation}{supp.\arabic{equation}}
\renewcommand{\theHfigure}{supp.\arabic{figure}}
\renewcommand{\theHtable}{supp.\arabic{table}}
\makeatother

\title{Supplemental Material for\\
``Beyond-quasiparticle transport with vertex correction:\\ self-consistent ladder formalism for electron-phonon interactions''}

\ifdefined\myroot\else  
\author{Jae-Mo Lihm}
\email{jaemo.lihm@gmail.com}
\affiliation{%
European Theoretical Spectroscopy Facility and Institute of Condensed Matter and Nanosciences, Université catholique de Louvain, Chemin des Étoiles 8, B-1348 Louvain-la-Neuve, Belgium
}%
\author{Samuel Ponc\'e}
\email{samuel.ponce@uclouvain.be}
\affiliation{%
European Theoretical Spectroscopy Facility and Institute of Condensed Matter and Nanosciences, Université catholique de Louvain, Chemin des Étoiles 8, B-1348 Louvain-la-Neuve, Belgium
}%
\affiliation{%
WEL Research Institute, avenue Pasteur 6, 1300 Wavre, Belgium
}%
\fi

\date{\today}

\maketitle

\section{Additional results} \label{sec:additional_results}

\begin{figure}[h]
\centering
\includegraphics[width=0.99\linewidth]{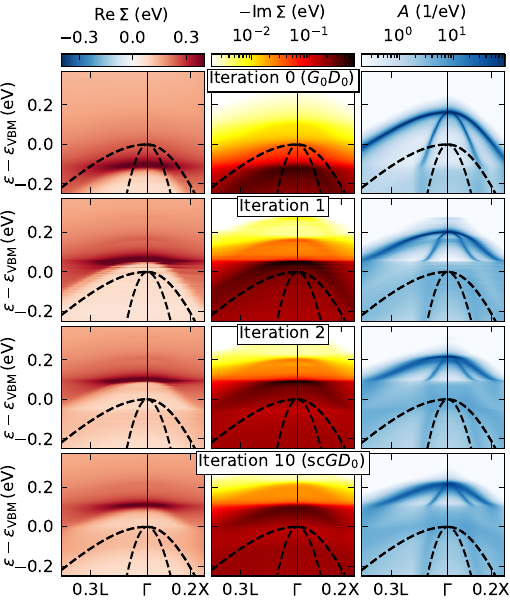}
\caption{
Self-energy and spectral function of AlN at 300~K at different steps of the \scGD iteration.
}
\label{fig:spectral_AlN_conv}
\end{figure}

\Figu{fig:spectral_AlN_conv} shows the convergence of the \scGD self-energy and spectral function for AlN at $300~\mathrm{K}$.
Starting from the \GD spectral function, the spectral function converges in 10 iterations to the \scGD solution.

\begin{figure}[tb]
\centering
\includegraphics[width=0.99\linewidth]{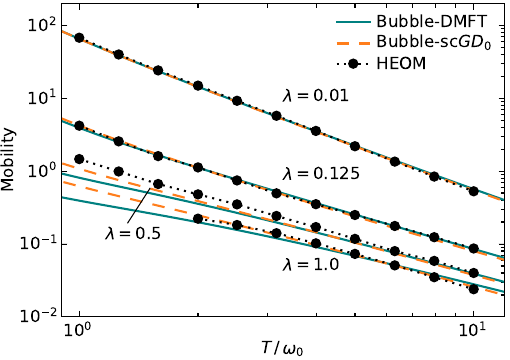}
\caption{
    Mobility of the 1D Holstein model with $t = \omega_0 = 1$ and  $\lambda = 0.01,\, 0.125,\, 0.5,\, 1.0$, computed using the bubble approximation with the DMFT~\cite{Jankovic2024Holstein} or the \scGD spectral functions, compared against the HEOM result~\cite{Jankovic2023}.
}
\label{fig:Holstein_DMFT}
\end{figure}

\Figu{fig:Holstein_DMFT} shows the mobility of the 1D Holstein model computed using the bubble approximation with the dynamical mean-field theory (DMFT) and \scGD spectral functions, compared with the HEOM result.
See Ref.~\cite{Mitric2022} for the details of DMFT for the Holstein model.

\begin{figure}[tb]
\centering
\includegraphics[width=0.99\linewidth]{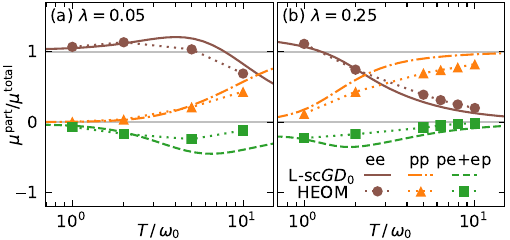}
\caption{
    Relative contribution of the electronic [ee, \Eq{eq:Lambda_def_ee}], phonon-assisted [pp, \Eq{eq:Lambda_def_pp}], and cross [pe+ep, \Eqs{eq:Lambda_def_pe} and \eqref{eq:Lambda_def_ep}] terms to the ladder-\scGD (L-\scGD) mobility of the Peierls model (parameters same as \Fig{fig:conductivity_models}(e, f)).
    The HEOM result is from Ref.~\cite{Jankovic2025a}.
}
\label{fig:conductivity_Peierls_assisted}
\end{figure}

\begin{figure*}[tb]
\centering
\includegraphics[width=0.99\linewidth]{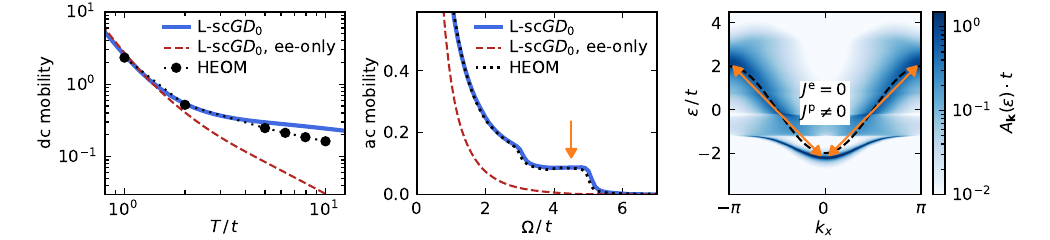}
\caption{
    Mobility and spectral functions of the 1d Peierls model.
    (a) dc mobility ($t = \omega_0 = 1$, $\lambda=0.25$) compared with the HEOM results from Ref.~\cite{Jankovic2025a}.
    (b) ac mobility ($t = \omega_0 = 1$, $\lambda=0.05$, $T=1.0$), compared with the HEOM results from Ref.~\cite{Jankovic2025a}.
    The orange arrow marks the plateau due to the phonon-assisted current.
    (c) Spectral function ($t = \omega_0 = 1$, $\lambda=0.5$, $T=0$).
    The orange arrow marks the indirect transition due to the phonon-assisted current, which leads to the plateau.
}
\label{fig:Peierls_ph_assisted}
\end{figure*}

\begin{figure}[tb]
\centering
\includegraphics[width=0.99\linewidth]{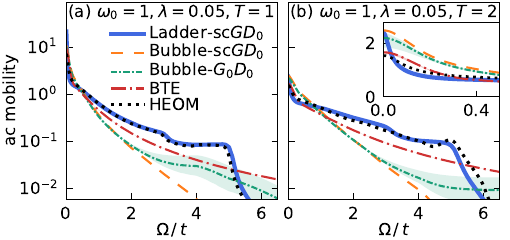}
\caption{
    Ac mobility of the Peierls model with $t = 1$ and $\omega_0 = 1.0$, compared with the HEOM results from Ref.~\cite{Jankovic2025a}.
}
\label{fig:Peierls_ac_omega1.0}
\end{figure}

\begin{figure}[tb]
\centering
\includegraphics[width=0.99\linewidth]{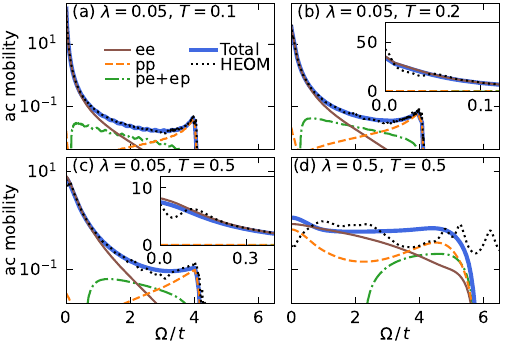}
\caption{
    Decomposition of the ladder-\scGD ac mobility of the Peierls model at $\omega_0 = 0.044$ shown in \Fig{fig:Peierls_ac} into the electronic [ee, \Eq{eq:Lambda_def_ee}], phonon-assisted [pp, \Eq{eq:Lambda_def_pp}], and cross [pe+ep, \Eqs{eq:Lambda_def_pe} and \eqref{eq:Lambda_def_ep}] contributions.
}
\label{fig:Peierls_ac_decomp}
\end{figure}

\Figu{fig:conductivity_models}(e) shows that the HEOM mobility of the Peierls model at intermediate coupling ($\lambda = 0.25$) decreases with temperature and flattens at high temperatures.
This temperature dependence is captured by the ladder-\scGD method, but not in other approximations.
To better understand this behavior, we decompose the total ladder-\scGD mobility into the electronic, phonon-assisted, and cross contributions.
\Figu{fig:conductivity_Peierls_assisted} shows that the temperature dependence in the relative contributions in the HEOM and ladder-\scGD results is qualitatively consistent.
At low temperatures, transport is dominated by the electronic contribution, while at higher temperatures and intermediate coupling, the phonon-assisted term prevails.
The cross term between the electronic and phonon-assisted currents is negative and cannot be neglected.
Neglecting these phonon-assisted and cross contributions leads to an inaccurate temperature dependence of the mobility, as shown in \Fig{fig:Peierls_ph_assisted}(a).

\begin{figure}[tb]
\centering
\includegraphics[width=0.99\linewidth]{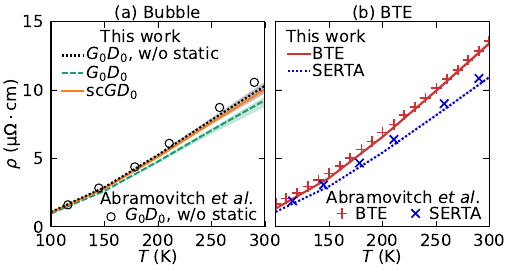}
\caption{
    Comparison of the resistivity of SrVO$_3$ calculated in this work and in Ref.~\cite{Abramovitch2024}.
    (a) Bubble resistivity. The result of Ref.~\cite{Abramovitch2024} corresponds to bubble-\GD without the static correction to the self-energy~\cite{AbramovitchPrivateCommunication}.
    (b) BTE and SERTA resistivities (without the $+U$ correction).
}
\label{fig:SrVO3_dc_comparison}
\end{figure}

In \Fig{fig:Peierls_ac} of the main text, we show the ac mobility of the Peierls model at $\omega = 0.044$.
\Figu{fig:Peierls_ac_omega1.0} shows analogous results at a larger phonon frequency $\omega = 1.0$.
The peak at $\Omega = 4t$ is broadened compared to the $\omega = 0.044$ case, but is still well captured by the ladder-\scGD method.

For the ac mobility of the Peierls model shown in \Fig{fig:Peierls_ac}, we aim to understand the origin of the peak at $\Omega \approx 4t$.
The decomposition in \Fig{fig:Peierls_ac_decomp}(a) shows that the kink arises from the $\Lambda^{\rm (pp)}$ term.
We specifically find that the kink arises from the phonon-assisted bubble.
The peak corresponds to the absorption of a $\Omega \approx 4t$ photon, with a hole created at $(\bk, \veps_\bk) \approx (0, -2t)$ and an electron at $(\bkq, \veps_\bkq) \approx (\pi, 2t)$, with the remaining momentum transferred to a phonon: see \Fig{fig:Peierls_ph_assisted}(b) and (c).
While the \eph coupling between $k_x = 0$ and $k_x + q_x = \pi$ is zero, $g(k_x=0, q_x=\pi) = 0$, its covariant derivative is maximal, $\frac{\dd g}{\dd k_x}(k_x=0, q_x=\pi) = 2i \sqrt{2\lambda \omega_0 t}$ [see \Eq{eq:peierls_def}], thus allowing for a large phonon-assisted current.

\begin{figure}[tb]
\centering
\includegraphics[width=0.99\linewidth]{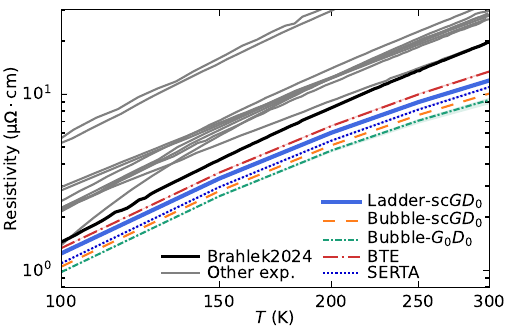}
\caption{
    Temperature dependence of the resistivity of SrVO$_3$, same as \Fig{fig:SrVO3_dc} but in logarithmic scale.
}
\label{fig:SrVO3_dc_logT}
\end{figure}

\Figu{fig:SrVO3_dc_comparison} compares the resistivity of SrVO$_3$ calculated in this work and in Ref.~\cite{Abramovitch2024} and shows good agreement between the two.
\Figu{fig:SrVO3_dc_logT} shows the temperature dependence of the resistivity of SrVO$_3$ shown in \Fig{fig:SrVO3_dc} in logarithmic scale.
We find that the curves obtained with different calculation methods are almost vertically shifted, and the slope of the resistivity is similar.

\begin{figure}[tb]
\centering
\includegraphics[width=0.99\linewidth]{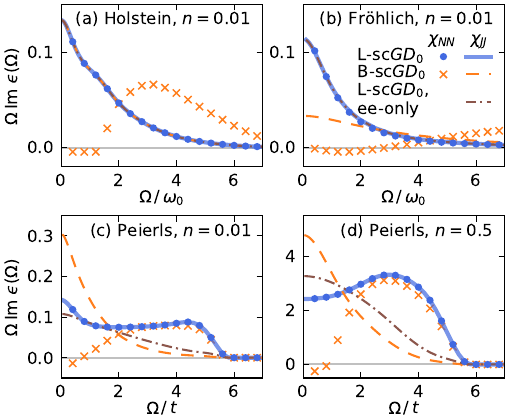}
\caption{
    Imaginary part of the dielectric functions of the (a) Holstein, (b) Fr\"ohlich, and (c, d) Peierls models.
    See \Fig{fig:dielectric_models} for the full description.
}
\label{fig:dielectric_models_imag}
\end{figure}

\Figu{fig:dielectric_models_imag} shows the imaginary part of the dielectric functions of the Holstein, Fr\"ohlich, and Peierls models, whose real part is shown in \Fig{fig:dielectric_models}.
As in the real part, we find that the ladder-\scGD yields dielectric functions consistent with optical conductivity, while the bubble-\scGD does not.
In addition, the phonon-assisted current is also necessary to satisfy the continuity equation for the Peierls model.

\section{Computational details}

\begin{table}[b]
\begin{tabular}{c|c|c}
\hline
Material &
$\veps^{\rm large}_{\rm min}$, $\veps^{\rm large}_{\rm max}$, $\veps^{\rm large}_{\rm step}$ &
$\veps^{\rm small}_{\rm min}$, $\veps^{\rm small}_{\rm max}$, $\veps^{\rm small}_{\rm step}$
\\ \hline
Si (CB)   & $-300$, $+300$, $1.0$ & $-100$, $+100$, $0.1$ \\
Si (VB)   & $-300$, $+350$, $1.0$ & $-100$, $+150$, $0.1$ \\
ZnO (CB, $T > 150~\mathrm{K}$) & $-400$, $+500$, $5.0$ & $-200$, $+300$, $1.0$ \\
ZnO (CB, $T \leq 150~\mathrm{K}$)  & $-400$, $+350$, $5.0$ & $-200$, $+150$, $0.5$ \\
SrVO$_3$  & $-350$, $+350$, $20.0$ & $-100$, $+100$, $1.0$ \\
AlN (VB)  & $-850$, $+400$, $20.0$ & $-200$, $+250$, $1.0$ \\
NaCl (CB) & $-400$, $+600$, $5.0$  & $-100$, $+300$, $1.0$ \\
\hline
\end{tabular}
\caption{
    Frequency mesh parameters used in the \scGD and ladder calculations in meV units.
    The energies are with respect to the CBM energy, VBM energy, or Fermi level for the calculations of the CB, VB, and metals, respectively.
}
\label{tab:params_freq}
\end{table}

\begin{table*}
\begin{tabular}{c|cccc|ccccc|cc}
\hline
Material & $a$ (bohr) & $E_{\rm cut}$ (Ry) & $N_\bk^{\rm DFT}$ & SOC & $N_{\bk, \bq}^{\rm Wannier}$ & Initial guess & $E^{\rm froz}$ (eV) & $E^{\rm dis}$ (eV) & $N^{\rm band}$ & $N_{\bk, \bq}$ & $[\veps^{\rm min}, \veps^{\rm max}]$ (eV)
\\ \hline
Si (CB)   & 10.336  & 40  & $20^3$ & X & $16^3$, $8^3$ & Si: sp$^3$      & 3.9 & -    & 14 & $120^3$ & $[0, 0.2]$ \\
Si (VB)   & 10.336  & 60  & $16^3$ & \checkmark & $12^3$, $6^3$ & bond-centered s & -   & -    & 8  & $120^3$ & $[-0.2, 0]$ \\
ZnO (CB)  & 6.14085 & 100 & $12^3$ & \checkmark & $12^3$, $6^3$ & Zn: s           & 3.4 & 9.4  & 12 & $360^3$ & $[0, 0.15]$ \\
SrVO$_3$  & 7.26033 & 70  & $24^3$ & X & $6^3$,  $6^3$ & V: $t_{2g}$     & -   & -    & 3  & $80^3$  & $[-0.15, 0.15]$ \\
AlN (VB)  & 8.258   & 70  & $12^3$ & X & $6^3$,  $6^3$ & Al, N: sp$^3$   & 4.7 & 19.7 & 14 & $80^3$  & $[-0.65, 0]$ \\
NaCl (CB) & 10.5636 & 80  & $20^3$ & X & $8^3$,  $8^3$ & Na: s           & 1.5 & 6.5  & 5  & $240^3$ & $[0, 0.4]$ \\
\hline
\end{tabular}
\caption{
    Computational parameters used in this work: lattice parameter $a$, kinetic energy cutoff $E^{\rm cut}$,
    size of the $\bk$-grid for DFT and DFPT calculations $N_\bk^{\rm DFT}$,
    inclusion of the spin-orbit coupling (SOC),
    size of the $\bk$- and $\bq$-grids for Wannierization $N_{\bk,\bq}^{\rm Wannier}$,
    initial guesses for constructing the Wannier functions,
    upper bound of the frozen (inner) window $E^{\rm froz}$,
    upper bound of the disentanglement (outer) window $E^{\rm dis}$,
    number of bands included for Wannierization $N^{\rm band}$,
    size of the $\bk$- and $\bq$-grids for the \scGD and ladder calculations,
    and the active space window $[\veps^{\rm min}, \veps^{\rm max}]$.
    We use different settings for the conduction band (CB) and valence band (VB) of Si, due to the importance of the SOC on the valence band.
    The frozen and disentanglement windows are defined with respect to the CBM energy.
    The active space window is defined with respect to the CBM energy for CB calculations, the VBM energy for VB calculations, and the Fermi level for SrVO$_3$.
    A hyphen denotes that the corresponding window was not set.
    The $c$ axis lattice parameter for ZnO was 9.8364~bohr.
    For ZnO, we use a larger window $[0, 0.3]$~eV and a coarser $\bk$ and $\bq$ grid of $240^3$ for calculations at $T \geq 150~\mathrm{K}$.
    For SrVO$_3$, we use a 0.02~eV cold smearing~\cite{Marzari1999smearing} for the DFT and DFPT calculations.
    For the spectral function visualization of SrVO$_3$, we use a wider active space window $[-0.3, 0.3]$~eV.
}
\label{tab:comp_params}
\end{table*}

We perform DFT and DFPT calculations using the \textsc{Quantum ESPRESSO} package~\cite{Giannozzi2017}.
We use norm-conserving fully-relativistic pseudopotentials~\cite{Hamann2013ONCVPSP} in the Perdew-Burke-Ernzerhof (PBE) exchange-correlation functional~\cite{Perdew1996} from \textsc{PseudoDojo} (v0.4)~\cite{Setten2018}.
We then use \textsc{Wannier90}~\cite{Pizzi2020W90} and EPW~\cite{Giustino2007EPW, Ponce2016EPW, Lee2023EPW} to construct localized Wannier functions and real-space matrix elements.
Finally, we perform Wannier interpolation and solve the \scGD and ladder equations using an in-house developed code \EPjl~\cite{EPjl} written in the Julia programming language~\cite{Bezanson2017Julia}.
We included the dipolar contribution to the long-range intertomic force constants and $e$-ph coupling, but did not include the quadrupolar~\cite{Brunin2020PRL, Brunin2020PRB, Jhalani2020, Park2020} or Berry connection~\cite{Ponce2023PRL, Ponce2023PRB} terms.

For the \scGD and ladder calculations, we use frequency meshes with nonuniform resolution.
A wide energy range $[\veps^{\rm large}_{\rm min}, \veps^{\rm large}_{\rm max}]$ is sampled with a coarse step $\veps^{\rm large}_{\rm step}$, and a smaller range $[\veps^{\rm small}_{\rm min}, \veps^{\rm small}_{\rm max}]$ is sub-sampled with a finer step $\veps^{\rm small}_{\rm step}$.
The frequency mesh parameters are shown in Table~\ref{tab:params_freq}.
We summarize other computational parameters in Table~\ref{tab:comp_params}.

\begin{table}[t]
    \centering
    
    \begin{tabular}{c||r|r|r|r|r}
    \hline
      \multirow{3}{*}{Materials} & \multicolumn{5}{c}{CPU hour}  \\ \cline{2-6} 
    & \multicolumn{1}{c|}{\multirow{2}{*}{DFPT}}
    & \multicolumn{1}{c|}{\multirow{2}{*}{BTE}}
    & \multicolumn{3}{c}{Ladder-\scGD} \\ \cline{4-6} 
    & \multicolumn{1}{c|}{}
    & \multicolumn{1}{c|}{}
    & \multicolumn{1}{c|}{Matrix elements}
    & \multicolumn{1}{c|}{\scGD}
    & \multicolumn{1}{c}{Ladder} \\ \hline\hline
    Si (CB)  &   49 &   3 &  10 &  160 &  150 \\
    Si (VB)  &  170 &   4 &  11 &  590 &  640 \\
    ZnO (CB) & 9400 &  34 &  52 & 2800 & 2000 \\
    SrVO$_3$ & 4200 & 100 & 180 &  800 &  980 \\ \hline
    \end{tabular}
    \caption{Summary of the computational cost for BTE and ladder-\scGD and transport calculations. 
    Since the computational costs of the BTE, \scGD, and ladder steps scale with the number of temperature values, we normalized them assuming a calculation for five temperatures.
    We note that the matrix element calculations for ladder-\scGD involve writing data to disk and hence take longer than the corresponding BTE calculation.}
\label{tab:computational_cost}
\end{table}

\begin{table}[t]
\begin{tabular}{c|c|c|c}
\hline
Material & \#$\nk$ (irr.) & \#$\nk$ (full) & \#$\bq$ \\ \hline
Si (CB)  &  589 &  20850 & 97844 \\
Si (VB)  & 1082 &  35478 & 68331 \\
ZnO (CB, $T > 150~\mathrm{K}$)    & 2422 & 45866 & 69164 \\
ZnO (CB, $T \leq 150~\mathrm{K}$) & 2577 & 49010 & 74941 \\
SrVO$_3$ & 3010 & 126916 & 398796 \\
\hline
\end{tabular}
\caption{
    The number of electronic states $\nk$ inside the active space in the irreducible and full Brillouin zones, and the number of phonon wavevectors $\bq$ connecting two electronic wavevectors.
}
\label{tab:params_numbers}
\end{table}

Table~\ref{tab:computational_cost} summarizes the computational cost of the BTE and ladder-\scGD calculations.
The numbers of selected electronic states inside the active-space window, which determines the computational cost, are shown in Table~\ref{tab:params_numbers}.
We utilized multithreading to parallelize the ladder-\scGD calculation over CPUs in a single node.
The number of iterations depends on the temperature, but both \scGD and ladder equations converged within 15 iterations for all systems at the considered temperature range.

\section{Derivation details} \label{sec:derivation_details}

\subsection{Calculation of the diamagnetic current} \label{sec:supp_diamagnetic}
Here, we derive the relation between the diamagnetic current and the (paramagnetic) current-current correlator in terms of the Kramers--Kronig relation.
Since the diamagnetic current is symmetric, we only consider the diagonal component.
The conductivity satisfies the Kramers--Kronig relation
\begin{equation}
    \Im \sigma_{\alpha\alpha}(\Omega)
    = -\frac{1}{\pi} \mc{P} \nint \dd\Omega' \frac{\Re \sigma_{\alpha\alpha}(\Omega')}{\Omega' - \Omega} \,.
\end{equation}
By substituting \Eq{eq:sigma_chiR} into this equation, we find
\begin{align}
    &\frac{1}{\Omega} \bigl[
        \Re \Lambda^{\rm R}_{\alpha\alpha}(\Omega) - \expval{\hat{J}^{\rm D}_{\alpha\alpha}}
    \bigr]
    \nnnl
    &= \frac{1}{\pi} \mc{P} \nint \dd\Omega' \, \frac{\Im  \Lambda^{\rm R}_{\alpha\alpha}(\Omega')}{\Omega' (\Omega' - \Omega)}
    \nnnl
    &= \frac{1}{\pi \Omega} \, \mc{P} \nint \dd\Omega' \, \biggl[
        \frac{\Im  \Lambda^{\rm R}_{\alpha\alpha}(\Omega')}{\Omega' - \Omega}
        - \frac{\Im  \Lambda^{\rm R}_{\alpha\alpha}(\Omega')}{\Omega'}
    \biggr] \,.
\end{align}
The first term on the last line is identical to the first term on the first line due to the Kramers--Kronig relation of $\Lambda^{\rm R}$.
Thus, we find
\begin{equation}
    \expval{\hat{J}^{\rm D}_{\alpha\alpha}}
    = \frac{1}{\pi} \, \mc{P} \nint \dd\Omega' \, \frac{\Im  \Lambda^{\rm R}_{\alpha\alpha}(\Omega')}{\Omega'} \,.
\end{equation}

\subsection{Electronic and phonon-assisted bubble conductivity} \label{sec:sigma_bubble_derivation}

Here, we derive \Eqs{eq:sigma_bubble} and \eqref{eq:sigma_pp_bubble}, which are the expressions for the electronic bubble and phonon-assisted bubble contributions to the longitudinal conductivity in the diagonal Green's function approximation.

First, for the electronic bubble term, we start from \Eq{eq:sigma_Lambda_ee_bubble} and apply the diagonal Green's function approximation [\Eq{eq:sc_G_diag_approx}] to find
\begin{multline}
    \Lambda^{{\rm (Bubble)}\gtrless}_{\alpha\beta}(\Omega)
    = -2\pi i \sum_{n_1 n_2} \intbk v^{\alpha*}_{n_1 n_2 \bk} v^\beta_{n_1 n_2 \bk}
    \\
    \times \nint \dd \veps \, f^\mp(\veps + \Omega) f^\pm(\veps) A_{n_1 \bk}(\veps + \Omega) A_{n_2 \bk}(\veps) \,.
\end{multline}
Substituting this result into \Eq{eq:sigma_chi} and applying
\begin{equation}
    \frac{f^-(\veps) f^+(\veps \!+\! \Omega) + f^+(\veps) f^-(\veps \!+\! \Omega)}{1 + 2n(\Omega)}
    = f^+(\veps) - f^+(\veps \!+\! \Omega)
\end{equation}
yields \Eq{eq:sigma_bubble}.

Next, we consider the phonon-assisted bubble correlator \Eq{eq:sigma_Lambda_pp_Bubble}.
Applying the diagonal approximation [\Eq{eq:sc_G_diag_approx}] to the greater and lesser components, we find
\begin{align} \label{eq:sigma_pp_bubble_der1}
    &\Lambda_{\alpha\beta}^{{\rm (pp\tbar Bubble)} \gtrless}(\Omega)
    \nnnl
    &= -2\pi i \! \sum_{m n \nu} \nsint{\bk \bq}
    (\mc{D}_\alpha g)^*_{mn\nu}(\bk, \bq) (\mc{D}_\beta g)_{mn\nu}(\bk, \bq)
    \nnnl
    &\ \times \nint \dd \veps \, f^\pm(\veps) A_\nk(\veps) \Bigl[
    \nnnl
    &\quad
        (n_\nuq \!+\! 1) \, f^\mp(\veps \!\mp\! \omega_\nuq \!+\! \Omega) \,
        A_\mkq(\veps \!\mp\! \omega_\nuq \!+\! \Omega)
    \nnnl
    &\quad +
        n_\nuq \, f^\mp(\veps \!\pm\! \omega_\nuq \!+\! \Omega) \,
        A_\mkq(\veps \!\pm\! \omega_\nuq \!+\! \Omega)
    \Bigr]
    \,.
\end{align}
For the greater correlator (upper signs), the first term in the square bracket corresponds to the excitation of an electron at $\nk$ to $\mkq$ with its energy changing from $\veps$ to $\veps - \omega_\nuq + \Omega$, while simultaneously creating a phonon at $\nuq$.
The second term corresponds to a similar process, while absorbing a phonon at $\nuq$ so that the final electron energy is $\veps + \omega_\nuq + \Omega$.
Consequently, the two terms have phonon occupation coefficients $n_\nuq + 1$ and $n_\nuq$, respectively.
The lesser correlator (lower signs) can be interpreted similarly.

Using
\begin{multline}
    \frac{
        (n_\nuq + \tfrac{1}{2} \pm \tfrac{1}{2}) f^\mp(\veps - \omega_\nuq + \Omega) f^\pm(\veps)
    }{
        n(\Omega) + \tfrac{1}{2} \pm \tfrac{1}{2}
    }
    \\
    = \bigl[ n_\nuq + f^+(\veps) \bigr] \bigl[ f^+(\veps \!-\! \omega_\nuq) - f^+(\veps \!-\! \omega_\nuq \!+\! \Omega) \bigr]
\end{multline}
and
\begin{multline}
    \frac{
        (n_\nuq + \tfrac{1}{2} \mp \tfrac{1}{2}) f^\mp(\veps + \omega_\nuq + \Omega) f^\pm(\veps)
    }{
        n(\Omega) + \tfrac{1}{2} \pm \tfrac{1}{2}
    }
    \\
    = \bigl[ n_\nuq + f^-(\veps) \bigr] \bigl[ f^+(\veps \!+\! \omega_\nuq) - f^+(\veps \!+\! \omega_\nuq \!+\! \Omega) \bigr] \,,
\end{multline}
we obtain
\begin{multline} \label{eq:sigma_pp_bubble_der2}
     \Lambda_{\alpha\beta}^{{\rm (pp\tbar Bubble)} >}(\Omega)
    + \Lambda_{\alpha\beta}^{{\rm (pp\tbar Bubble)} <}(\Omega) \\
    = -2\pi i \bigl[ 1 \!+\! 2n(\Omega) \bigr] \! \sum_{m n \nu} \nsint{\bk \bq}
    (\mc{D}_\alpha g)^*_{mn\nu}(\bk, \bq) (\mc{D}_\beta g)_{mn\nu}(\bk, \bq) \\
     \times \sum_\pm \nint \dd \veps \, \Bigl\{
    \bigl[ n_\nuq \!+\! f^\pm(\veps) \bigr]
    \bigl[ f^+(\veps \!\mp\! \omega_\nuq) \!-\! f^+(\veps \!\mp\! \omega_\nuq \!+\! \Omega) \bigr] \\
     \times  A_\mkq(\veps \mp \omega_\nuq + \Omega) A_\nk(\veps)
    \Bigr\} \,.
\end{multline}
By substituting this into \Eq{eq:sigma_chi}, we obtain \Eq{eq:sigma_pp_bubble}.

\subsection{Ward identity and continuity equation for multiband systems} \label{sec:ward_identity_multiband}

In this subsection, we discuss the Ward--Takahashi identity and the continuity equation for multiband systems.
Contrary to the single-band case considered in \App{app:Ward}, care must be taken in defining the long-wavelength limit $\bQ \to \mb{0}$ as the Green's functions and response functions are gauge dependent.
Therefore, we fix the gauge of $\bkQ$-dependent quantities by mapping them to the eigenstates at $\bk$.
The gauge is fixed by multiplying the overlap matrix $N_\mnk(\bQ) = \braket{u_\mkQ | u_\nk}$.
In the long-wavelength limit, this matrix element expands as
\begin{equation} \label{eq:N_expansion}
    N_{\mnk}(\bQ) = \delta_{mn} + i \bQ \cdot \mb{\xi}_\mnk + O(Q^2) \,,
\end{equation}
where the $O(\bQ)$ term involves the Berry connection $\mb{\xi}_\bk$ [\Eq{eq:current_xi_def}].

For a $\bk$-dependent matrix $A_{mn\bk}$, such as the Green's function, self-energy, velocity, and energy, we define the gauge-fixed quantity as
\begin{equation}
    \tilde{A}_{mn \bkQ}
    \equiv \sum_{m'n'} N^\dagger_{m m' \bk}(\bQ) A_{m' n' \bkQ} N_{n'n\bk}(\bQ) \,,
\end{equation}
where we denote gauge-fixed quantities with a tilde.
The $O(\bQ)$ coefficient of gauge-fixed quantities is their covariant derivative:
\begin{align}
    \tilde{A}&_{ mn \bkQ} - A_\mnk
    \nnnl
    &= \bQ \cdot \biggl[ \frac{\partial A_{mn \bkQ}}{\partial \bQ} \Big|_{\bQ=0}
    \nnnl
    &\quad
    -i \sum_{m'} \bigl( \mb{\xi}_{m m' \bk} A_{m' n \bk} - A_{m m' \bk} \mb{\xi}_{m'n\bk} \bigr) \biggr]
    + O(\bQ^2)
    \nnnl
    &= \bQ \cdot (\mb{\mc{D}} A)_\mnk  + O(\bQ^2) \,.
\end{align}
We used \Eq{eq:N_expansion} to expand $N(\bQ)$ in terms of the Berry connection $\mb{\xi}$.
The covariant derivative $\mb{\mc{D}}$ is defined in \Eq{eq:gauge_cov_der_def}.

The gauge-fixed \eph coupling is defined as
\begin{multline}
    \tilde{g}_{mn\nu}(\bkQ, \bq)   \\
    \equiv \sum_{m' n'} N^\dagger_{m m' \bkq}(\bQ) g_{m'n'\nu}(\bkQ, \bq) N_{n'n\bk}(\bQ) \,.
\end{multline}
The finite-difference of \eph coupling, defined in \Eq{eq:ward_1band_Delta_def}, is gauge-fixed as
\begin{align}
    (\Delta_\bQ \tilde{g}_{mn\nu})(\bk, \bq)
    &= \tilde{g}_{mn\nu}(\bkQ, \bq) - g_{mn\nu}(\bk, \bq)
    \nnnl
    &= \bQ \cdot (\mb{\mc{D}} g)_{mn\nu}(\bk, \bq) + O(\bQ^2) \,.
\end{align}

For response and vertex functions, which have the left index with momentum $\bkQ$ and the right index with momentum $\bk$ [see \Eq{eq:lr_R_def}], we only need the gauge fixing for the left index.
The gauge-fixed vertex function is defined as
\begin{equation}
    \delta_Y^c \tilde{\Sigma}_{12}(k, Q)
    \equiv \sum_{n_1'} N^\dagger_{n_1 n_1' \bk}(\bQ) \, \delta_Y^c \Sigma^{c_1 c_2}_{n_1' n_2}(k, Q) \,,
\end{equation}
and the gauge-fixed response function is defined analogously.
For example, the bare density vertex function [\Eq{eq:ward_1band_Sigma_N_def}] in the multiband reads
\begin{align}
    \delta^{c}_{N} \Sigma_{0, 12}(k, Q)
    &\equiv N_{n_1 n_2 \bk}(\bQ) \delta_{c_1 c} \delta_{c_2 c} \,.
\end{align}
After fixing the gauge, we find
\begin{align}
    \delta^{c}_{N} \tilde{\Sigma}_{0, 12}(k, Q)
    &\equiv (N_\bk^\dagger N_\bk)_{n_1 n_2}(\bQ) \delta_{c_1 c} \delta_{c_2 c}
    \nnnl
    &= \delta_{n_1 n_2} \delta_{c_1 c} \delta_{c_2 c} \,.
\end{align}
Special care is required for the bare electronic current vertex function to properly define the gauge-fixed energy difference term.
We generalize its definition \Eq{eq:ward_1band_Sigma_Delta_e_def} to the multi-band case as
\begin{multline}
    \delta^{c}_{\Delta^{\rm e}} \Sigma_{0, 12}(k, Q)
    = i \bigl[ \veps_{n_1 \bkQ} N_{n_1 n_2 \bk}(\bQ)
    \\
    - N^\dagger_{n_1 n_2 \bk}(\bQ) \veps_{n_2 \bk} \bigr]
    \delta_{c_1 c} \delta_{c_2 c} \,.
\end{multline}
After gauge fixing, we find
\begin{equation}
    \delta^{c}_{\Delta^{\rm e}} \tilde{\Sigma}_{0, 12}(k, Q)
    = i (\tilde{\veps}_{n_1 n_2 \bkQ} - \veps_{n_1 n_2 \bk}) \delta_{c_1 c} \delta_{c_2 c} \,,
\end{equation}
which yields the correct long-wavelength limit
\begin{equation}
    \tilde{\veps}_{n_1 n_2 \bkQ} - \veps_{n_1 n_2 \bk}
    = \bQ \cdot \mb{v}_{n_1 n_2 \bk} + O(\bQ^2) \,.
\end{equation}

The derivation of the Ward--Takahashi identity and the continuity equation in \App{app:Ward} of the main text can be generalized to the multiband case by replacing the $\bkQ$-dependent quantities with their gauge-fixed counterparts.
For example, the Ward--Takahashi identity [\Eq{eq:ward_1band_continuity}] becomes
\begin{equation}
    Z^{c} \tilde{P}^c_{12}(k ,Q)
    = i \delta^{c}_{\Delta} \tilde{G}_{12}(k, Q)
    - \Omega \, \delta^{c}_N \tilde{G}_{12}(k, Q) \,,
\end{equation}
where
\begin{equation}
    \tilde{P}^c_{12}(k, Q)
    = \tilde{G}_{12}(k+Q) \delta_{c_2 c}
    - \tilde{G}_{12}(k) \delta_{c_1 c}
    \,.
\end{equation}
Note that $\tilde{G}(k) = G(k)$.
The single-particle Ward identity \Eq{eq:ward_1band_G_ward}] becomes
\begin{equation}
    (\mb{\mc{D}} G)_{12}(k)
    = - \sum_{c} Z^{c} \, \delta^{c}_{\mb{J}} G_{12}(k, \Omega=0) \,,
\end{equation}

The only place where the derivation needs to be modified is \Eqs{eq:ward_bubble_final} and \eqref{eq:ward_bubble_final2}.
With the gauge-fixed quantities, \Eqs{eq:ward_bubble1} and \eqref{eq:ward_bubble2} read
\begin{multline}
    \sum_{3'} \tilde{\Pi}_{123'4}(k, Q) (\veps + \Omega - \tilde{\veps}_{n_{3'} n_3 \bkQ}) Z^{c_{3'} c_3}
    \\
    = \delta_{13} \tilde{G}_{42}(k)
    + \sum_5 \tilde{\Pi}_{1254}(k, Q) \tilde{\Sigma}_{53}(k+Q) \,,
\end{multline}
and
\begin{multline}
    \sum_{4'} (\veps - \veps_{n_4 n_{4'} \bk}) Z^{c_4 c_{4'}} \tilde{\Pi}_{1234'}(k, Q)
    \\
    = \tilde{G}_{13}(k+Q) \delta_{42}
    + \sum_5 \tilde{\Pi}_{1235}(k, Q) \tilde{\Sigma}_{45}(k) \,.
\end{multline}
We subtract these equations, set $c_3 = c_4 = c$, $n_3 = n_4 = n$, and sum over $n$ to obtain
\begin{align} \label{eq:ward_multiband_bse_1}
    & Z^c \sum_{n n'} \tilde{\Pi}_{n_1 n_2 n n'}^{c_1 c_2 cc} (k, Q)
    (\tilde{\veps}_{n n' \bkQ} - \veps_{n n' \bk} - \Omega)
    \nnnl
    &= \tilde{G}_{12}(k+Q) \delta_{c_2 c} - \tilde{G}_{12}(k) \delta_{c_1 c}
    \nnnl
    & + \sum_{n n' c'} \Bigl[ \tilde{\Pi}_{n_1 n_2 n n'}^{c_1 c_2 c c'}(k, Q) \tilde{\Sigma}_{n n'}^{c c'}(k)
    \nnnl
    &\hspace{4em} - \tilde{\Pi}_{n_1 n_2 n' n}^{c_1 c_2 c' c}(k, Q) \tilde{\Sigma}_{n' n}^{c' c}(k+Q) \Bigr]
    \nnnl
    &= \tilde{P}^c_{12}(k, Q)
    - \sum_{56} \tilde{\Pi}_{1256}(k, Q) \bigl( \tilde{\Sigma}_{56}(k+Q) \delta_{c_6 c}
    \nnnl
    &\hspace{14em} - \tilde{\Sigma}_{56}(k) \delta_{c_5 c} \bigr)
    \,.
\end{align}
The first line of this equation can be rewritten as
\begin{multline} \label{eq:ward_multiband_bse_2}
    Z^c \sum_{n n'} \tilde{\Pi}_{n_1 n_2 n n'}^{c_1 c_2 cc} (k, Q)
    (\tilde{\veps}_{n n' \bkQ} - \veps_{n n' \bk} - \Omega)
    \\
    =
    i Z^{c} \bigl[ \tilde{\Pi} \circ (\delta^{c}_{\Delta^{\rm e}} \tilde{\Sigma}_{0}
    - \Omega \, \delta^{c}_N \tilde{\Sigma}_{0} ) \bigr]_{12}(k, Q) \,,
\end{multline}
By substituting \Eq{eq:ward_multiband_bse_2} into \Eq{eq:ward_multiband_bse_1}, we obtain the multiband version of \Eq{eq:ward_bubble_final2}.

\subsection{Proof of the asymptotic limit of the cumulant function} \label{sec:supp_cumulant_proof}

Here, we prove \Eq{eq:cumulant_C_limit}, the long-time limit of the cumulant function.
First, using
\begin{align}
    \lim_{t \to \infty} \frac{e^{-i\veps t} - 1}{\veps}
    &= -i \lim_{t \to \infty} \nbint{0}{t} e^{-i\veps t'} \dd t'
    \nnnl
    &= -i \lim_{t \to \infty} \lim_{\eta \to 0^+} \nbint{0}{t} e^{-i\veps t'} e^{-\eta t'} \dd t'
    \nnnl
    &= -i \lim_{\eta \to 0^+} \nbint{0}{\infty} e^{-i(\veps - i\eta) t'} \dd t'
    \nnnl
    &= - \lim_{\eta \to 0^+} \frac{1}{\veps - i\eta}
    \nnnl
    &= -\mc{P} \frac{1}{\veps} - i\pi\delta(\veps) \,,
\end{align}
we find
\begin{align}
    &\quad \lim_{t \to \infty} \nbint{-\infty}{\infty} \dd\veps \, \beta(\veps + \veps_\nk) \frac{e^{-i\veps t} - 1}{\veps}
    \nnnl
    &= -\mc{P} \nbint{-\infty}{\infty} \dd\veps \, \frac{\beta(\veps + \veps_\nk)}{\veps} -i \pi \beta(\veps_\nk)
    \nnnl
    &= \Sigma(\veps_\nk) \,.
\end{align}
The principal value term gives the real part of the self-energy via the Kramers--Kronig relation [\Eq{eq:cumulant_KK}], and the delta function term gives the imaginary part.

Then, for the cumulant function defined in \Eq{eq:cumulant_C_beta}, we first subtract the linear part and take the long-time limit to find
\begin{align}
    &\quad \lim_{t \to \infty} \bigl[ C^{\rm R}(t) + i \Sigma(\veps_\nk) t \bigr]
    \nnnl
    &= \nbint{-\infty}{\infty} \dd\veps \, \beta(\veps + \veps_\nk) 
    \Bigl[ \frac{e^{-i\veps t} + i\veps t - 1}{\veps^2} 
    + it \frac{e^{-i\veps t} - 1}{\veps}
    \Bigr]
    \nnnl
    &= \nbint{-\infty}{\infty} \dd\veps \, \beta(\veps + \veps_\nk) 
    \frac{(1 + i \veps t) e^{-i\veps t} - 1}{\veps^2}
    \nnnl
    &= - \nbint{-\infty}{\infty} \dd\veps \, \beta(\veps + \veps_\nk) 
    \frac{\dd}{\dd\veps} \Bigl( \frac{e^{-i\veps t} - 1}{\veps} \Bigr) \,.
\end{align}
Integrating by parts, we find
\begin{align}
    &\quad \lim_{t \to \infty} \bigl[ C^{\rm R}(t) + i \Sigma(\veps_\nk) t \bigr]
    \nnnl
    &= \nbint{-\infty}{\infty} \dd\veps \, \frac{\dd \beta(\veps + \veps_\nk)}{\dd \veps}
    \frac{e^{-i\veps t} - 1}{\veps}
    \nnnl
    &= \frac{\dd}{\dd \veps_\nk} \nbint{-\infty}{\infty} \dd\veps \, \beta(\veps + \veps_\nk)
    \frac{e^{-i\veps t} - 1}{\veps}
    \nnnl
    &= \frac{\dd \Sigma(\veps) }{\dd \veps} \biggr|_{\veps = \veps_\nk} \,.
\end{align}
This concludes the proof of \Eq{eq:cumulant_C_limit}.

\ifdefined\myroot\else
\FloatBarrier

\end{document}
\fi

\end{document}